\documentclass[12pt]{article}

\usepackage[latin1]{inputenc}
\usepackage[affil-it]{authblk}
\usepackage{amsmath,amsfonts,amssymb,bm,bbm} 
\usepackage{pifont}
\usepackage{graphicx}
\usepackage{titlesec}
\usepackage[table]{xcolor}
\usepackage{multirow}
\usepackage{booktabs}
\usepackage[labelfont={small,bf},textfont={small},labelsep=space,tableposition=top]{caption}
\usepackage{subcaption}
\usepackage[ruled,linesnumbered,vlined]{algorithm2e}
\usepackage[frozencache=true,cachedir=_minted-main]{minted}
\usepackage[numbers]{natbib}
\usepackage[pagebackref=true,breaklinks=true,colorlinks,bookmarks=false]{hyperref}

\usepackage{pgf,tikz,pgfplots,calc}
\usetikzlibrary[calc]
\usetikzlibrary{positioning}
\pgfplotsset{compat=1.15}
\usepackage{mathrsfs}
\usepackage{relsize}

\usepackage{jour} 

\title{\Large\bf Factorizer: A Scalable Interpretable Approach to Context Modeling for Medical Image Segmentation}
\author[a,d]{\normalsize Pooya Ashtari \footnote{Corresponding author. Email: \url{pooya.ashtari@esat.kuleuven.be}}}
\author[b]{\normalsize Diana M. Sima}
\author[c]{\normalsize Lieven De Lathauwer}
\author[d]{\normalsize Dominique Sappey-Marinier}
\author[e]{\normalsize Frederik Maes}
\author[a]{\normalsize Sabine Van Huffel}

\affil[a]{\footnotesize Department of Electrical Engineering (ESAT), STADIUS Center, KU Leuven, Leuven, Belgium}
\affil[b]{\footnotesize icometrix, Research and Development, Leuven, Belgium}
\affil[c]{\footnotesize Group Science, Engineering and Technology, KU Leuven Kulak, Kortrijk, Belgium}
\affil[d]{CREATIS (CNRS UMR5220 \& INSERM U1294), Université Claude-Bernard Lyon 1, Lyon, France}
\affil[e]{\footnotesize Department of Electrical Engineering (ESAT), PSI, KU Leuven, Leuven, Belgium}

\date{}

\begin{document}
\maketitle

\begin{abstract}
Convolutional Neural Networks (CNNs) with U-shaped architectures have dominated medical image segmentation, which is crucial for various clinical purposes. However, the inherent locality of convolution makes CNNs fail to fully exploit global context, essential for better recognition of some structures, e.g., brain lesions. Transformers have recently proven promising performance on vision tasks, including semantic segmentation, mainly due to their capability of modeling long-range dependencies. Nevertheless, the quadratic complexity of attention makes existing Transformer-based models use self-attention layers only after somehow reducing the image resolution, which limits the ability to capture global contexts present at higher resolutions. Therefore, this work introduces a family of models, dubbed Factorizer, which leverages the power of low-rank matrix factorization for constructing an end-to-end segmentation model. Specifically, we propose a linearly scalable approach to context modeling, formulating Nonnegative Matrix Factorization (NMF) as a differentiable layer integrated into a U-shaped architecture. The shifted window technique is also utilized in combination with NMF to effectively aggregate local information. Factorizers compete favorably with CNNs and Transformers in terms of accuracy, scalability, and interpretability, achieving state-of-the-art results on the BraTS dataset for brain tumor segmentation and ISLES'22 dataset for stroke lesion segmentation. Highly meaningful NMF components give an additional interpretability advantage to Factorizers over CNNs and Transformers. Moreover, our ablation studies reveal a distinctive feature of Factorizers that enables a significant speed-up in inference for a trained Factorizer without any extra steps and without sacrificing much accuracy. The code and models are publicly available at \href{https://github.com/pashtari/factorizer}{https://github.com/pashtari/factorizer}.
\end{abstract}


\section{Introduction} \label{sec: introduction}

Medical image segmentation is an essential prerequisite for the analysis of anatomical structures for various clinical purposes, including diagnosis and treatment planning. In recent years, the vast majority of effective segmentation models are based on Convolutional Neural Networks (CNNs), particularly, U-Net \citep{ronneberger2015u}, consisting of encoder and decoder parts with skip connections in between. In a typical U-Net \citep{ronneberger2015u, cciccek20163d}, the encoder learns a low-resolution contextual representation consisting of progressively downsampled feature maps, while the decoder progressively upsamples the low-resolution feature maps to propagate contextual information to the higher-resolution layers. Moreover, skip connections between encoder and decoder layers of equal resolution help to recover spatial information lost during downsampling.

Currently, U-Net models mostly rely on convolution operations with small receptive fields, which are capable of exploiting only local context at each resolution. Hence, they generally fail to effectively model long-range spatial dependencies, often necessary, for example, for better recognition of brain lesions, which can be very infiltrative, extensive, and thus dramatically vary in shape and size. Moreover, capturing even small focal tumors within the receptive field is extremely difficult without any notion about the global context of normal brain anatomy since such tumors can occur anywhere in the brain. Several works \citep{chen2017deeplab, gu2019net} have employed dilated convolution for expanding the receptive fields. Nevertheless, the learning capabilities of convolutional layers are still limited due to their inherent locality. As a solution, integrating self-attention modules into CNNs \citep{wang2018non, fu2019dual} has been proposed to enhance the capability of modeling non-local context.

Transformers \cite{vaswani2017attention} have achieved state-of-the-art performance on various natural language processing tasks. The attention mechanism enables Transformers to effectively model the pairwise interactions between the words in a sentence. Recently, Transformer-based models have been applied to vision tasks and demonstrated promising results. Specifically, Vision Transformer (ViT) \citep{dosovitskiy2020image} outperformed state-of-the-art CNNs on image recognition by large-scale pre-training and fine-tuning a pure Transformer. Unlike CNNs, ViT encodes images as a sequence of 1D patch embeddings (known as tokens) and dynamically highlights the important tokens using self-attention layers, which in turn increases the capability of learning long-range dependencies. Due to lack of locality inductive bias, ViT is data-hungry and generally requires a larger dataset to perform as effectively as its CNN counterparts, leading to poor performance when trained on insufficient data, which is usually the case in medical imaging. Furthermore, the quadratic complexity of self-attention makes Transformers computationally intractable on long sequences of patches. Therefore, existing models use self-attention layers only after somehow reducing the image resolution, thereby failing to fully exploit the global context at the higher resolution.  

This work proposes a family of architectures, dubbed as Factorizer, which leverages the power of low-rank matrix approximation (LRMA) to construct an end-to-end medical image segmentation model. Among LRMA methods, Nonnegative Matrix Factorization (NMF) has demonstrated a remarkable ability to compress data and automatically extract easy-to-interpret sparse factors \citep{gillis2014and, gillis2020nonnegative}. Hence, we propose a linearly scalable alternative to self-attention by formulating an NMF algorithm as a differentiable layer. Moreover, a series of matricization operations is introduced, which enables NMF to effectively exploit both global and local contexts. The Factorizer block is constructed by replacing the self-attention layer of a ViT block with our NMF-based modules and then integrated into a U-shaped architecture with skip connections.

We evaluated the effectiveness of our approach for the segmentation of brain tumors and stroke lesions in MRI data. Factorizers achieved competitive results on the BraTS \citep{brats1, brats2} and ISLES'22 \citep{petzsche2022isles} datasets, having outperformed state-of-the-art methods based on CNN and Transformer. Our experiments showed that NMF components are highly meaningful, which gives a great advantage to Factorizers over CNNs and Transformers in terms of interpretability. Furthermore, our ablation studies revealed a distinctive interesting feature of Factorizers that enables us to easily speed up the inference for a trained Factorizer model with no extra steps and without sacrificing much accuracy.

\paragraph{Contributions} The main contributions of this work are as follows:
\begin{itemize}
	\item To the best of our knowledge, this work presents the first end-to-end deep model with matrix factorization layers for medical image segmentation.
	\item A differentiable NMF layer is constructed using a block coordinate descent solver to efficiently model contextual information.
	\item Shifted Window (SW) Matricize operation is introduced and combined with NMF to fully exploit local contexts.
	\item Scalable interpretable U-shaped segmentation models based on NMF are proposed.
	\item The proposed models achieve state-of-the-art results on the BraTS and ISLES'22 datasets.
\end{itemize}

\paragraph{Notation} 
We denote vectors by boldface lower-case letters, e.g., $ \mathbf{x} $, matrices by boldface upper-case letters, e.g., $ \mathbf{X} $, and tensors by boldface calligraphic letters, e.g., $ \bm{\mathcal{X}} $. Elements in a matrix (tensor) are denoted by $ \mathbf{X}[i,j] $ ($ \bm{\mathcal{X}}[i_1, \dots, i_N] $). The $ i $th row and $ j $th column of a matrix is denoted by $ \mathbf{X}[i,:] $ and $ \mathbf{X}[:,j] $, respectively. A sequence of $ N $ vectors (a.k.a. \textit{tokens}) is denoted by $ (\mathbf{x}_n)_{n=1}^{N} $. We use $ [\mathbf{x}_1|\dots|\mathbf{x}_N] $ to denote a matrix $ \mathbf{X} $ created by stacking $ \mathbf{x}_i $s along the columns. We show the inner product between matrices by $ \langle \mathbf{X}, \mathbf{Y} \rangle = \sum_{i,j} \mathbf{X}[i,j] \mathbf{Y}[i,j] $ and $ \ell _{2} $-norm of a matrix by $ \| \mathbf{X} \| = \sqrt{\langle \mathbf{X}, \mathbf{X}  \rangle}$.

\section{Related Work} \label{sec: background}

\paragraph{CNN-based Segmentation Models}
Convolutional neural networks (CNNs) have dominated medical image segmentation. Particularly, following an encoder-decoder architecture with skip connections, U-Net \citep{ronneberger2015u} has achieved state-of-the-art on various medical image datasets. The simplicity and effectiveness of a U-shaped architecture have led to the emergence of numerous U-Net variants in the field. \citet{cciccek20163d} extended U-Net by replacing all 2D operations with their 3D counterparts. UNet++ \citep{zhou2018unet++} follows a deeply-supervised encoder-decoder network consisting of sub-networks connected through a series of nested, dense skip connections. nnU-Net \citep{isensee2021nnu} proved effective in various medical image segmentation tasks by only making minor modifications to the standard 3D U-Net \citep{cciccek20163d} and defining a recipe to automatically configure key design choices. \citet{myronenko20183d} proposed a U-Net-like architecture with ResNet blocks, a.k.a. ResSegNet, which ranked first in the Brain Tumor Segmentation Challenge (BraTS) 2018. \citet{ashtari2020low} proposed a lightweight CNN for glioma segmentation, with low-rank constraints being imposed on the kernel weights of the convolutional layers in order to reduce overfitting.  

Despite their success, these networks generally fail to effectively model long-range spatial dependencies, often necessary for better recognition of some region semantics such as tumors, since they rely on convolution operations with small kernel sizes, aggregating only local information in an image.

\paragraph{Visual Transformers}
Transformers with attention mechanisms \citep{vaswani2017attention}, introduced originally for language modeling, have recently proven promising on computer vision tasks. Particularly, the pioneering Vision Transformer (ViT) model \citep{dosovitskiy2020image} outperformed state-of-the-art CNNs on image recognition by large-scale pre-training and fine-tuning a pure Transformer applied to sequences of image patches. In contrast to CNNs, ViT lacks any inductive bias such as locality, and therefore, generally shows poorer performance than its CNN counterparts (e.g., ResNets \citep{he2016deep}) when trained from scratch on small-size or mid-size datasets, which is usually the case for medical imaging. 

Efforts have been made to mitigate this limitation. For example, Tokens-to-Token ViT (T2T-ViT) \citep{yuan2021tokens} introduces a hierarchical architecture to ViT by progressively combining neighboring tokens into a single token to reduce the sequence length and aggregate local context. \citet{liu2021swin} proposed a hierarchical Transformer, called Swin Transformer, adopting the shifted windowing scheme, which brings more efficiency by limiting self-attention computation to non-overlapping local windows while also allowing for cross-window connection. As another exemplification of a hierarchical Transformer, Pyramid vision Transformer (PvT) \citep{wang2021pyramid} significantly reduces computational and memory overhead by reducing the sequence length at each stage through non-overlapping patch embedding and learning low-resolution key-value pairs via spatial-reduction attention (SRA) in each block. Convolutional vision Transformer (CvT) \citep{wu2021cvt} incorporates depthwise convolutions into self-attention layers and uses strided convolution for simultaneously tokenizing and downsampling the image, exploiting the excellent capability of convolution at capturing low-level local features.

Transformer-based methods were recently proposed to deal with the task of 2D image segmentation. Segmentation Transformer (SETR) \citep{zheng2021rethinking} uses a ViT encoder and a decoder with progressive upsampling (which alternates Conv layers and upsampling operations) and multi-level feature Aggregation. SegFormer \citep{xie2021segformer} consists of a PvT-based encoder and a lightweight Multilayer Perceptron (MLP) decoder with upsampling operations. \citet{chen2021transunet} proposed a model for multi-organ segmentation by incorporating ViT into the bridge of a 2D convolutional U-Net architecture. \citet{zhang2021transfuse} proposed to combine a shallow CNN with a Transformer in a parallel style. \citet{valanarasu2021medical} proposed a Transformer model with an axial attention mechanism for the segmentation of 2D medical images. \citet{cao2021swin} proposed a U-shaped architecture based purely on Swin Transformer.

For 3D medical image segmentation, \citet{xie2021cotr} proposed a model comprising a CNN backbone to extract features, a Transformer to model long-range dependencies, and a CNN decoder to construct the segmentation map. More recently, \citet{hatamizadeh2022unetr} proposed UNETR, which utilizes ViT as the main encoder but directly connects it to the convolutional decoder via skip connections, as opposed to using a Transformer only in the bridge. nnFormer \citep{zhou2021nnformer} uses an initial convolutional tokenizer and interleaves local and global self-attention blocks with convolutional downsamplers. Since self-attention is prohibitively expensive on long sequences, all these models apply Transformer on a low-resolution stage after either patch embedding or a CNN backbone, making them fail to fully exploit the global context at the higher resolutions. In contrast, our proposed approach based on NMF offers a scalable alternative to the attention mechanism, which enables the exploitation of the global context at the highest-resolution stage of a 3D network.

\paragraph{Matrix Factorization Models} 
In the context of machine learning, low-rank matrix factorization (MF) methods have proven extremely useful for representation learning, dimensionality reduction, and collaborative filtering. NMF has been used for unsupervised and semi-automated segmentation of brain tumors on multiparametric MRI data \citep{sauwen2016comparison, sauwen2017semi}. However, only a few works have incorporated MF into an end-to-end deep model to perform a computer vision task. Most notably, \citet{geng2021attention} proposed a framework, called Hamburger, where the global context is modeled as solving a low-rank matrix completion problem by suitable optimization algorithms that guide the design of layers able to capture global information. They demonstrated the effectiveness of a Hamburger model based on NMF with a multiplicative update (MU) solver for semantic segmentation. Our approach to context modeling is based on NMF with a Hierarchical Alternating Least Squares (HALS) solver and introduces a series of matricization operations which enable NMF to effectively exploit both global and local contexts. Moreover, our proposed block is inspired by the overall design of the ViT block and incorporated into a U-shaped architecture.

\section{Method} \label{sec: method}

\subsection{Matrix Factorization for Context Modeling} \label{sec: motivation}

Here we provide the motivation behind incorporating matrix factorization into deep learning by first presenting an alternative view of the attention mechanism and then showing how it relates to the matrix factorization approach to modeling contextual information.

\paragraph{Revisiting Attention Mechanism}
The attention mechanism is the key component that enables Transformers to model complex dependencies between the elements of a sequence. Consider an input sequence of $C$-dimensional tokens $ (\mathbf{x}_n)_{n=1}^{N} $, stacked into the rows of matrix $ \mathbf{X} = [\mathbf{x}_1|\dots|\mathbf{x}_N]^T $. In a self-attention layer, the input is first projected onto three learnable weight matrices $ \mathbf{W}^Q, \mathbf{W}^K, \bm{W}^V \in \mathbb{R}^{C \times E} $ to get three different matrices: \textit{Query} $ \mathbf{Q} = [\mathbf{q}_1|\dots|\mathbf{q}_N]^T = \mathbf{X} \mathbf{W}^Q $, \textit{Key}  $ \mathbf{K} = [\mathbf{k}_1|\dots|\mathbf{k}_N]^T = \mathbf{X} \mathbf{W}^K $, and \textit{Value} $ \mathbf{V} = [\mathbf{v}_1|\dots|\mathbf{v}_N]^T = \mathbf{X} \bm{W}^V $. The output is then defined by
\begin{equation}
	\label{eq: attention}
	A(\bm{Q}, \bm{K}, \bm{V}) = \text{Softmax}\bigg( \frac{\bm{Q} \bm{K}^{T} }{\sqrt{E}} \bigg) \bm{V},
\end{equation}
where Softmax is taken row-wise. Note that attention takes $ \mathcal{O}(N^2 E) $ time and needs $ \mathcal{O}(N^2) $ memory to store the attention map, scaling quadratically with the sequence length, which is prohibitively expensive for large inputs, such as high-resolution or 3D images. 

Taking a closer look at equation \eqref{eq: attention}, we notice that attention can be viewed as a special case of nonparametric regression. To reveal this, let's consider key-value pairs $ \{ (\mathbf{k}_n, \mathbf{v}_n) \}_{n=1}^{N} $ as a training set and queries $ \{ \mathbf{q}_n \}_{n=1}^{N} $ as a test set; where $ \mathbf{k}_n $s and $ \mathbf{q}_n $s are feature vectors, and $ \mathbf{v}_n $s are vectors of target variables. The predictions for the queries using a Nadaraya-Watson kernel regression \citep{hardle2004nonparametric} are given by
\begin{equation}
	\label{eq: kernel regression}
	\hat{f}(\mathbf{q}_n) = \frac{\sum_{m=1}^{N} \varphi(\mathbf{q}_n, \mathbf{k}_m) \mathbf{v}_n}{\sum_{m=1}^{N} \varphi(\mathbf{q}_n, \mathbf{k}_m)},
\end{equation}
where kernel $ \varphi(.,.) $ is, in general, a similarity measure. Setting $ \varphi(\mathbf{a}, \mathbf{b}) = \exp(\mathbf{a}^T \mathbf{b}/\sqrt{E}) $ simply yields the attention formula. 

\paragraph{Low-Rank Matrix Factorization}
Formulating attention as a kernel regression with softmax kernel suggests that we can probably use other regression methods on queries, keys, and values to model interactions between the sequence elements. Notably, one can take a matrix completion approach to regress on the queries via low-rank approximation of a block matrix of queries, keys, and values with missing entries, that is
\begin{equation}
	\label{eq: matrix completion}
	\left(\begin{array}{c c}
		\mathbf{K} & \mathbf{V} \\
		\mathbf{Q} & ? \\
	\end{array}\right) \approx 	
	\left(\begin{array}{c}
		\mathbf{B} \\
		\mathbf{F} \\
	\end{array}\right)
	\left(\begin{array}{c}
		\mathbf{G} \\
		\mathbf{H} \\
	\end{array}\right)^T,
\end{equation}
where $ \mathbf{B}, \mathbf{F} \in \mathbb{R}^{N \times R} $ and $ \mathbf{G}, \mathbf{H} \in \mathbb{R}^{E \times R} $; and $ R \leq \min(N, E) $ is the rank. The reconstructed matrix $ \mathbf{F}\mathbf{H}^T $ then gives an estimation of the missing block as the output. This can be viewed as a joint matrix factorization of $ \mathbf{Q} $, $ \mathbf{K} $, and $ \mathbf{V} $, i.e., $ \mathbf{K} \approx \mathbf{B}\mathbf{G}^T, \ \mathbf{V} \approx \mathbf{B}\mathbf{H}^T, \ \mathbf{Q} \approx \mathbf{F}\mathbf{G}^T $.

In this work, we further simplify the procedure by considering only a single linear map to generate a single matrix $ \mathbf{Z} = \mathbf{X} \bm{W} $, where $ \bm{W} \in \mathbb{R}^{C \times E} $ is a learnable weight matrix, and using a regular matrix factorization rather than a joint one, that is $ \mathbf{Z} \approx \mathbf{F}\mathbf{G}^T $, then the output is the reconstructed matrix $ \mathbf{F}\mathbf{G}^T $. Note that this is an unsupervised scheme, unlike the joint factorization, which forms a regression model. Depending on the matrix factorization algorithm, a suitable activation function may be applied before the factorization to constrain the input. Particularly, in the case of NMF, ReLU of the matrix must be first taken to make all of the entries nonnegative. As we will see in Section \ref{sec: experiments}, our results suggest that NMF can potentially be an efficient yet effective alternative to attention-based context modeling for medical image segmentation.

\subsection{Overall Architecture} \label{sec: architecture}

\begin{figure}[t]
	\centering
	\includegraphics[width=.9\linewidth]{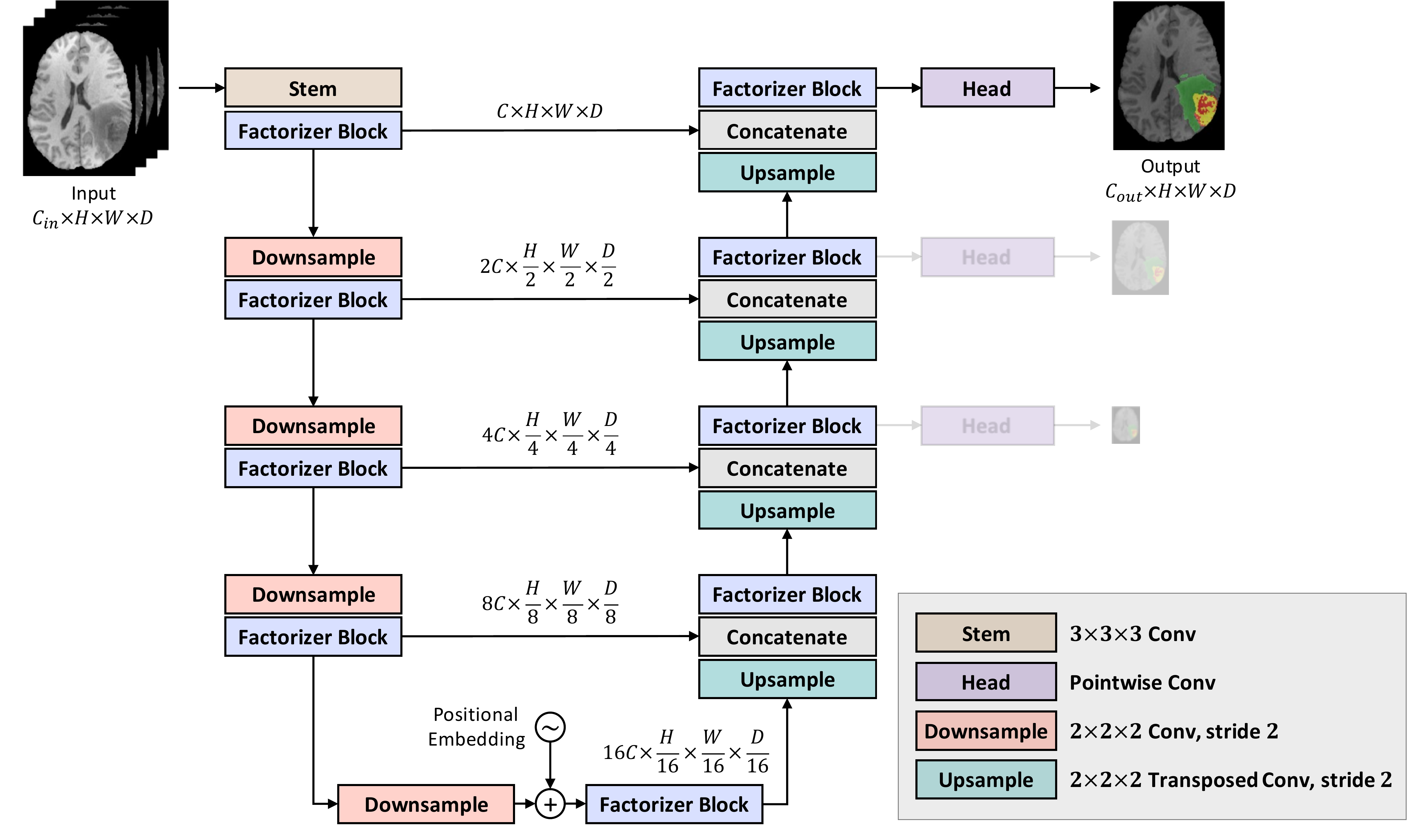}	
	\caption{The overall architecture of Factorizer.}
	\label{fig: factorizer}
\end{figure}

As shown in Figure \ref{fig: factorizer}, a Factorizer model follows a U-Net-style architecture consisting of encoder and decoder parts with skip connections in between at equal resolutions. Given an input image $ \bm{\mathcal{X}} \in \mathbb{R}^{C_{in} \times H \times W \times D} $ with $ C_{in} $ channels and resolution $ (H, W, D) $, the network outputs a \textit{logit map} of size $ (C_{out}, H, W, D) $, where $ C_{out} $ is the number of foreground classes. A single 3D convolution with a kernel size of $ (3, 3, 3) $ is used as the stem to increase the number of channels to $ C = 32 $. However, note that in contrast to ViT, Factorizer does not flatten the spatial dimensions at the initial stage to generate a sequence of tokens.

The network has four stages, with the resolution decreasing to $ 1/16 $ in the bridge. At each stage of the encoder (decoder), the input tensor is downsampled (upsampled) by a factor of two while the number of channels is doubled (halved). Convolution (transposed convolution) with a kernel size of $ (2, 2, 2) $ and stride of 2 is used for downsampling (upsampling). In the bridge, learnable position embeddings are added to the input right after downsampling. We use deep supervision \citep{lee2015deeply} at the three highest resolutions in the decoder, applying pointwise convolutions (i.e., convolution with a kernel size of $ (1, 1, 1) $) to get the output and two auxiliary low-resolution logit tensors.

\subsection{Factorizer Block} \label{sec: factorizer block}

\begin{figure}[t]
	\centering
	\begin{subfigure}{.16\textwidth}
		\centering
		\includegraphics[width=.85\textwidth]{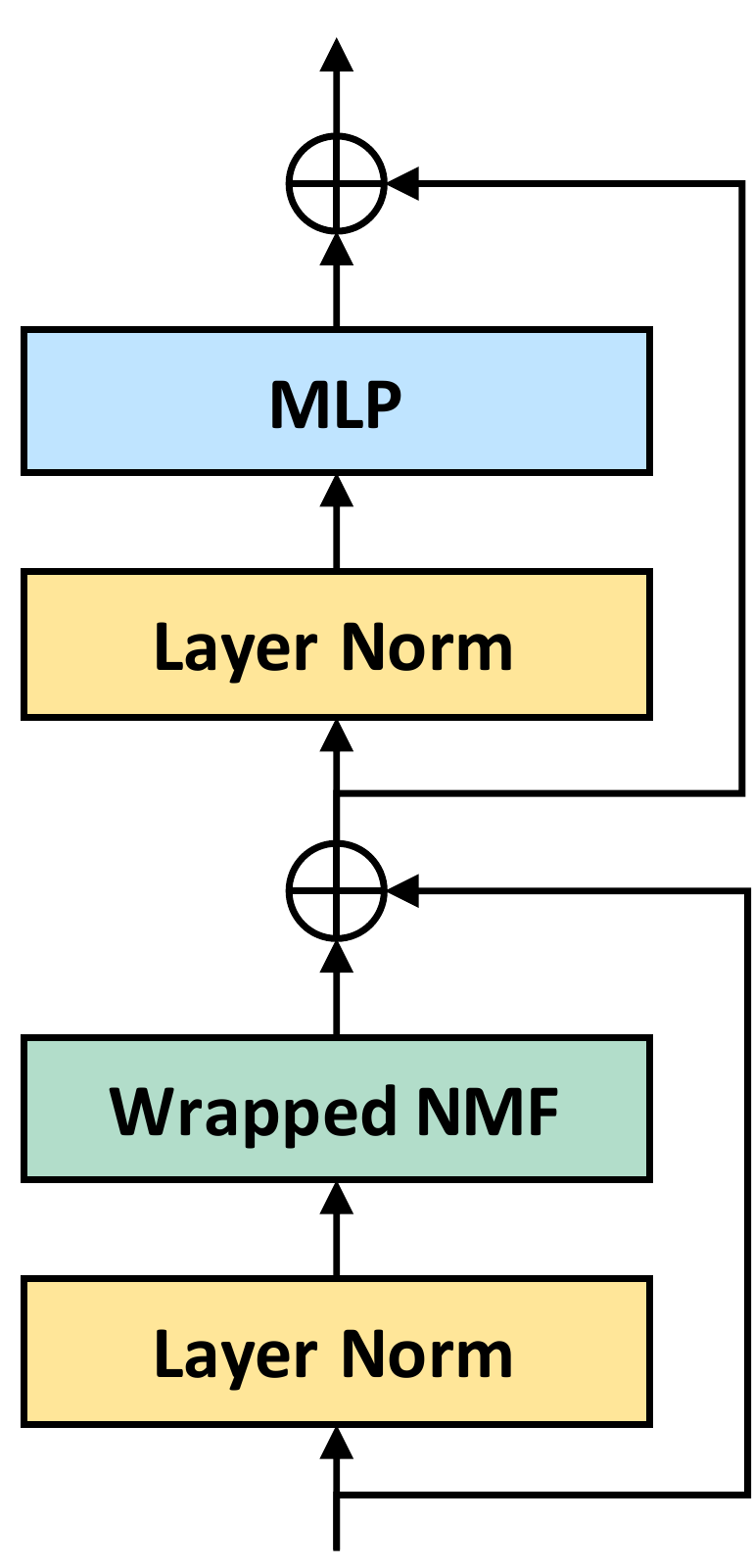}
		\caption{}
		\label{fig: factorizer block}
	\end{subfigure}
	\begin{subfigure}{.15\textwidth}
		\centering
		\includegraphics[width=.85\textwidth]{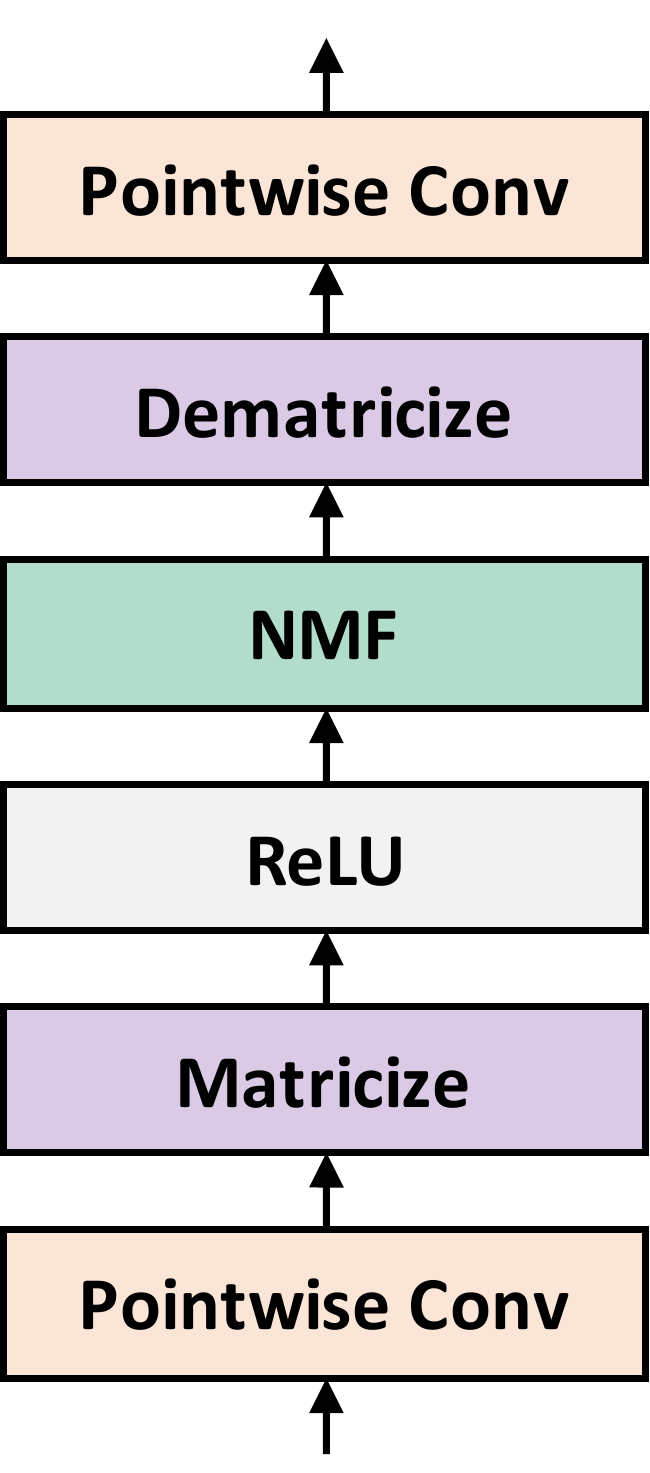}
		\caption{}
		\label{fig: wrapped nmf}
	\end{subfigure}
	\begin{subfigure}{.67\textwidth}
		\centering
		\includegraphics[width=.97\textwidth]{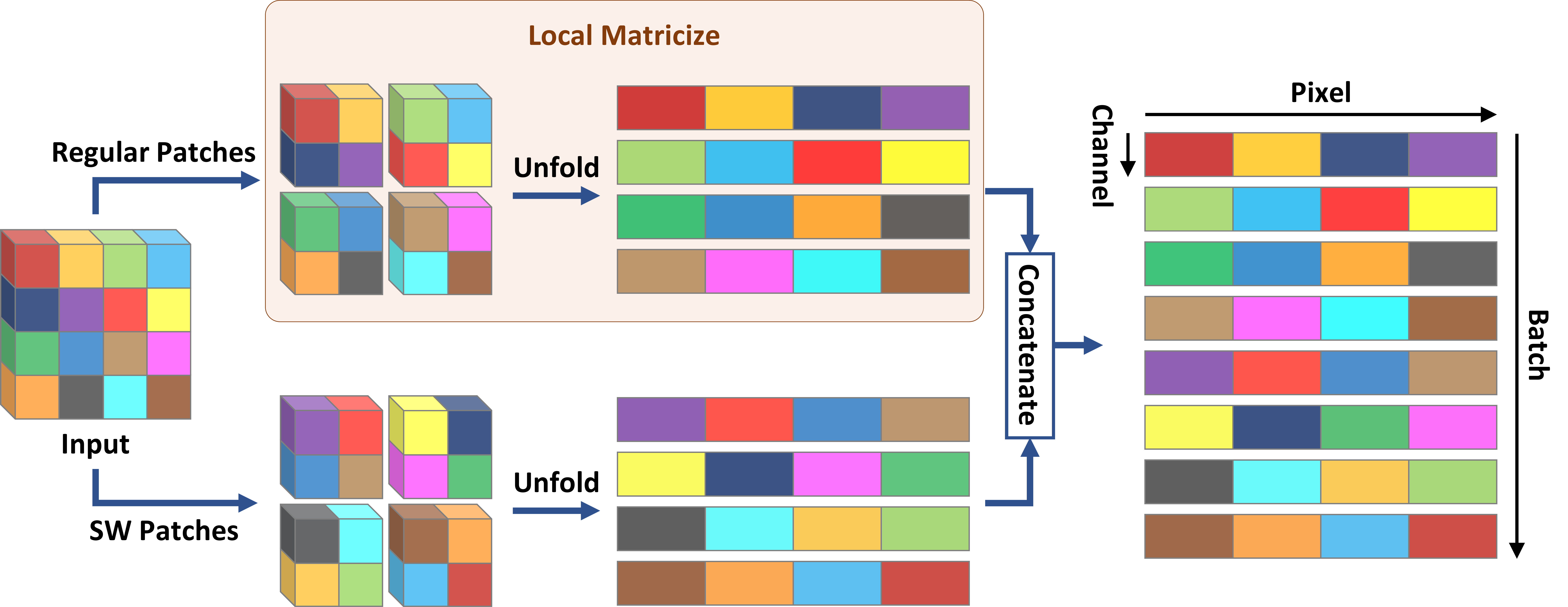}
		\caption{}
		\label{fig: sw-matricize}
	\end{subfigure}
	\caption{An overview of the Factorizer block and its components. (a) overall Factorizer block (b) Wrapped NMF module (c) An illustration of SW Matricize on a 2D toy example.}
	\label{fig: factorizer block and its components}
\end{figure}

A Factorizer block is constructed by replacing the multi-head self-attention module in a ViT block \citep{dosovitskiy2020image} with a Wrapped NMF module (described in Section \ref{sec: wrapped nmf}). As shown in Figure \ref{fig: factorizer block}, a Factorizer block comprises NMF module and MLP, each of which comes after Layer Normalization and before a residual connection, that is,
\begin{align}
	\label{eq: factorizer block}
	& \bm{\mathcal{Y}} = \text{WrappedNMF}(\text{LayerNorm}(\bm{\mathcal{X}})) + \bm{\mathcal{X}}, \nonumber \\
	& \bm{\mathcal{Z}} = \text{MLP}(\text{LayerNorm}(\bm{\mathcal{Y}})) + \bm{\mathcal{Y}},
\end{align}
where MLP has two linear layers with a Gaussian Error Linear Unit (GELU) nonlinearity in between:
\begin{equation}
	\label{eq: MLP}
	\text{MLP}(\bm{\mathcal{X}}) = \text{PointwiseConv}(\text{GELU}(\text{PointwiseConv}(\bm{\mathcal{X}}))).
\end{equation}
The number of input and output channels are the same, but the number of inner channels is double that of input channels.

\subsection{Wrapped NMF Module} \label{sec: wrapped nmf}

The major component of Factorizer is the Wrapped NMF module, which relies on matricization (i.e., an operation that turns a tensor into a matrix) and NMF. As shown in Figure \ref{fig: wrapped nmf}, a Wrapped NMF subblock first applies a pointwise convolution to linearly project each voxel. The output is then reshaped into a batch of matrices using a \textit{Matricize} operation, which is detailed later on in Section \ref{sec: matricize}. The resulting matrices are passed through ReLU to clamp all their elements into nonnegative values and then low-rank approximated using NMF. The reconstructed matrices are reshaped back to their original size using the \textit{Dematricize} operation, i.e., the inverse of \textit{Matricize}. Finally, another pointwise convolution is applied, yielding the output. More formally, Wrapped NMF can be computed as follows:
\begin{align}
	\label{eq: wrapped nmf}
	& \bm{\mathcal{X}}^1 = \text{PointwiseConv}(\bm{\mathcal{X}}), \nonumber \\
	& \bm{\mathcal{X}}^2 = \text{Matricize}(\bm{\mathcal{X}}^1), \nonumber \\
	& \bm{\mathcal{X}}^3 = \text{NMF}(\text{ReLU}(\bm{\mathcal{X}}^2)), \nonumber \\
	& \bm{\mathcal{X}}^4 = \text{Dematricize}(\bm{\mathcal{X}}^3), \nonumber \\
	& \text{WrappedNMF}(\bm{\mathcal{X}}) = \text{PointwiseConv}(\bm{\mathcal{X}}^4),
\end{align}
where $ \text{NMF}(\cdot) $ is the NMF layer, described in \ref{sec: nmf}. Intermediate tensors $ \bm{\mathcal{X}}^i $s and the output have the same size as the input $ \bm{\mathcal{X}} $.

\subsubsection{Matricize} \label{sec: matricize}

Before applying any matrix factorization method, an input batch of multi-dimensional images, denoted by $ \bm{\mathcal{X}} \in \mathbb{R}^{B \times C \times H \times W \times D} $, must be turned into a batch of matrices, say $ \bm{\mathcal{Z}} \in \mathbb{R}^{B' \times M \times N} $. This operation is called \textit{Matricize}. In this work, we propose three Matricize operations: i) Global Matricize, ii) Local Matricize, and iii) Shifted Window (SW) Matricize. Depending on which operation is used, different variants for Factorizer are obtained: i) Global Factorizer, ii) Local Factorizer, and iii) \textbf{S}hifted \textbf{Win}dow (Swin) Factorizer. The Matricize operations are described in the following.

\paragraph{Global Matricize}
This operation simply flattens the spatial dimensions and divides the channels into multiple groups (analogous to heads of Multi-Head Self-Attention). More specifically, Global Matricize reshapes a batch of $ C $-channel 3D images $ \bm{\mathcal{X}} \in \mathbb{R}^{B \times C \times H \times W \times D} $ into a batch of matrices denoted by a 3D tensor $ \bm{\mathcal{Z}} \in \mathbb{R}^{B(C/E) \times E \times HWD} $, where $ E $ is called the head dimension, i.e., the number of channels per head (or matrix). This operation is obviously suitable for modeling global context and imposes no locality inductive bias.

\paragraph{Local Matricize}
Global Matricize lacks a notion of locality which is typically useful for images, especially in low-data regimes. Local Matricize is proposed to mitigate this shortcoming by splitting an input $ \bm{\mathcal{X}} \in \mathbb{R}^{B \times C \times H \times W \times D} $ into a grid of non-overlapping patches of size $ (E, P, P, P) $. These patches are flattened spatially and then concatenated along the batch dimension, yielding a batch of matrices presented by the tensor $ \bm{\mathcal{Z}} \in \mathbb{R}^{BCHW/(EP^3) \times E \times P^3} $. The entire procedure can be summarized by Einstein notation:
\begin{equation}
	\label{eq: local matricize}
	\bm{\mathcal{Z}}[b g_c g_h g_w g_d, e, p_e p_w p_d] = \bm{\mathcal{X}}[b, g_c e, g_h p_h, g_w p_w, g_d p_d]
\end{equation}
where $ (g_c, g_h, g_w, g_d) $ and  $ (e, p_h, p_w, p_d) $ indices correspond to grid and patch dimensions, respectively. In PyTorch and TensorFlow, such Einstein operations can be simply implemented using \texttt{rearrange} function from \texttt{einops} library. Once NMF is applied to the resulting batch of matrices, the inverse operation of Local Matricize, called Local Dematricize, must be applied to transform them back to the initial shape. Local Dematricize can be easily obtained by composing the inverses of sub-operations in reverse order. In practice, Local Dematricize can also be formulated and implemented as an Einstein operation. A PyTorch-style pseudocode of Local (De)matricize module is provided in Algorithm \ref{alg: local-swin-matricize}.

Note that while Local Matricize seems to reshape an image in a similar way to the input patchifier of ViT, it concatenates patches along batch dimension rather than channel dimension. In fact, Local Matricize can be used when modeling within-patch interactions is desirable, which is different from ViT-based approaches, where interactions between embedded patches are typically modeled. 

\paragraph{Shifted Window Matricize}
While Local Matricize introduces locality to a model, voxels close to the boundaries of partitioning windows are not represented effectively. Since patches are low-rank approximated independently later on in an NMF layer, two neighboring boundary voxels from two adjacent patches are very likely to end up having excessively different feature maps in the output of the factorizer block, which in turn degrades the prediction performance for such voxels. To mitigate this problem, we utilize a shifted window approach proposed by \citet{liu2021swin} and introduce a Shifted Window (SW) Matricize operation, making the output feature maps smoother around boundaries. 

An illustration of SW Matricize is provided in Figure \ref{fig: sw-matricize}. Here, in addition to regular patches similar to those extracted by Local Matricize, shifted window (SW) patches are also included. Let $ \bm{\mathcal{X}} \in \mathbb{R}^{B \times C \times H \times W \times D} $ be the input and $ (P, P, P) $ the patch size. To extract SW patches, the input must be first shifted by the offset of $ (P/2, P/2, P/2) $ along spatial dimensions such that voxels shifted beyond the boundaries of the images are re-introduced at the first position, yielding a tensor of the same size $ (B, C, H, W, D) $. We call this operation \textit{Roll} (which can be implemented using \texttt{roll} function in PyTorch and TensorFlow). The resulting rolled tensor is then reshaped by Local Matricize to get SW patches. Finally, both the batches of regular and SW patches are concatenated along the batch dimension. Formally, SW Matricize is computed as follows:
\begin{align}
	\label{eq: sw-matricize}
	& \bm{\mathcal{X}}^1 = \text{LocalMatricize}_{P}(\bm{\mathcal{X}}), \nonumber \\
	& \bm{\mathcal{X}}^2 = \text{LocalMatricize}_{P}(\text{Roll}_{P/2}(\bm{\mathcal{X}})), \nonumber \\
	& \text{SWMatricize}(\bm{\mathcal{X}}) = \text{Concatenate}(\bm{\mathcal{X}}^1, \bm{\mathcal{X}}^2),
\end{align}
where $ \text{LocalMatricize}_{P} $ denotes Local Matricize with a patch size of $ (P, P, P) $, and $ \text{Roll}_{P/2} $ denotes the roll operator with a shift $ (P/2, P/2, P/2) $. To build SW Dematricize, we need to reconstruct the image from both regular and SW patches independently then compute their average to achieve smoother and more accurate feature maps. 

Further details are provided in \ref{app: matricize implementation details}, which includes a PyTorch implementation of SW (De)matricize module presented in Algorithm \ref{alg: local-swin-matricize}.

\subsubsection{Nonnegative Matrix Factorization} \label{sec: nmf}

Once the input is somehow transformed into a batch of matrices, and its negative elements are clipped to zero by ReLU, it is ready to be low-rank approximated by Nonnegative Matrix Factorization (NMF). This is the main component of a Factorizer model that contributes most to modeling local or global context in an image.  

NMF \citep{lee1999learning} seeks to approximate some given nonnegative matrix $ \mathbf{X} \in \mathbb{R}_{\ge 0}^{M \times N} $ by 
\begin{equation}
	\label{eq: nmf}
	\mathbf{X} \approx \mathbf{F} \mathbf{G}^T,
\end{equation}
where $ \mathbf{F} \in \mathbb{R}_{\ge 0}^{M \times R} $ and $ \mathbf{G} \in \mathbb{R}_{\ge 0}^{N \times R} $ are factor matrices, and the positive integer $ R \leq \min(M, N) $ is the rank. Once the factors $ \mathbf{F} $ and $ \mathbf{G}^T $ are somehow approximated, the NMF layer in a Factorizer block outputs the reconstructed matrix $ \hat{\mathbf{X}} = \mathbf{F} \mathbf{G}^T $. Note that similar to self-attention, the NMF layer can be viewed as an adaptive filter, meaning that the computation of factors involves the input, as opposed to convolution, where kernel weights are fixed and independent from the input and do not change after training. Various loss functions have been used to form the objective function and measure the quality of an approximation. Depending on the loss function, constraints, and regularization, many variants of NMF have been proposed. In this work, we use a standard NMF with the squared error, which is the most widely used variant for images, to find factors by solving the following problem:
\begin{equation}
	\label{eq: nmf objective}
	\underset{\mathbf{F} \geq 0, \mathbf{G} \geq 0}{\text{minimize}} \ \lVert \mathbf{X} - \mathbf{F} \mathbf{G}^T \rVert^2.
\end{equation}
In general, problem \eqref{eq: nmf objective} is nonconvex, and finding global minima is NP-hard. However, numerous iterative algorithms have been proposed to find a ``good local minimum". The majority of existing methods are based on a block coordinate descent (BCD) scheme (a.k.a. alternating optimization), where the objective function is iteratively minimized with respect to one factor while the other factor is kept fixed. That is, the convex subproblems
\begin{align}
	\label{eq: nmf bcd subproblems} 
	\mathbf{F} \gets \underset{\mathbf{F} \geq 0}{\argmin} \ \lVert \mathbf{X} - \mathbf{F} \mathbf{G}^T  \rVert^2, \qquad & 
	\mathbf{G} \gets \underset{\mathbf{G} \geq 0}{\argmin} \ \lVert \mathbf{X}^T - \mathbf{G} \mathbf{F}^T \rVert^2,
\end{align}
are exactly or approximately solved alternately. This ensures that the objective function value does not increase after each update and guarantees convergence to a stationary point under some mild conditions \citep{grippo2000convergence}.

\begin{algorithm}[t]	
	\caption{Multiplicative update for NMF ($ \odot $ and $ \frac{\ \cdot\ }{\ \cdot\ } $ denote element-wise matrix product and division, respectively). \label{alg: multiplicative update}}
	\DontPrintSemicolon
	\KwIn{$ \mathbf{X} \in \mathbb{R}_{\ge 0}^{M \times N} $ (input matrix), $ R $ (rank); $ T $ (\# iterations)}
	\KwOut{$ \mathbf{\hat{X}} \in \mathbb{R}_{\ge 0}^{M \times N} $ (low-rank matrix approximation)}
	Initialize factors $ \mathbf{F} \in \mathbb{R}_{\ge 0}^{M \times R} $ and $ \mathbf{G} \in \mathbb{R}_{\ge 0}^{N \times R} $\;
	\For{$ t = 1, \dots, T $}{
		\vspace{5pt}		
		$ \displaystyle \mathbf{F} \gets \mathbf{F} \odot \frac{\mathbf{X} \mathbf{G}}{\mathbf{F} \mathbf{G}^T \mathbf{G}} $\;
		\vspace{5pt}
		$ \displaystyle \mathbf{G} \gets \mathbf{G} \odot \frac{\mathbf{X}^T \mathbf{F}}{\mathbf{G} \mathbf{F}^T \mathbf{F}} $\;
		\vspace{5pt}		
	}
	\Return $ \displaystyle \mathbf{\hat{X}} = \mathbf{F} \mathbf{G}^T $
\end{algorithm}

Among numerous BCD-based algorithms for NMF, Multiplicative Update (MU) \citep{lee1999learning} is the best-known due to the advantage of being easy-to-implement and scalable. MU enforces a nonnegativity constraint by updating the previous values of a factor matrix by multiplication with a nonnegative scale factor. The pseudocode of MU is outlined in Algorithm \ref{alg: multiplicative update}. However, slow convergence of MU has been pointed out \citep[Chapter~8]{gillis2020nonnegative}, and hence more effective algorithms with faster convergence such as Hierarchical Alternating Least Squares (HALS) \citep{cichocki2009fast} have been introduced. HALS updates a factor, say $ \mathbf{F} = [\mathbf{f}_1 | \dots | \mathbf{f}_R] $, with inner iterations, in which the columns $ \mathbf{f}_r $s are updated by solving the following subproblem
\begin{equation}
	\label{eq: hals subproblem} 
	\mathbf{f}_r \gets \underset{\mathbf{f}_r \geq 0}{\argmin} \ \lVert \mathbf{E}_r - \mathbf{f}_r \mathbf{g}_r^T \rVert^2,
\end{equation}
where $ \mathbf{E}_r = \mathbf{X} - \sum_{\ell \neq r}^{R} \mathbf{f}_\ell \mathbf{g}_\ell^T $ is the residual matrix, which is, in fact, approximated by a rank-one matrix. An encouraging aspect of HALS is that each subproblem \eqref{eq: hals subproblem} can be easily shown to have a closed-form solution:
\begin{equation}
	\label{eq: hals subproblem solution} 
	\mathbf{f}_r^* \gets \max(0, \frac{\mathbf{E}_r \mathbf{g}_r}{\lVert \mathbf{g}_r \rVert^2}).
\end{equation}
Similarly, the update formula for columns of $ \mathbf{G} $ can be derived. Note that HALS is a $ 2R $-block coordinate descent procedure, where at each outermost iteration, first the columns of $ \mathbf{F} $ and then the columns of $ \mathbf{G} $ are updated. Algorithm \ref{alg: hals} provides pseudocode of HALS (further details on MU and HALS can be found in \citep[Chapter~8]{gillis2020nonnegative}). 

\begin{algorithm}[t]	
	\caption{Hierarchical alternating least squares for NMF. \label{alg: hals}}
	\DontPrintSemicolon
	\KwIn{$ \mathbf{X} \in \mathbb{R}_{\ge 0}^{M \times N} $ (input matrix), $ R $ (rank); $ T $ (\# iterations)}
	\KwOut{$ \mathbf{\hat{X}} \in \mathbb{R}_{\ge 0}^{M \times N} $ (low-rank matrix approximation)}
	Initialize factors $ \mathbf{F} \in \mathbb{R}_{\ge 0}^{M \times R} $ and $ \mathbf{G} \in \mathbb{R}_{\ge 0}^{N \times R} $\;
	\For{$ t = 1, \dots, T $}{
		$ \mathbf{A} =  \mathbf{X} \mathbf{G} $, $ \mathbf{B} = \mathbf{G}^T \mathbf{G} $\;
		\For{$ r = 1, \dots, R $}{				
			$ \displaystyle \mathbf{F}[:,r] \gets \max \left(0, \frac{\mathbf{A}[:,r] - \sum_{\ell \neq r} \mathbf{B}[\ell,r] \mathbf{F}[:,\ell]}{\lVert \mathbf{G}[:,r] \rVert^2} \right) $
		}
		$ \mathbf{A} =  \mathbf{X}^T \mathbf{F} $, $ \mathbf{B} = \mathbf{F}^T \mathbf{F} $\;
		\For{$ r = 1, \dots, R $}{					
			$ \displaystyle \mathbf{G}[:,r] \gets \max \left(0, \frac{\mathbf{A}[:,r] - \sum_{\ell \neq r} \mathbf{B}[\ell,r] \mathbf{G}[:,\ell]}{\lVert \mathbf{F}[:,r] \rVert^2} \right) $
		}
	}
	\Return $ \displaystyle \mathbf{\hat{X}} = \mathbf{F} \mathbf{G}^T $
\end{algorithm}

A special case of NMF is $ R = 1 $; i.e., $ \mathbf{X} \approx \mathbf{f} \mathbf{g}^T $, where $ \mathbf{f} \in \mathbb{R}_{\ge 0}^M $ and $ \mathbf{g} \in \mathbb{R}_{\ge 0}^N $; for which one can easily derive that both MU and HALS are simplified to the same update rule:
\begin{align}
	\label{eq: rank-one update formulas} 
	\mathbf{f} \gets \frac{\mathbf{X} \mathbf{g}}{\lVert \mathbf{g} \rVert^2}, \qquad &  \mathbf{g} \gets \frac{\mathbf{X} \mathbf{f}}{\lVert \mathbf{f} \rVert^2}.
\end{align}
In this paper, all Factorizer models are trained with $ R = 1 $, and compression ratios (and indirectly reconstruction errors) are controlled by adjusting the head dimension (i.e., the number of rows in a matrix, as discussed in \ref{sec: matricize}) to sufficiently small values. This means that a matricize operation transforms an image into a batch of such fat matrices (i.e., the number of columns is much larger than the number of rows) that rank-one approximation would suffice in practice. This greatly simplifies a factorizer model and improves interpretability while yielding better segmentation performance. However, in one of our ablation studies (see Section \ref{sec: inference ablation studies}), we experimented with both MU and HALS for $ R > 1 $ to investigate the impact of rank in the inference phase. The computational complexity of both MU and HALS is $ \mathcal{O}(MNR) $ per iteration \citep[Chapter~8]{gillis2020nonnegative}, making the Wrapped NMF layer scale linearly and be much cheaper than self-attention with quadratic complexity (which is computationally intractable on long sequences) and even than Performer \citep{choromanski2020rethinking}, as an efficient approximation of attention. 

It is worth mentioning that not all NMF algorithms and their settings can be used in the NMF layer of a Factorizer block. The selected algorithm must have some properties so that we can ultimately train the Factorizer model successfully in an end-to-end fashion using a gradient descent-based optimizer on GPU(s) through an existing deep learning framework, such as PyTorch. Firstly, the algorithm should be backpropagation-friendly and amenable to automatic differentiation, that is $ \frac{\partial \hat{\mathbf{X}}}{\partial \mathbf{X}} $ should not only be well-defined and somewhat smooth but also computable by means of an existing deep learning framework, such as PyTorch, so that we can practically train the Factorizer model in an end-to-end fashion using a gradient descent optimizer. Another related aspect is that the gradient $ \frac{\partial \hat{\mathbf{X}}}{\partial \mathbf{X}} $, as explained in \citep{geng2021attention}, starts to vanish during backpropagation after some iterations. Therefore, the number of outer iterations $ T $ in an NMF algorithm should be limited in order to have stable gradients. For MU and HALS, $ T = 5 $ is a reasonable choice in practice. Finally, update rules should be also friendly to GPU parallel processing for exploiting GPUs to train Factorizers in a reasonable amount of time. Taking all these factors into account, MU and HALS are appropriate choices. While HALS has better convergence properties, MU is more favorable for GPU training due to unparallelizable inner iterations of HALS.

\section{Experiments} \label{sec: experiments}

\subsection{Datasets} 

We evaluate the effectiveness of our models on the Brain Tumour Segmentation (BraTS) dataset \citep{brats1, brats2} from Medical Segmentation Decathlon \citep{antonelli2021medical} and Ischemic Stroke Lesion Segmentation (ISLES) 2022 dataset \citep{petzsche2022isles} from a MICCAI 2022 challenge.

\paragraph{BraTS}
This dataset consists of 484 multiparametric MRI (mpMRI) scans from patients diagnosed with either low-grade glioma or high-grade glioma (glioblastoma). Each scan comes with four 3D MRI sequences, namely T2 Fluid-Attenuated Inversion Recovery (FLAIR), native T1-weighted (T1), post-Gadolinium contrast T1-weighted (T1Gd), and T2-weighted (T2). Once images are preprocessed (i.e., rigidly co-registered to the same anatomical template, resampled to the same voxel spacing $ 1 \text{mm}^3 $, and skull-stripped), the ground truths are manually created by experts who label each voxel as enhancing tumor (ET), edema (ED), necrotic and non-enhancing tumor (NCR/NET), or everything else. However, for evaluation, the 3 nested subregions, namely enhancing tumor (ET), tumor core (TC--i.e., the union of ED and NCR/NET), and whole tumor (WT) are used (see the sample ground truths in Figure \ref{fig: brats qualitative results}).

\paragraph{ISLES'22}
This dataset is from the ISLES'22 challenge, which aims to evaluate automated methods of acute and sub-acute stroke lesion segmentation in 3D multiparametric MRI data, namely  DWI, Apparent Diffusion Coefficient (ADC), and FLAIR sequences. The DWI and ADC images of a patient are aligned while the FLAIR image in its native space has a different voxel size and must be registered to the DWI space. As DWI and ADC are the most informative modalities for stroke lesions, FLAIR is ignored in this paper to avoid the complication of FLIAR-DWI registration and simplify the pipelines. The dataset consists of 250 cases, each is skull-stripped and includes an expert-level annotation of the stroke lesions.

\subsection{Setup} \label{sec: setup}

All the models were implemented using PyTorch \citep{PyTorch2019} and MONAI \citep{monai} frameworks and trained on NVIDIA P100 GPUs. We followed the same training workflow in all the experiments. In the following, we first provide the details of this workflow and baseline models, then present the evaluation protocol and the results.

\paragraph{Preprocessing} 
For each scan in a dataset, a multi-channel 3D image as the input was first constructed by concatenating the modalities--i.e., FLAIR, T1, T1Gd, and T2 for BraTS, and DWI and ADC for ISLES'22. The image and its ground truth were then cropped with a minimal box filtering out zero regions of the image. The image was normalized channel-wise using a z-score to have intensities with zero mean and unit variance. Random patches of size $ (128, 128, 128) $ for BraTS and $ (64, 64, 64) $ for ISLES'22 were extracted during training. To reduce overfitting, we used data augmentation techniques, including random affine transform, random flip along each spatial dimension, additive Gaussian noise, random Gaussian smoothing, random intensity scaling,  random intensity shifting, and random gamma transform. Further details are provided in \ref{app: data augmentation pipeline}.

\paragraph{Training}
All models were trained for 100000 steps with a batch size of 2 (one sample per GPU) using AdamW optimizer with a base learning rate of $ 10^{-4} $, weight decay of $ 10^{-2} $, warmup of 2000 steps, and cosine annealing scheduler. The loss $ \mathcal{L}_{\text{total}} $ is computed by incorporating the three deep supervision outputs and the corresponding downsampled ground truths according to
\begin{equation}
	\label{eq: total loss}
	\mathcal{L}_{\text{total}} = \lambda_1 \mathcal{L}(\mathbf{G}_1, \mathbf{P}_1) + \lambda_2 \mathcal{L}(\mathbf{G}_2, \mathbf{P}_2) + \lambda_3 \mathcal{L}(\mathbf{G}_3, \mathbf{P}_3),
\end{equation}
where $ \lambda_1 = 1 $, $ \lambda_2 = 0.5 $, and $ \lambda_3 = 0.25 $; $ \mathbf{G}_i $ and $ \mathbf{P}_i $ correspond to the deep supervision at the $ i $th highest resolution; and the loss function $ \mathcal{L}(\cdot, \cdot) $ is a combination of \textit{soft Dice loss} \citep{milletari2016v} and cross-entropy loss, defined as
\begin{equation}
	\label{eq: loss}
	\mathcal{L}(\mathbf{G}, \mathbf{P}) = \mathcal{L}_{\text{Dice}}(\mathbf{G}, \mathbf{P}) + \mathcal{L}_{\text{CE}}(\mathbf{G}, \mathbf{P}),
\end{equation}
where
\begin{align}
	\label{eq: dice and cross-entropy loss}
	\mathcal{L}_{\text{Dice}}(\mathbf{G}, \mathbf{P}) = 1 - \frac{2 \langle \mathbf{G}, \mathbf{P} \rangle + \epsilon}{\lVert\mathbf{G}\rVert^2 + \lVert\mathbf{P}\rVert^2 + \epsilon}, \qquad & \mathcal{L}_{\text{CE}}(\mathbf{G}, \mathbf{P}) =  -\frac{1}{N}\langle \mathbf{G}, \log(\mathbf{P}) \rangle,
\end{align}
where $ \mathbf{G} \in \{0, 1\}^{J \times N} $ and $ \mathbf{P} \in [0, 1]^{J \times N} $ represent the one-hot encoded ground truth and the predicted probability map for each voxel, respectively, with $ J $ denoting the number of foreground classes and $ N $ denoting the number of voxels in the patch. The small constant $ \epsilon = 10^{-5} $ is commonly used to smooth the soft Dice loss and avoid division by zero.

\paragraph{Inference} 
A test image in the inference was first subjected to z-score intensity normalization, then the prediction was made using a sliding window approach with a 50\% overlap and a window size of $ 128 \times 128 \times 128 $ (same as the patch size used in training). Finally, the resulting probabilities were thresholded by 0.5 to obtain a binary segmentation map.

\paragraph{Evaluation Metrics}
The Dice score and Hausdorff Distance 95\% (HD95) were used as metrics to assess the performance of models in our experiments. For each segmentation region, the Dice score measures the voxel-wise overlap between the ground truth and the prediction, defined as
\begin{equation}
	\label{eq: dice metric}
	\text{Dice}(\mathbf{g}, \mathbf{y}) = \frac{2 \sum_{n=1}^{N} g_n y_n}{\sum_{n=1}^{N} g_n + \sum_{n=1}^{N} y_n}
\end{equation}	
where $ g_n \in \{0, 1\} $ and $ y_n \in \{0, 1\} $ represent the ground truth and the binary prediction for a voxel, respectively, and $ N $ is the number of voxels. If both the ground truth and the prediction do not have any nonzero values, that is the denominator of equation \eqref{eq: dice metric} is zero, the Dice score is defined as 1. Hausdorff Distance (HD) evaluates the distance between the boundaries of ground truth and prediction. HD is defined as follows:
\begin{equation}
	\label{eq: hd metric}
	\text{HD}(G, Y) = \max\{ \underset{\mathbf{g} \in G}{\max}\ \underset{\mathbf{y} \in Y}{\min}\ \lVert \mathbf{g} - \mathbf{y} \rVert, \underset{\mathbf{y} \in Y}{\max}\ \underset{\mathbf{g} \in G}{\min}\  \lVert \mathbf{y} - \mathbf{g} \rVert \},
\end{equation}
where $ G $ and $ Y $ denote the set of all voxels on the surface of ground truth and prediction, respectively. HD95 is a more robust version to outliers, which calculates the 95\% quantile rather than the maximum of surface distances.

\paragraph{Models}
In all the experiments, for Factorizer models with the overall architecture illustrated in Figure \ref{fig: factorizer}, the number of output channels of the stem was $ C = 32 $. For Local and Swin Factorizer, we used a large window size of $ (8, 8, 8) $ and $ (4, 4, 4) $ on the BraTS and ISLES'22 datasets, respectively, to aggregate local information, which is opposed to typical CNNs comprising convolutions mostly with a small receptive field of size $ (3, 3, 3) $. For all Factorizer models, a head dimension of $ E = 8 $ on BraTS and $ E = 4 $ on ISLES'22 was used. In NMF modules, the factor matrices were initialized with uniform distribution $ \mathcal{U}(0, 1) $. In training, we used a rank-one approximation ($ R = 1 $) with $ T = 5 $ outer iterations of HALS (which is equivalent to MU for $ R = 1 $), as described in Section \ref{sec: nmf}.

We compare Factorizers against seven baseline models, among which nnU-Net, Res-U-Net, and Performer follow the same overall architecture and setup as those of Factorizers except that each has a different encoder/decoder block, allowing us to better assess the impact of blocks rather than architectures by eliminating the effect of architectural variability. These three baselines are detailed below: 

\begin{itemize}
	\item nnU-Net: This model is based on nnU-Net \citep{isensee2018no}, i.e., a standard 3D U-Net \citep{cciccek20163d} with minor modifications. Each encoder (and decoder) block is composed of two convolutions with a kernel size of $ (3, 3, 3) $. Group Normalization (with a group size of 8) is adopted right after each convolution and before LeakyReLU nonlinearity. This model does not have any positional embedding in its bridge since CNNs already have some notion of position.
	
	\item Res-U-Net: ResNet block \citep{he2016deep} is the cornerstone of Res-U-Net. This block is similar to that of nnU-Net, except that it has a residual connection after the last Group Normalization. This model is similar to SegResNet \citep{myronenko20183d}.
	
	\item Performer: The encoder and decoder blocks of this model are based on ViT (note that the input is first flattened into a sequence of voxels before feeding it to a ViT block). However, attention scales quadratically with the number of voxels, hence is prohibitively expensive in our case, where the input can have up to $ 128^3 $ voxels. Therefore, we replace original attention layers of a ViT block with FAVOR+ (Fast Attention Via positive Orthogonal Random features) used in Performer, recently proposed by \citet{choromanski2020rethinking} as a linearly scalable alternative to Transformer. FAVOR+ gives an unbiased estimate of attention using only linear (as opposed to quadratic) time and space complexity. Note that except for the attention layers, the rest of the components of this baseline are the same as those of Factorizer.
\end{itemize}

We also use four state-of-the-art Transformer-based baselines, namely TransBTS \citep{wang2021transbts}, UNETR \citep{hatamizadeh2022unetr}, Swin UNETR \citep{hatamizadeh2022swin}, and nnFormer \citep{zhou2021nnformer}, each of which has a different overall architecture than that of Figure \ref{fig: factorizer}. These models follow U-shaped architectures but apply Transformer blocks only to low-resolution images after somehow downsizing the input using patchifying or convolutional tokenizers, thereby avoiding the computational intractability of self-attention on long sequences. In contrast, Global Factorizer and Performer exploit the global context at all stages of their architectures, from the lowest to highest resolution.

\subsection{Results and Discussion} \label{sec: results and discussion}

\subsubsection{Brain Tumor Segmentation (BraTS)}  \label{sec: results and discussion: brats}

\begin{figure}[t]
	\centering
	
	\begin{subfigure}{.5\textwidth}
		\centering
		\includegraphics[width=.95\textwidth]{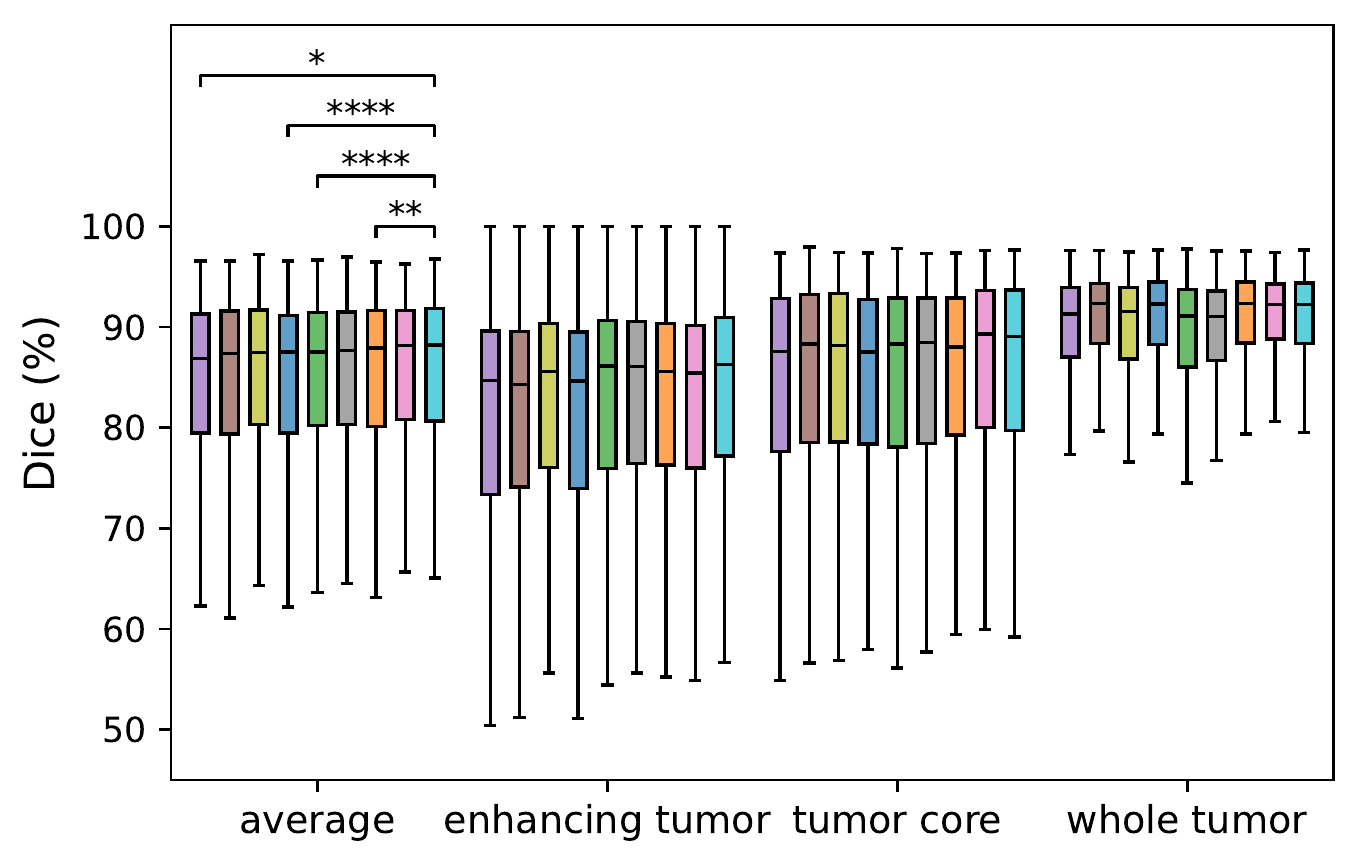}
		\caption{}
		\label{fig: brats dice boxplot}
	\end{subfigure}%
	\begin{subfigure}{.5\textwidth}
		\centering
		\includegraphics[width=.95\textwidth]{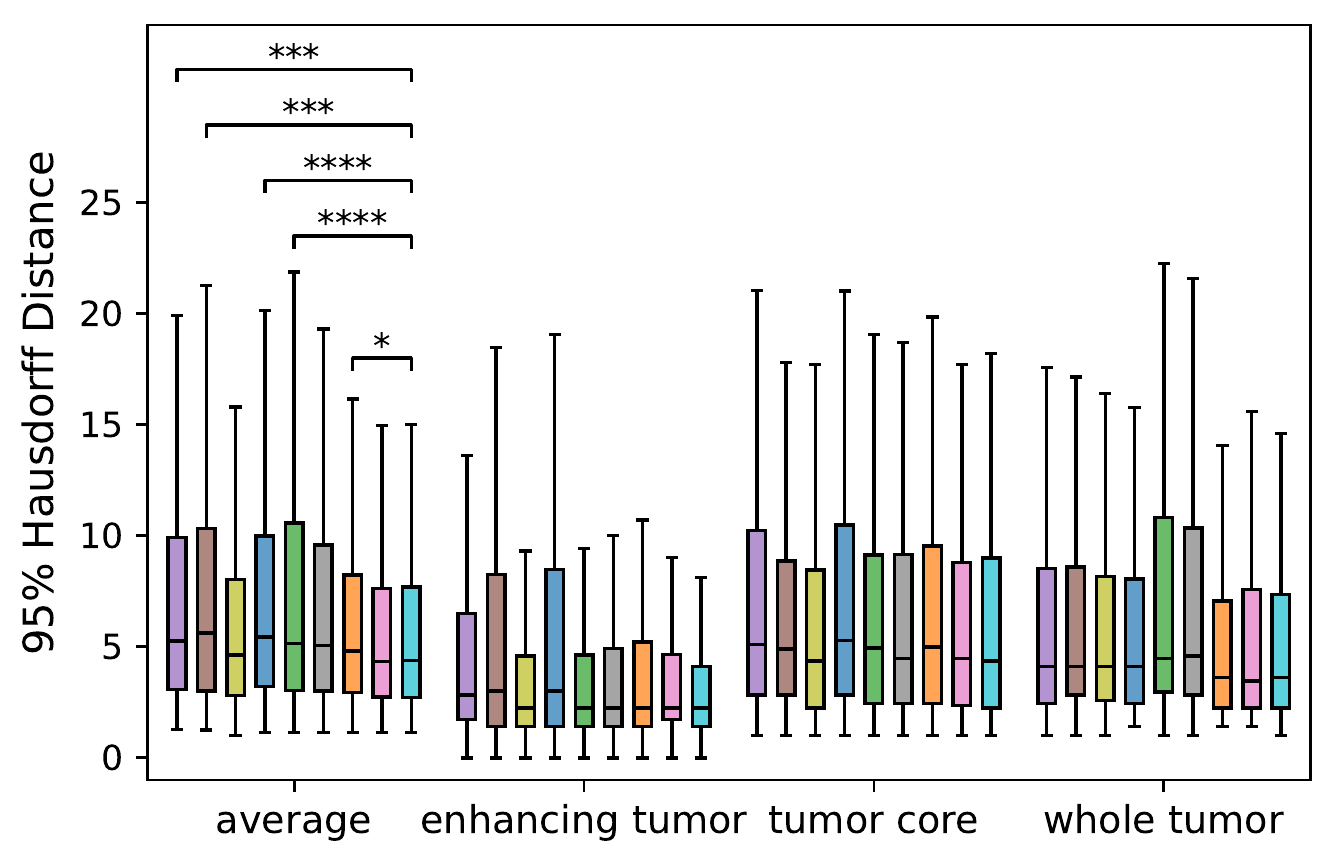}
		\caption{}
		\label{fig: brats hd boxplot}
	\end{subfigure}
	
	\begin{subfigure}{.5\textwidth}
		\centering
		\includegraphics[width=.95\textwidth]{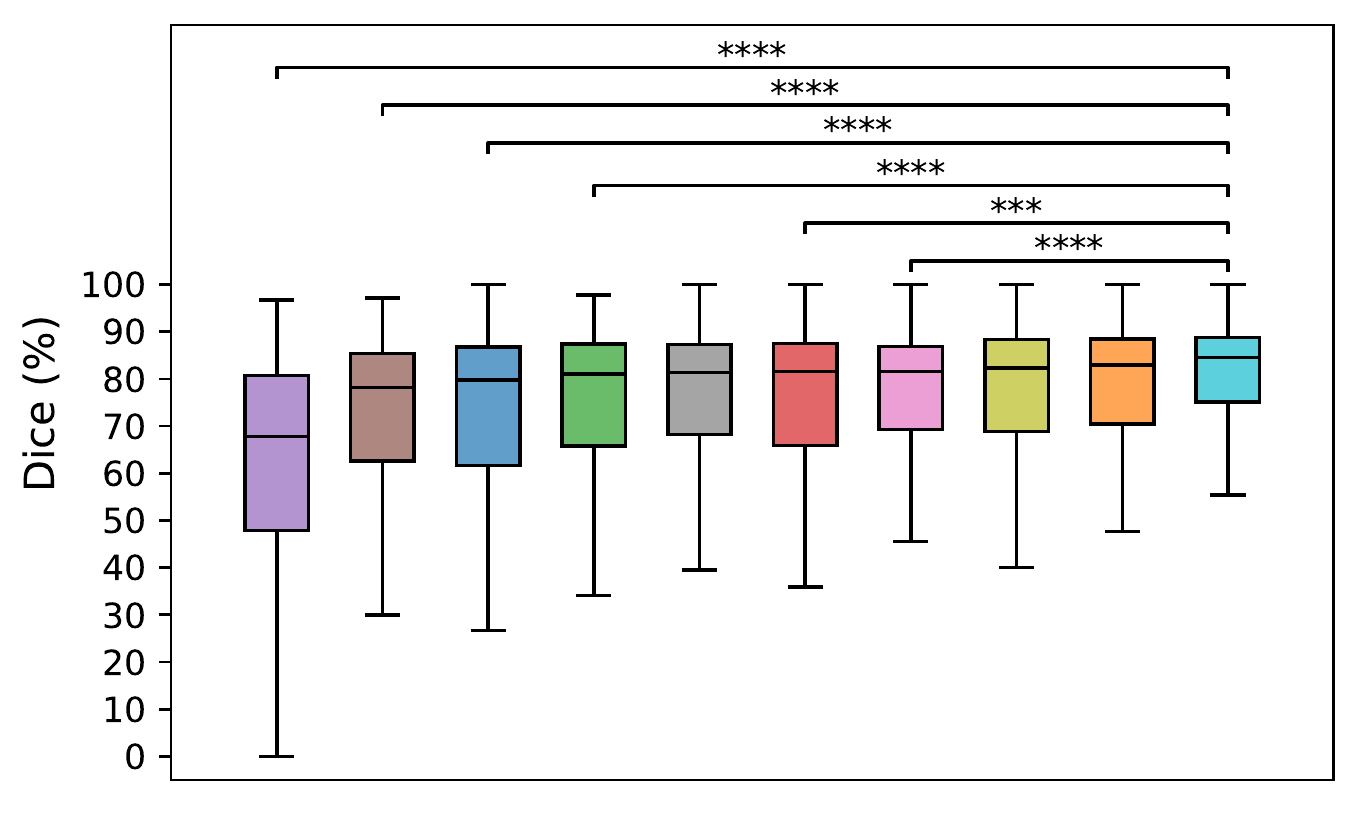}
		\caption{}
		\label{fig: isles dice boxplot}
	\end{subfigure}%
	\begin{subfigure}{.5\textwidth}
		\centering
		\includegraphics[width=.95\textwidth]{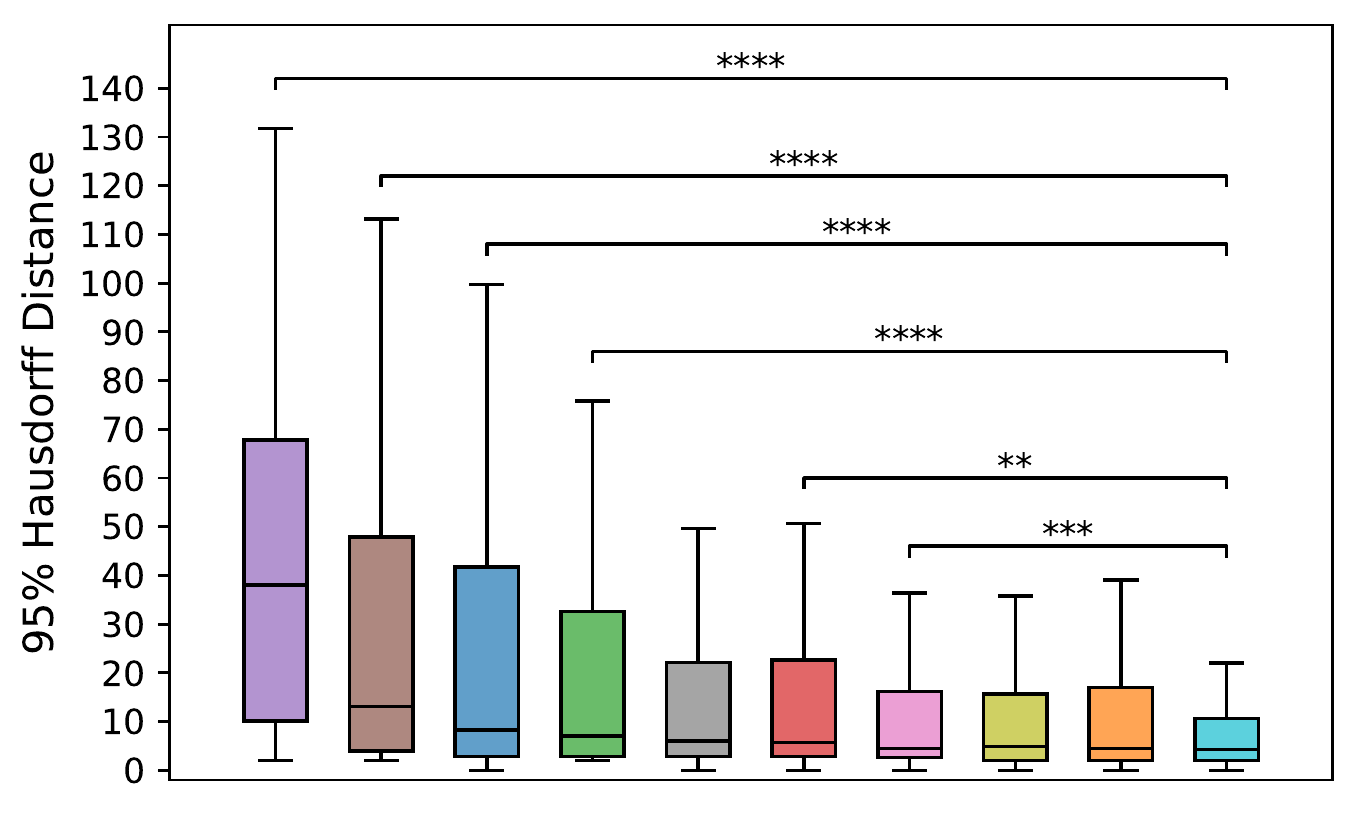}
		\caption{}
		\label{fig: isles hd boxplot}
	\end{subfigure}
	
	\begin{subfigure}{\textwidth}
		\centering
		\vspace{5pt}
		\includegraphics[width=.95\textwidth]{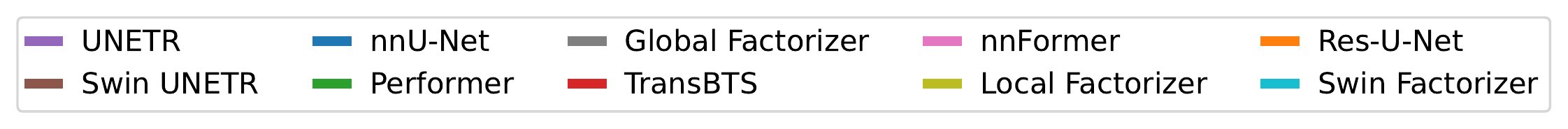}
	\end{subfigure}
	\caption{The Dice score and 95\% Hausdorff distance of different models on the BraTS dataset. The asterisks indicate statistical significance according to pairwise Wilcoxon signed-rank test (*: $\text{p-value} < 0.05 $; **: $ \text{p-value} < 0.01 $; ***: $ \text{p-value} < 0.001 $; and ****: $ \text{p-value} < 0.0001 $).}
	\label{fig: scores boxplot}
\end{figure}

\paragraph{Quantitative Evaluation}
For all the experiments, we performed 5-fold cross-validation to estimate how capable our models are in generalizing to unseen data. The results on the BraTS dataset are reported in Table \ref{tab: brats validation results} and illustrated by box plots in Figure \ref{fig: brats dice boxplot} and \ref{fig: brats hd boxplot}, where pairwise Wilcoxon signed-rank tests were used for comparing the performance of our best model, Swin Factorizer, with that of the baselines.

\begin{table}[t]
	\caption{Comparison of different models on the BraTS dataset. Scores are obtained by 5-fold cross-validation. The best results are in \textbf{boldface} and second best ones are \underline{underlined}.}
	\label{tab: brats validation results}
	\newcolumntype{P}[1]{>{\centering\arraybackslash}p{#1}}
	\def\arraystretch{1.3}
	\resizebox{\textwidth}{!}{	
		\begin{tabular}{m{.2\textwidth} | P{.1\textwidth} P{.1\textwidth} | *{4}{P{.08\textwidth}} | *{4}{P{.08\textwidth}}}		
			\toprule
			
			\multirow{2}{*}{Model} & \multirow{2}{*}{\#Params} & \multirow{2}{*}{FLOPs} & \multicolumn{4}{c|}{Dice (\%) \textuparrow}                                                                     & \multicolumn{4}{c}{HD95 (mm) \textdownarrow}  \\ \cline{4-11} 	
			&                           &                        & ET                   & TC                   & WT 				  & \multicolumn{1}{|c|}{Avg.} & ET & TC & WT & \multicolumn{1}{|c}{Avg.} \\ 
			
			\hline
			
			nnU-Net                  & 28.7M                     & 1152.8G                & 75.89 &    81.94 &    90.09 & \multicolumn{1}{|c|}{82.64}     &  8.61 &   9.13 &   9.31  &  \multicolumn{1}{|c}{9.19}    \\
			Res-U-Net              & 28.9M                     & 1168.6G               &         77.95 &    82.49 &    \textbf{90.39}  & \multicolumn{1}{|c|}{83.61}     &   6.08 &   8.18 &   \underline{8.06}    &  \multicolumn{1}{|c}{7.52} \\
			
			\hline
			
			Performer              & 6.7M                      & 222.7G                 &   78.43                   &       82.34               &   88.72  & \multicolumn{1}{|c|}{83.16}     &  6.72  & 9.78  & 12.93   &  \multicolumn{1}{|c}{10.21}     \\
			TransBTS            & 33.0M                      & 653.0G                 &   73.46 &    80.00 &    85.42  & \multicolumn{1}{|c|}{79.63}     &  15.48 &  13.77 &  20.73 &  \multicolumn{1}{|c}{17.12}     \\
			UNETR              & 111.5M                     & 1124.9G                &   76.40 &    82.06 &    89.37  & \multicolumn{1}{|c|}{82.61}     &  8.09 &   9.41 &  10.59   &  \multicolumn{1}{|c}{9.56}     \\
			Swin UNETR              & 62.2M                      & 1549.8G                 &   76.50 &    82.65 &    90.10  & \multicolumn{1}{|c|}{83.08}     &  9.69 &   8.75 &  11.04   &  \multicolumn{1}{|c}{9.91}     \\ 
			nnFormer             & 149.4M                      & 489.9G                 &   77.71 &    \textbf{83.41} &    90.05  & \multicolumn{1}{|c|}{\underline{83.72}}     &  5.42 &   \underline{7.45} &   \textbf{7.68}   &  \multicolumn{1}{|c}{\underline{6.93}}     \\    
			
			\hline
			
			Global Factorizer      & 5.9M                      & 170.0G                 &         78.20 &    82.86 &    88.65    & \multicolumn{1}{|c|}{83.24}     &   6.67 &   9.53 &  11.69  &  \multicolumn{1}{|c}{9.71}      \\
			Local Factorizer       & 5.9M                      & 170.0G                 &          \underline{78.67} &    82.85 &    89.30    & \multicolumn{1}{|c|}{83.61}     &  \underline{5.29} &   7.77 &   8.91  &  \multicolumn{1}{|c}{7.41}      \\
			Swin Factorizer        & 5.9M                      & 174.2G                &       \textbf{79.33} &    \underline{83.14} &    \underline{90.16}    & \multicolumn{1}{|c|}{\textbf{84.21}}     &   \textbf{4.91} &   \textbf{7.31} &   8.23    &  \multicolumn{1}{|c}{\textbf{6.89}}      \\ 
			
			\bottomrule
		\end{tabular}
	}
\end{table}

\begin{figure}[t]
	\centering
	\begin{subfigure}{.5\textwidth}
		\centering
		\includegraphics[width=.95\textwidth]{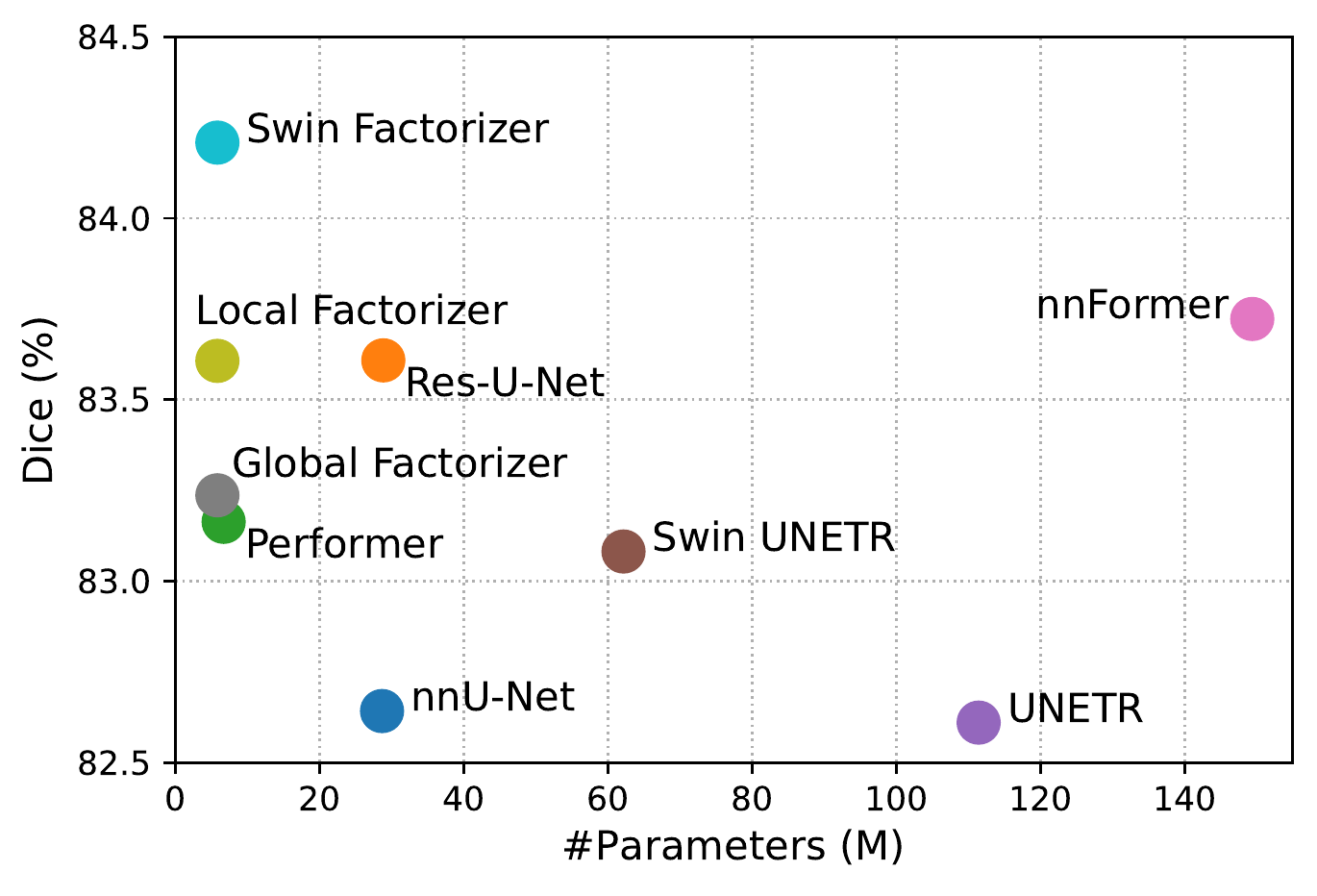}
		\caption{}
	\end{subfigure}%
	\begin{subfigure}{.5\textwidth}
		\centering
		\includegraphics[width=.95\textwidth]{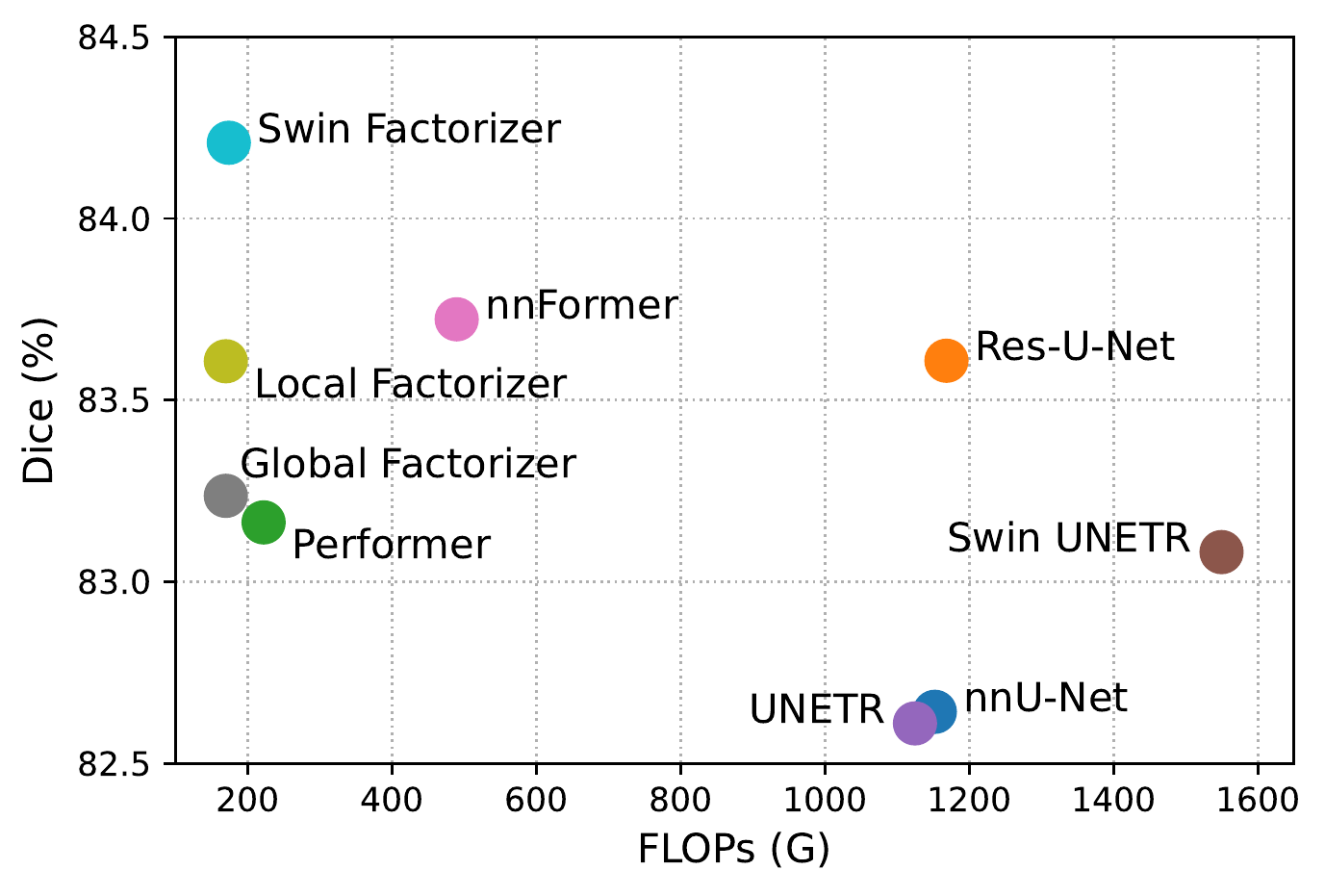}
		\caption{}
	\end{subfigure}
	\caption{Performance of Factorizer models in terms of average Dice on BraTS, versus the number of parameters (left) and versus their speed in terms of the number of FLOPs (right).}
	\label{fig: dice_vs_flops}
\end{figure}

Swin Factorizer is the clear overall winner in the brain tumor segmentation task. With an average Dice score of 84.21\% and HD95 of 6.89 mm, Swin Factorizer significantly outperformed Res-U-Net, the best CNN-based and second-best overall baseline, with p-values of $ < 0.01 $ for the average Dice score and $ < 0.05 $ for the average HD95, while requiring six times fewer computations. Swin Factorizer was the best-performing model on ET and the second best-performing model on TC and WT, particularly yielding the highest Dice score of 79.33\% on ET. Despite having over 95\% fewer parameters and requiring over 60\% fewer FLOPs, Swin Factorizer still outperformed nnFormer, the best-performing baseline, in terms of average Dice and HD95. With an average Dice score of 83.61\%, Local Factorizer yielded comparable results to Res-U-Net, and both Global and Local Factorizer demonstrated improved performance compared to Performer, nnU-Net, TransBTS, UNETR, and Swin UNETR. As shown in Figure \ref{fig: dice_vs_flops}, while achieving competitive performance over the baselines, Factorizer models have much fewer parameters and are significantly cheaper.

Performer is the Transformer-based counterpart of Global Factorizer in the sense that they both follow the same overall architecture without imposing any locality inductive bias in their blocks. However, while having lower computational complexity, Global Factorizer, with an average Dice score of 83.24\% and an average HD95 of 9.71 mm, marginally outperformed Performer, with an average Dice score of 83.16\% and an average HD95 of 10.21 mm. Local Factorizer improved the performance of Global Factorizer by exploiting locality, particularly with the average HD95 significantly dropping from 9.71 mm to 7.41 mm ($ \text{p-value} < 0.0001 $). As expected, Swin Factorizer improved all the scores of Local Factorizer, which is consistent with the fact that Swin Factorizer modifies Local Factorizer by better representing the boundary voxels in Local Matricize.

\begin{figure}[t]
	\newcommand{\imagewithdice}[2]{
		\begin{tikzpicture}[every node/.style={inner sep=0, outer sep=0}]
			\node (image) at (0,0) {\includegraphics[width=1.0\textwidth]{#1}};
			\node[white, anchor=west] (dice) at (-1.15, -1.3) {\fontfamily{phv}\scriptsize\textbf{#2}};
		\end{tikzpicture}
		\vspace*{-15pt}	
	}
	\centering
	
	\begin{subfigure}{1.0\textwidth}
		\centering
		\includegraphics[width=.6\textwidth]{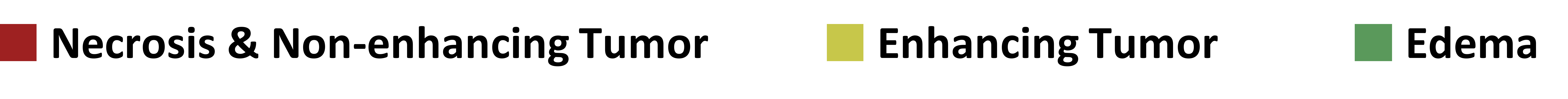}
	\end{subfigure}

	\begin{subfigure}{.167\textwidth}
		\centering
		\imagewithdice{figures/qualitative_results/gt_BraTS_451.pdf}{}
	\end{subfigure}%
	\begin{subfigure}{.167\textwidth}
		\centering
		\imagewithdice{figures/qualitative_results/swin-factorizer_pred_BraTS_451.pdf}{95.6}
	\end{subfigure}%
	\begin{subfigure}{.167\textwidth}
		\centering
		\imagewithdice{figures/qualitative_results/global-factorizer_pred_BraTS_451.pdf}{94.8}
	\end{subfigure}%
	\begin{subfigure}{.167\textwidth}
		\centering
		\imagewithdice{figures/qualitative_results/unetr_pred_BraTS_451.pdf}{85.1}
	\end{subfigure}%
	\begin{subfigure}{.167\textwidth}
		\centering
		\imagewithdice{figures/qualitative_results/nnformer_pred_BraTS_451.pdf}{84.2}
	\end{subfigure}%
	\begin{subfigure}{.167\textwidth}
		\centering
		\imagewithdice{figures/qualitative_results/unet_pred_BraTS_451.pdf}{91.6}
	\end{subfigure}

	\begin{subfigure}{.167\textwidth}
		\centering
		\imagewithdice{figures/qualitative_results/gt_BraTS_481.pdf}{}
	\end{subfigure}%
	\begin{subfigure}{.167\textwidth}
		\centering
		\imagewithdice{figures/qualitative_results/swin-factorizer_pred_BraTS_481.pdf}{90.5}
	\end{subfigure}%
	\begin{subfigure}{.167\textwidth}
		\centering
		\imagewithdice{figures/qualitative_results/global-factorizer_pred_BraTS_481.pdf}{84.7}
	\end{subfigure}%
	\begin{subfigure}{.167\textwidth}
		\centering
		\imagewithdice{figures/qualitative_results/unetr_pred_BraTS_481.pdf}{84.1}
	\end{subfigure}%
	\begin{subfigure}{.167\textwidth}
		\centering
		\imagewithdice{figures/qualitative_results/nnformer_pred_BraTS_481.pdf}{81.1}
	\end{subfigure}%
	\begin{subfigure}{.167\textwidth}
		\centering
		\imagewithdice{figures/qualitative_results/unet_pred_BraTS_481.pdf}{77.2}
	\end{subfigure}

	\begin{subfigure}{.167\textwidth}
		\centering
		\imagewithdice{figures/qualitative_results/gt_BraTS_103.pdf}{}
		\caption*{\scriptsize\textbf{Ground Truth}}
	\end{subfigure}%
	\begin{subfigure}{.167\textwidth}
		\centering
		\imagewithdice{figures/qualitative_results/swin-factorizer_pred_BraTS_103.pdf}{86.6}
		\caption*{\scriptsize\textbf{Swin Factorizer}}
	\end{subfigure}%
	\begin{subfigure}{.167\textwidth}
		\centering
		\imagewithdice{figures/qualitative_results/global-factorizer_pred_BraTS_103.pdf}{78.2}
		\caption*{\scriptsize\textbf{Global Factorizer}}
	\end{subfigure}%
	\begin{subfigure}{.167\textwidth}
		\centering
		\imagewithdice{figures/qualitative_results/unetr_pred_BraTS_103.pdf}{85.8}
		\caption*{\scriptsize\textbf{UNETR}}
	\end{subfigure}%
	\begin{subfigure}{.167\textwidth}
		\centering
		\imagewithdice{figures/qualitative_results/nnformer_pred_BraTS_103.pdf}{88.3}
		\caption*{\scriptsize\textbf{nnFormer}}
	\end{subfigure}%
	\begin{subfigure}{.167\textwidth}
		\centering
		\imagewithdice{figures/qualitative_results/unet_pred_BraTS_103.pdf}{74.7}
		\caption*{\scriptsize\textbf{nnU-Net}}
	\end{subfigure}	
	
	\caption{Qualitative results on brain tumor segmentation. All the examples are from the validation sets of the 5-fold cross-validation. TC is the union of red (NCR/NET) and yellow (ET) regions, and WT is the union of green (ED), red, and yellow regions. The patient-wise average Dice score is presented for each case.}
	\label{fig: brats qualitative results}
\end{figure}

\paragraph{Qualitative Comparisons}
Qualitative comparisons of glioma segmentation models are presented in Figure \ref{fig: brats qualitative results}. Swin Factorizer demonstrates superior performance in segmenting TC and ET. This capability is evident in row 1 (row 2), where Swin Factorizer more successfully delineates TC compared to the other models, which do not as accurately distinguish ED (normal tissues) from TC. Particularly, nnU-Net misclassifies a significantly larger part of ED (normal tissues) as NCR/NET (ET). Row 3 exemplifies a successful detection of TC by Swin Factorizer, while the other models miss a fairly large NCR/NET region.

\begin{figure}[t]
	\newcommand{\imagewithtext}[2]{
		\begin{tikzpicture}[every node/.style={inner sep=0, outer sep=0}]
			\node (image) at (0,0) {\includegraphics[width=.8\textwidth]{#1}};
			\node[white, anchor=west] (modality) at (-1.2, 1.2) {\fontfamily{phv}\scriptsize\textbf{#2}};
		\end{tikzpicture}
	}
	\centering
	\resizebox{\textwidth}{!}{
		\begin{tabular}{l !{\vrule width 2pt} c r}
			\begin{subfigure}{.2\textwidth}
				\begin{subfigure}{\textwidth}
					\centering
					\imagewithtext{figures/nmf_components/gt_BraTS_217.pdf}{GT}
				\end{subfigure}
				
				\begin{subfigure}{\textwidth}
					\centering
					\imagewithtext{figures/nmf_components/t1_BraTS_217.pdf}{T1}
				\end{subfigure}
				
				\begin{subfigure}{\textwidth}
					\centering
					\imagewithtext{figures/nmf_components/t1gd_BraTS_217.pdf}{T1Gd}
				\end{subfigure}
				
				\begin{subfigure}{\textwidth}
					\centering
					\imagewithtext{figures/nmf_components/t2_BraTS_217.pdf}{T2}
					\caption*{ }
				\end{subfigure}
				\caption{Modalities}
				\label{fig: modalities}
			\end{subfigure}&
			\begin{subfigure}{.4\textwidth}
				\centering
				\begin{subfigure}{.5\textwidth}
					\centering
					\includegraphics[width=.8\textwidth]{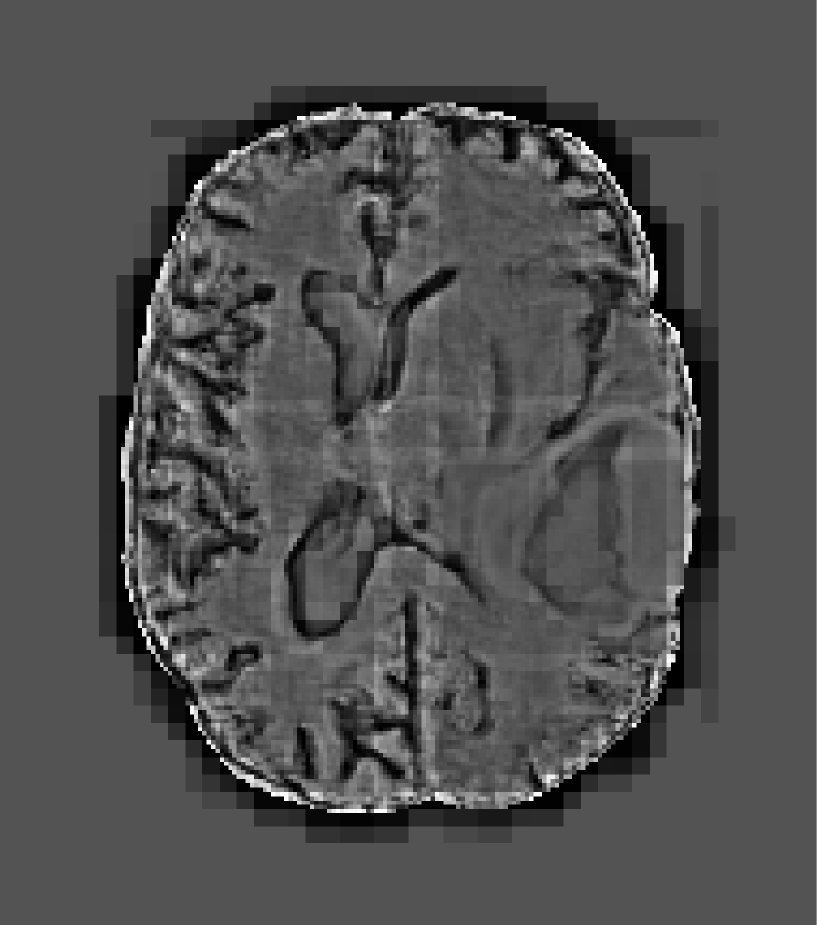}
				\end{subfigure}%
				\begin{subfigure}{.5\textwidth}
					\centering
					\includegraphics[width=.8\textwidth]{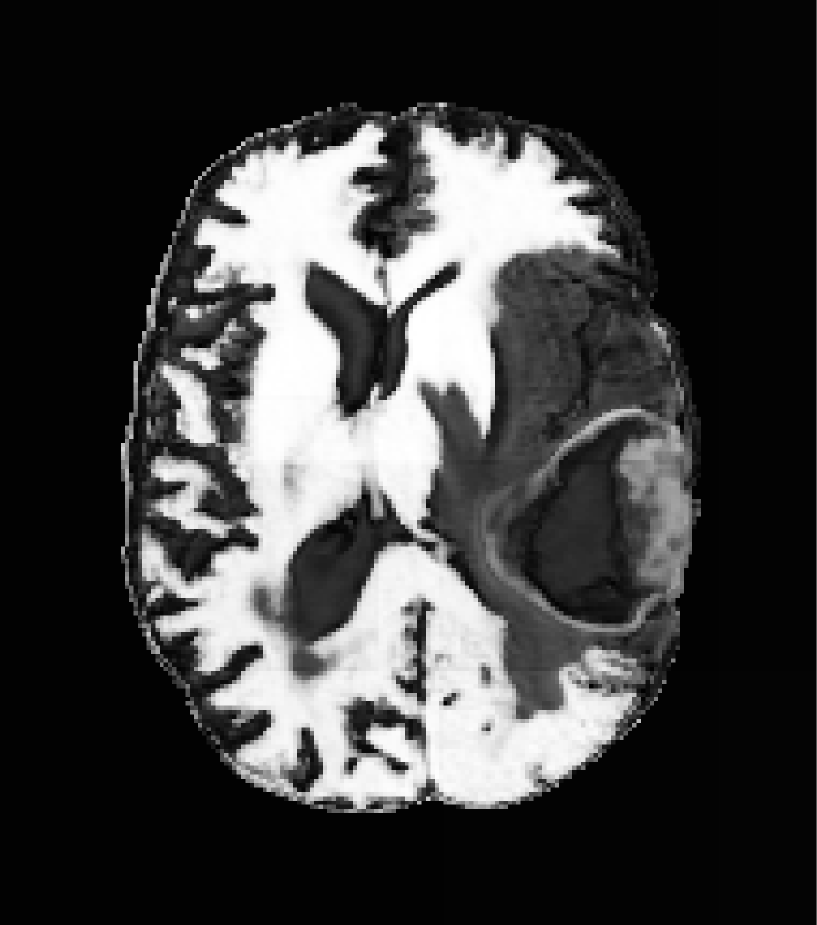}
				\end{subfigure}
				
				\begin{subfigure}{.5\textwidth}
					\centering
					\includegraphics[width=.8\textwidth]{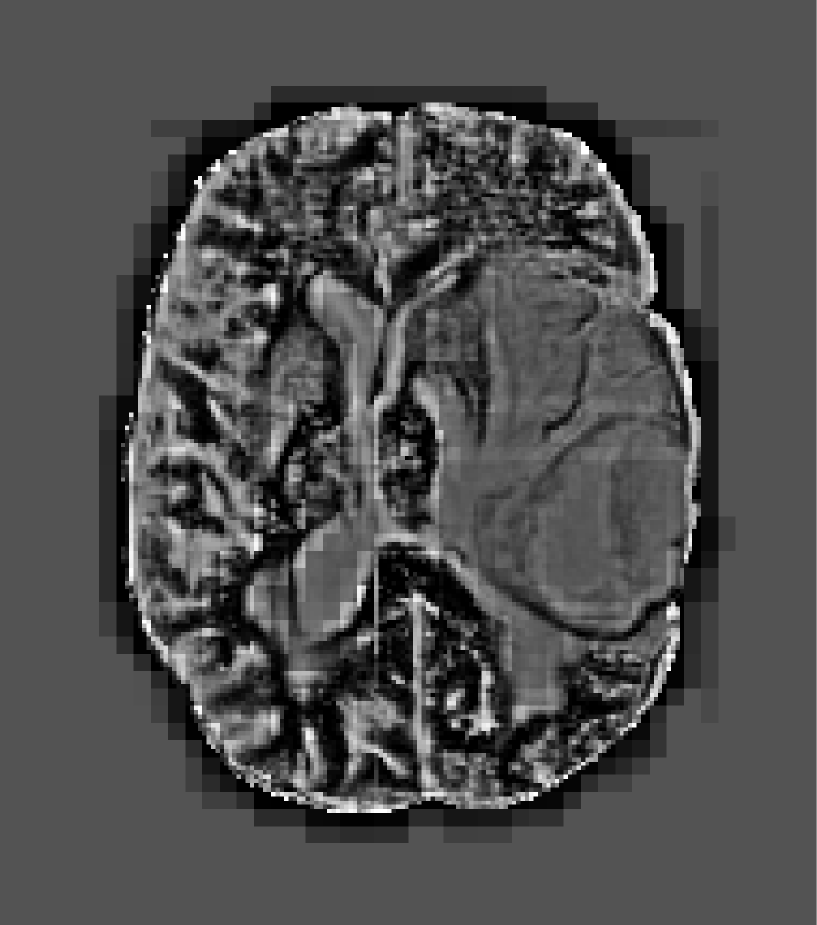}
				\end{subfigure}%
				\begin{subfigure}{.5\textwidth}
					\centering
					\includegraphics[width=.8\textwidth]{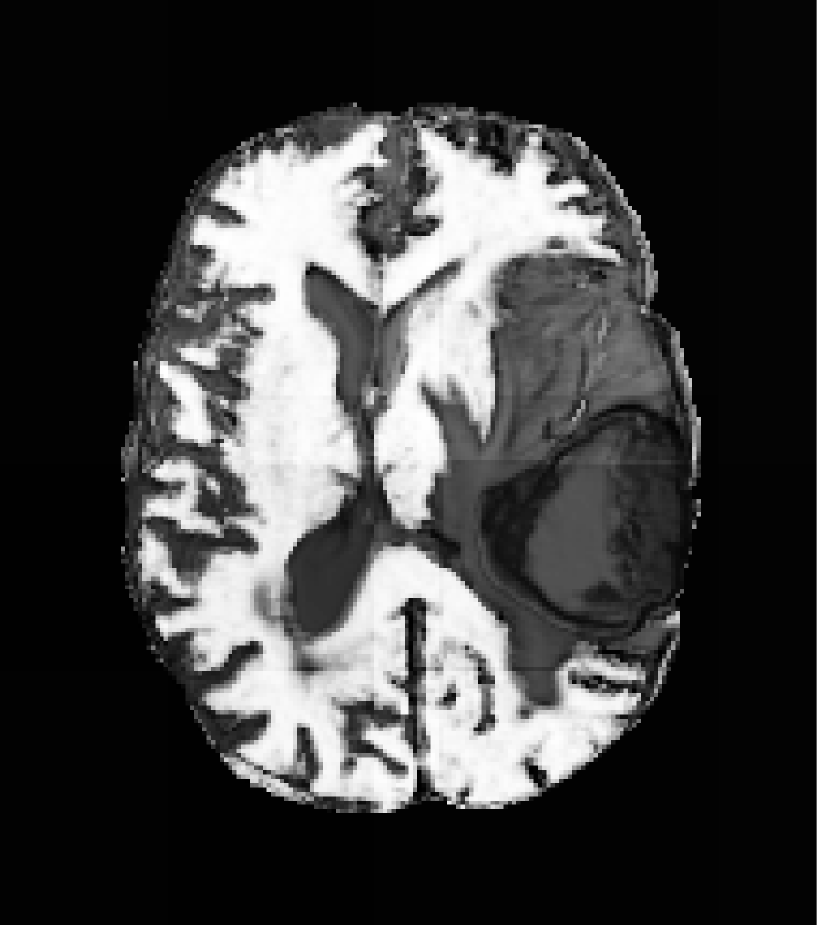}
				\end{subfigure}
				
				\begin{subfigure}{.5\textwidth}
					\centering
					\includegraphics[width=.8\textwidth]{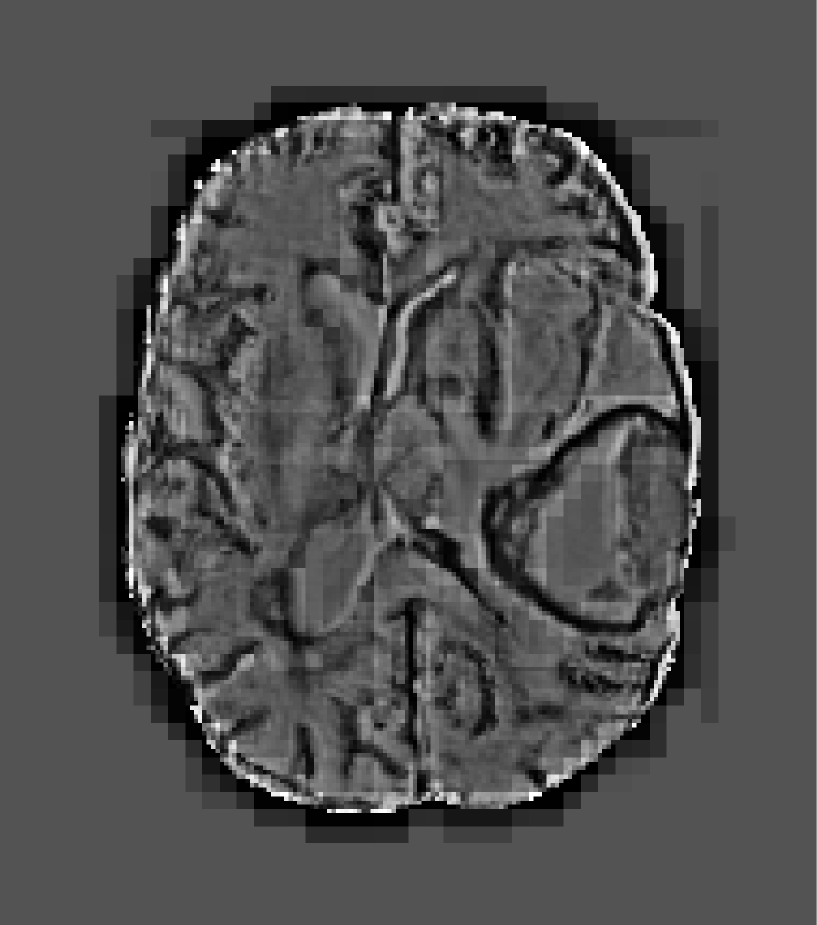}
				\end{subfigure}%
				\begin{subfigure}{.5\textwidth}
					\centering
					\includegraphics[width=.8\textwidth]{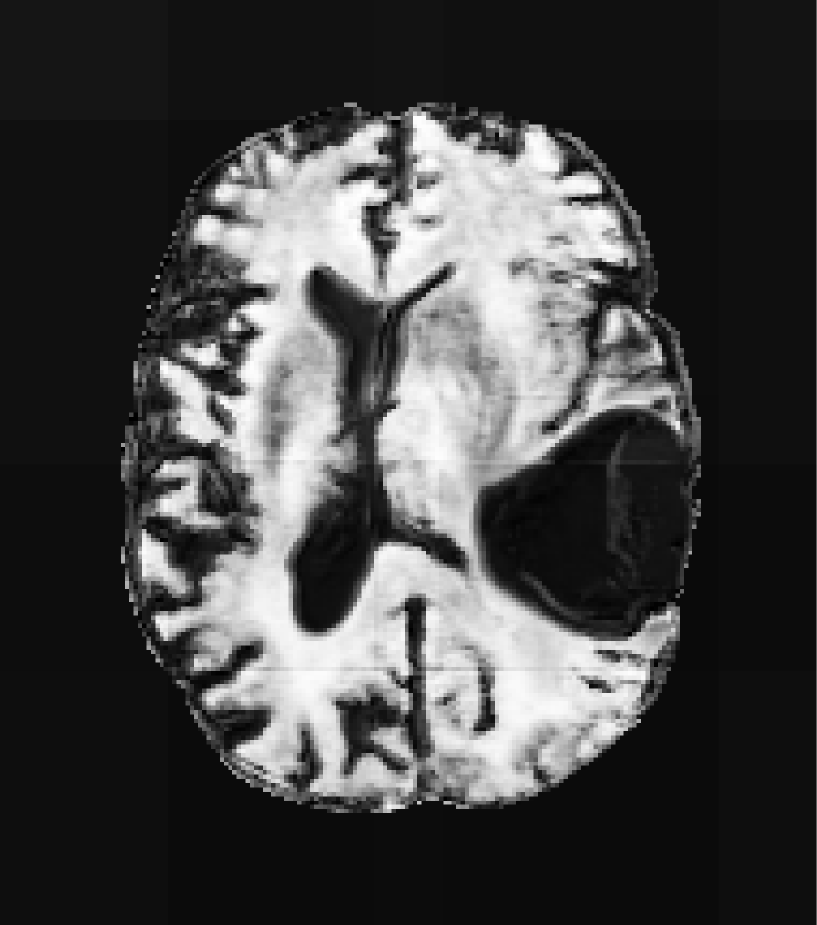}
				\end{subfigure}
				
				\begin{subfigure}{.5\textwidth}
					\centering
					\includegraphics[width=.8\textwidth]{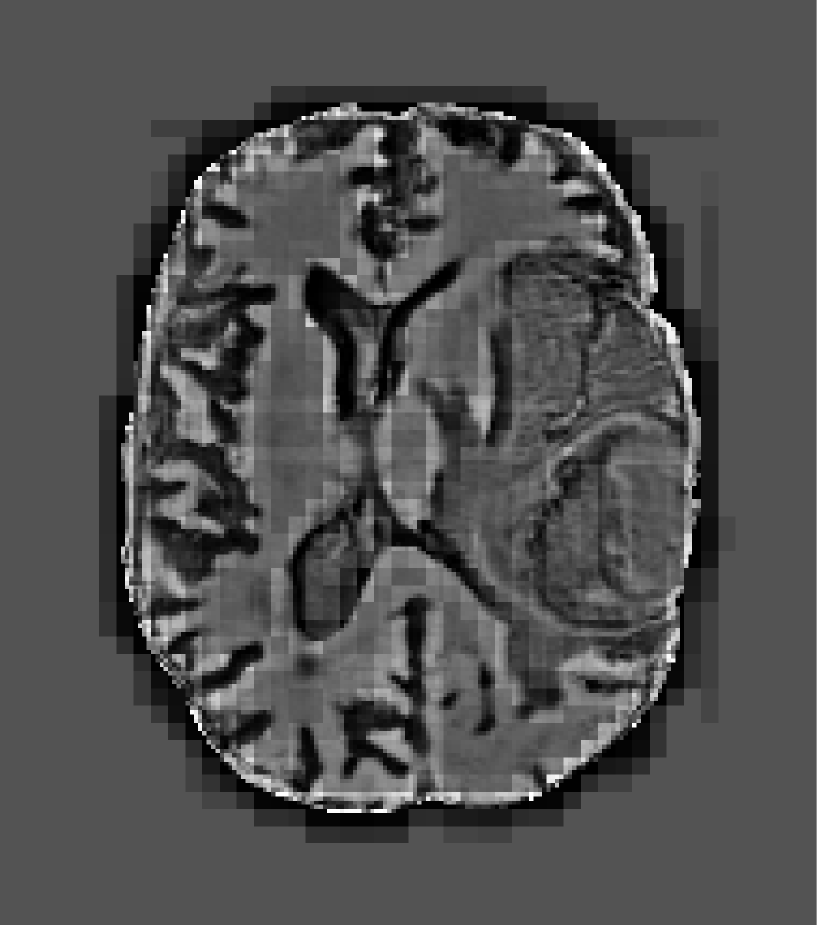}
					\caption*{\footnotesize\textbf{Swin Factorizer}}
					\label{fig: fist nmf layer components of swin factorizer}
				\end{subfigure}%
				\begin{subfigure}{.5\textwidth}
					\centering
					\includegraphics[width=.8\textwidth]{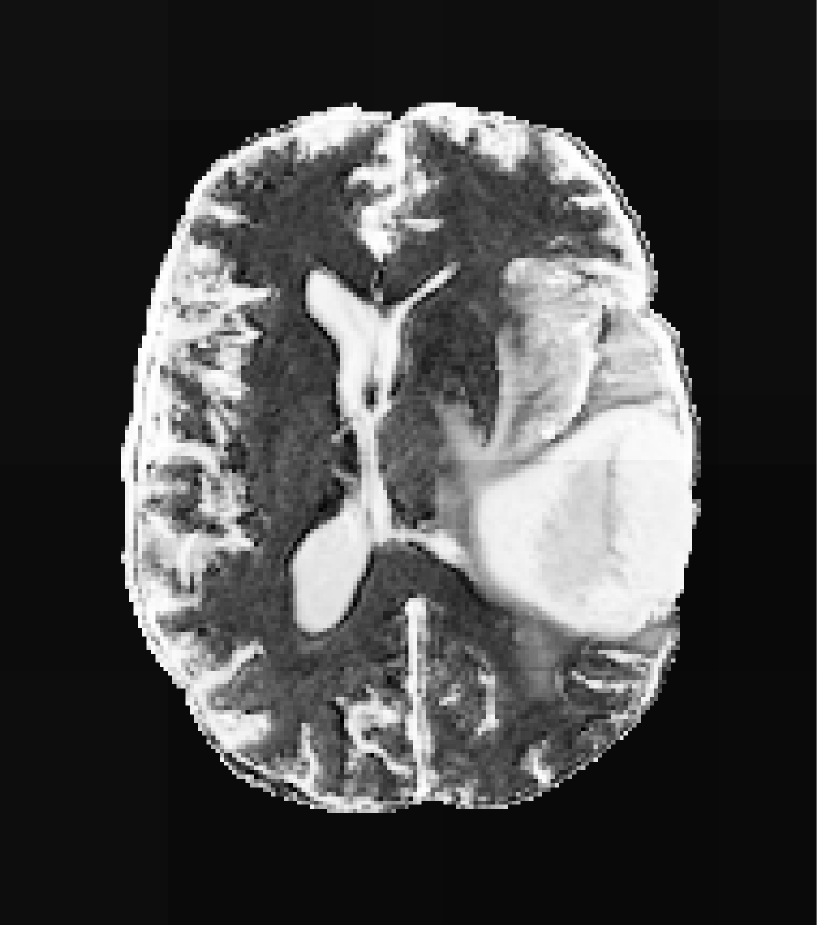}
					\caption*{\footnotesize\textbf{Global Factorizer}}
					\label{fig: first nmf layer components of global factorizer}
				\end{subfigure}
				\caption{Components of first NMF layer}
				\label{fig: first nmf layer components}
			\end{subfigure}&
			\begin{subfigure}{.4\textwidth}
				\centering
				\begin{subfigure}{.5\textwidth}
					\centering
					\includegraphics[width=.8\textwidth]{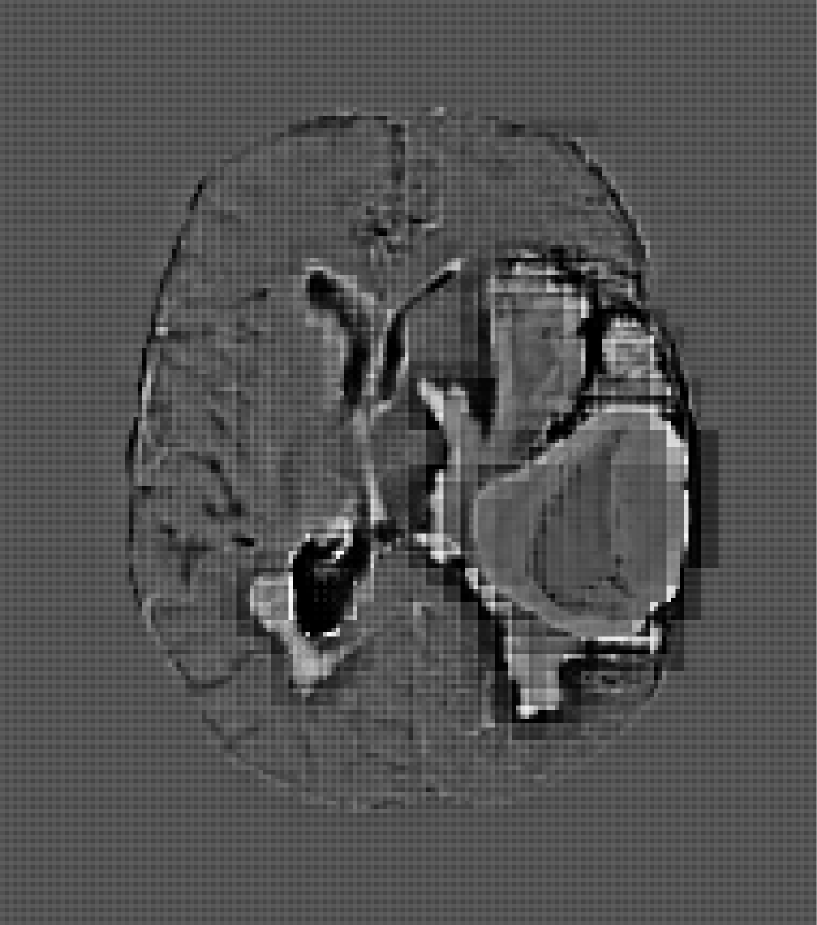}
				\end{subfigure}%
				\begin{subfigure}{.5\textwidth}
					\centering
					\includegraphics[width=.8\textwidth]{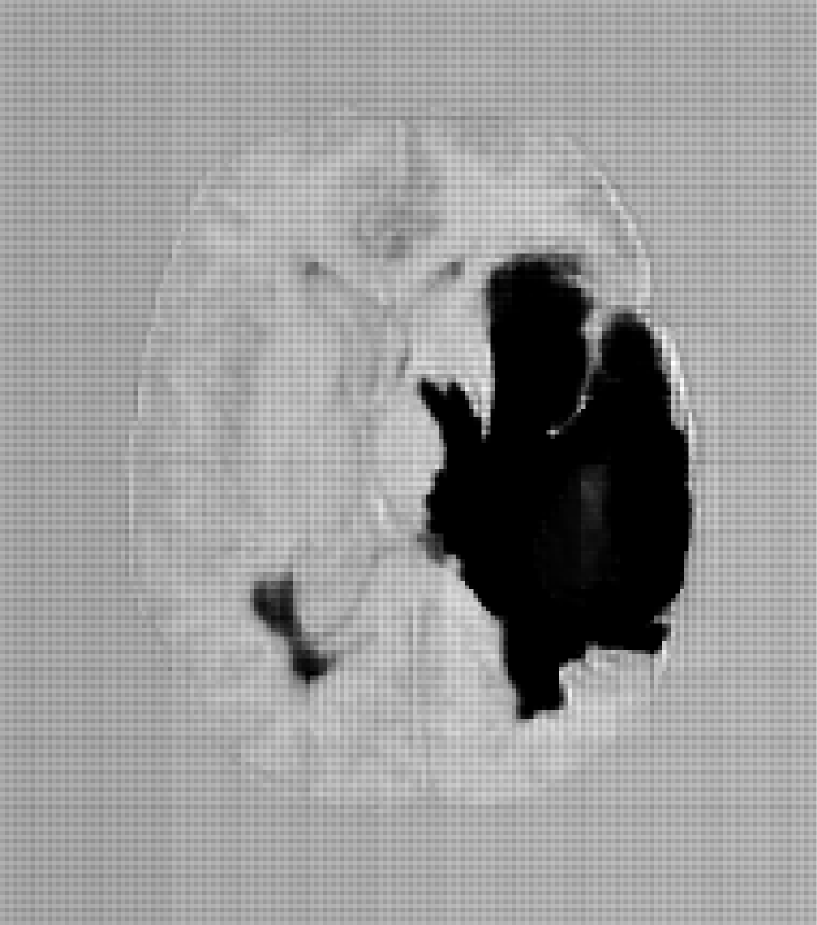}
				\end{subfigure}
				
				\begin{subfigure}{.5\textwidth}
					\centering
					\includegraphics[width=.8\textwidth]{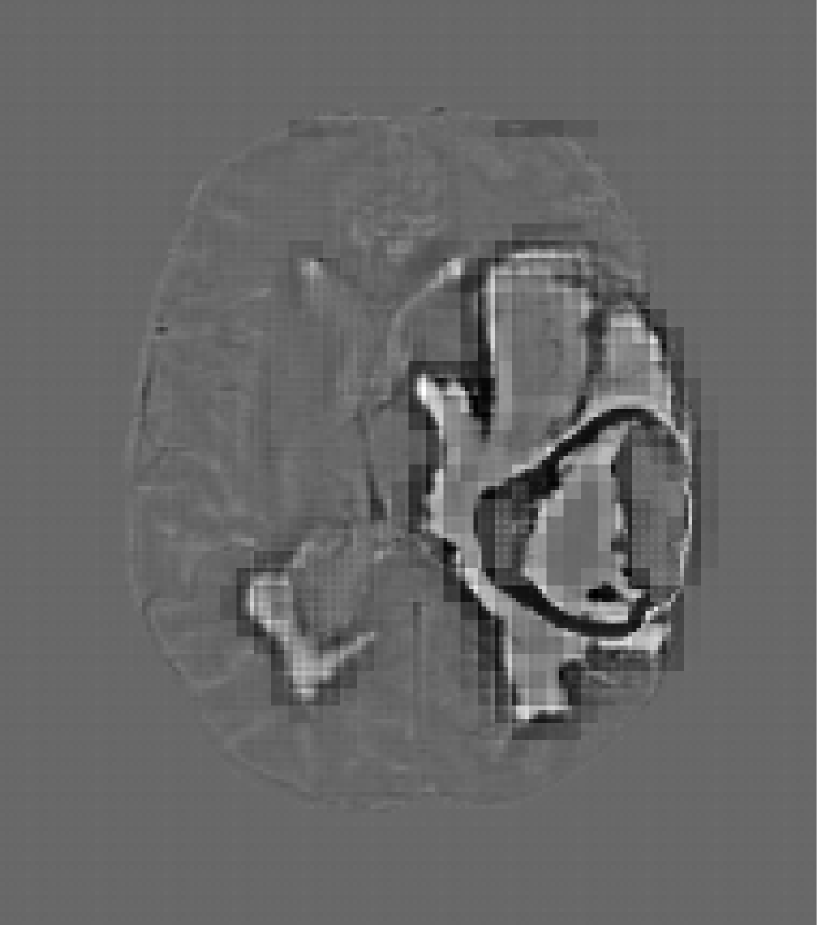}
				\end{subfigure}%
				\begin{subfigure}{.5\textwidth}
					\centering
					\includegraphics[width=.8\textwidth]{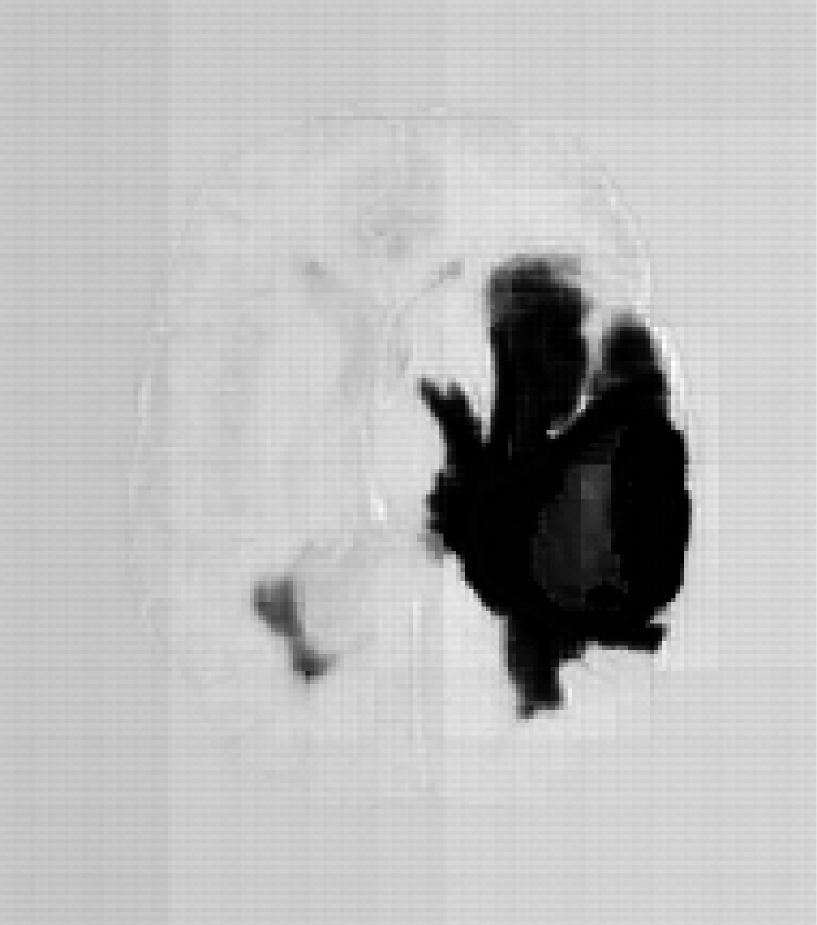}
				\end{subfigure}
				
				\begin{subfigure}{.5\textwidth}
					\centering
					\includegraphics[width=.8\textwidth]{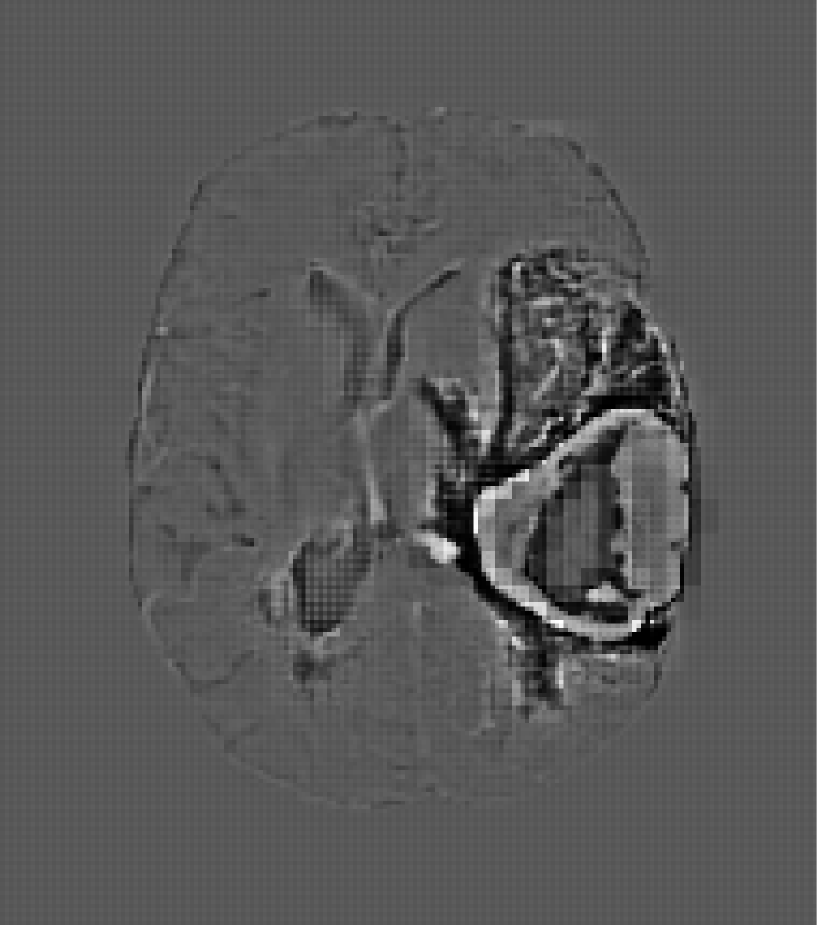}
				\end{subfigure}%
				\begin{subfigure}{.5\textwidth}
					\centering
					\includegraphics[width=.8\textwidth]{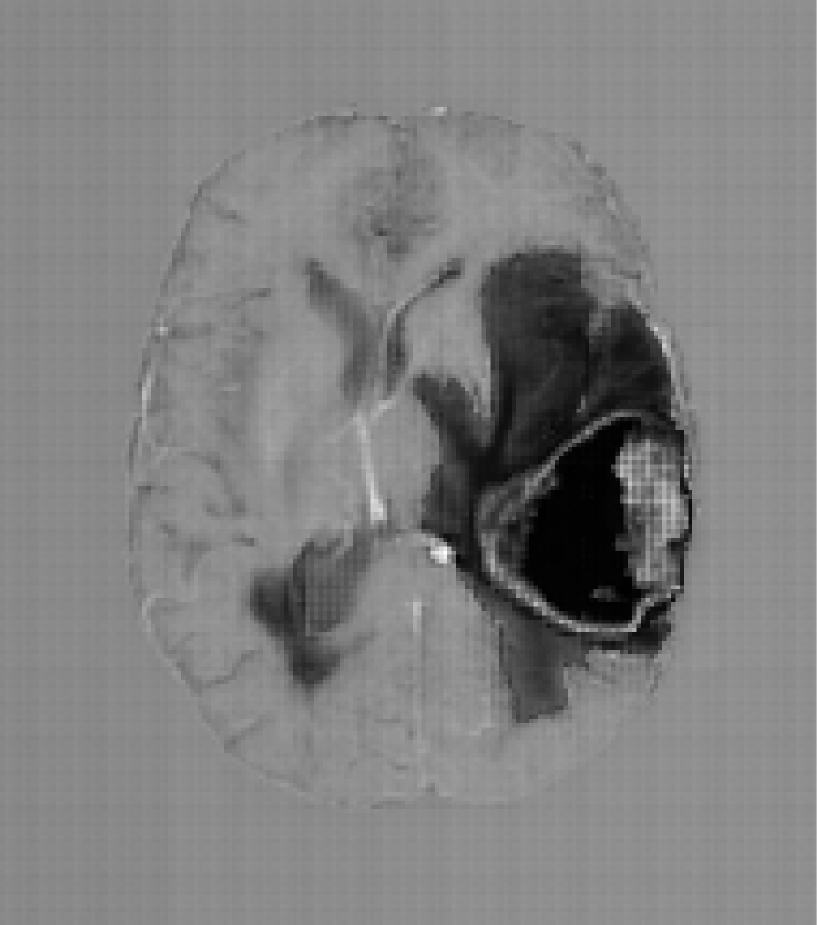}
				\end{subfigure}
				
				\begin{subfigure}{.5\textwidth}
					\centering
					\includegraphics[width=.8\textwidth]{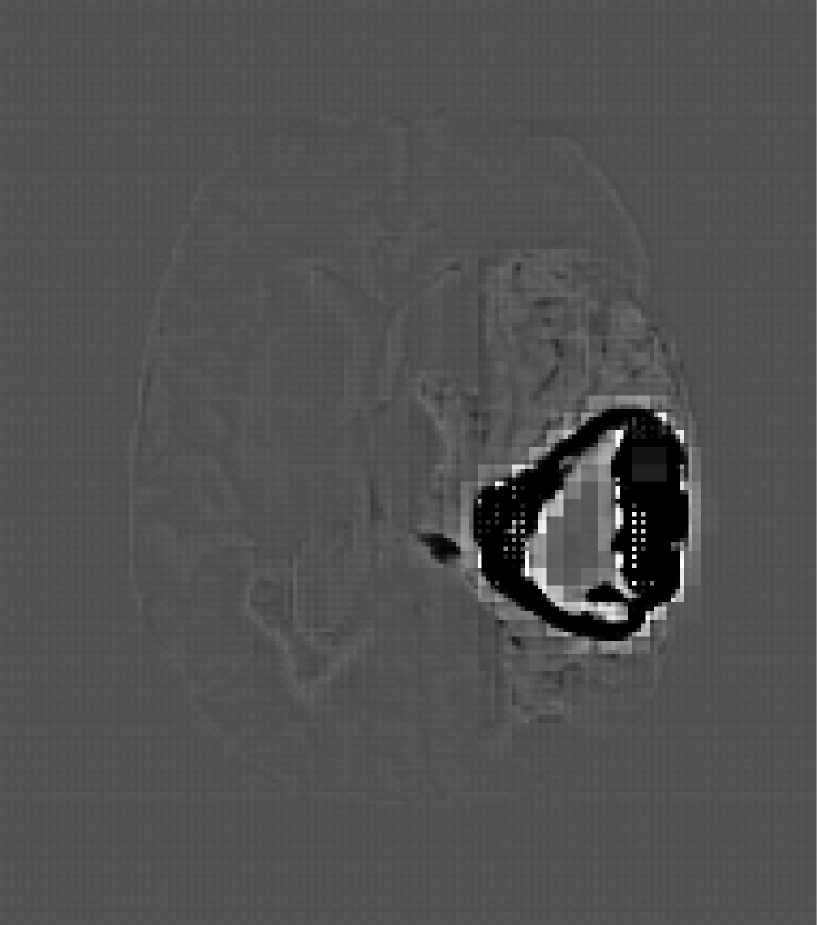}
					\caption*{\footnotesize\textbf{Swin Factorizer}}
					\label{fig: last nmf layer components of swin factorizer}
				\end{subfigure}%
				\begin{subfigure}{.5\textwidth}
					\centering
					\includegraphics[width=.8\textwidth]{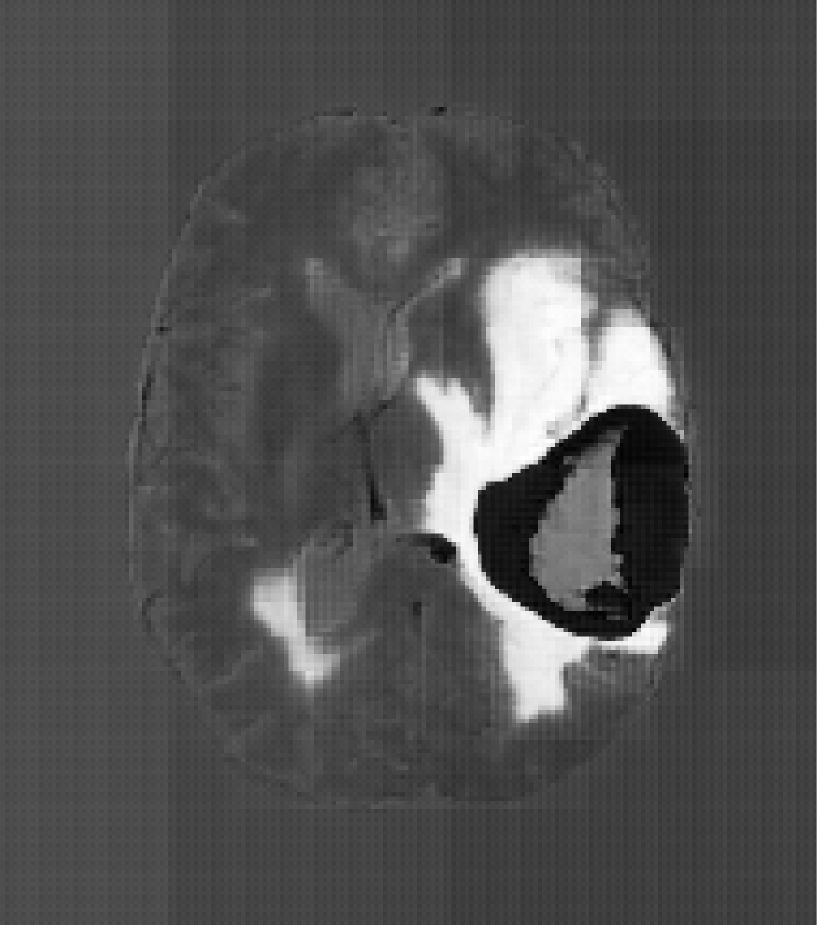}
					\caption*{\footnotesize\textbf{Global Factorizer}}
					\label{fig: last nmf layer components of global factorizer}
				\end{subfigure}
				\caption{Components of last NMF layer}
				\label{fig: last nmf layer components}
			\end{subfigure}
		\end{tabular}
	}
	
	\caption{NMF Components of Factorizer models on BraTS. The example is from a validation set of the 5-fold cross-validation. Panel (a) shows the modalities and ground truth for a high-grade glioma case. Panel (b) and (c) present the NMF spatial components in the first and last block, respectively.}
	\label{fig: nmf components}
\end{figure}

\paragraph{NMF Components Interpretability}
One additional advantage of Factorizer over Transformer and CNN models is its higher interpretability resulting from meaningful components of NMF in the sense that each component represents specific image semantics in practice. Note that both the first and last Factorizer blocks (i.e.,  the high-resolution blocks of the encoder and the decoder, which are just after the stem layer and just before the head layer, respectively) have $ C = 32 $ channels, divided into groups of 8-channel heads during matricization, where the NMF of each head has only a single rank-one term, i.e., $ R = 1 $. As a result, in total, there are $ CR/8 = 4 $ components in both the first and last blocks. Figure \ref{fig: first nmf layer components} and \ref{fig: last nmf layer components} show the components of the first and last NMF layers on a high-grade glioma case for Swin Factorizer and Global Factorizer. More precisely, each component illustrates the factor matrix corresponding to spatial dimensions after dematricization.

Interestingly, the components appear highly meaningful and interpretable in the sense that each component gives an interpretation by differentiating one region from another. For instance, as observed in Figure \ref{fig: first nmf layer components}, the first (row 1) and second (row 2) components discriminate roughly WT while the third (row 3) and fourth (row 4) components capture TC. For Swin Factorizer, NCR is distinguished more clearly in the third and fourth components whereas for Global Factorizer NCR is more recognizable in the second component. As we get to deeper layers, components become even more meaningful such that in the last layer, each component clearly detects some regions. For example, as seen in Figure \ref{fig: last nmf layer components}, the first (row 1) and second (row 2) components very clearly discriminate WT while the third (row 3) and fourth (row 4) components also capture TC and NCR. 

Another interesting observation is that Global Factorizer yields more discriminative and higher-level components in the first layer, which can be attributed to the fact Global Factorizer models long-range dependencies through all of its blocks, including the first one; however, the receptive field of Swin Factorizer in the first stage of the encoder is small but progressively increases as it passes through downsampling layers. Therefore, Swin Factorizer extracts lower-level features in shallower layers and higher-level ones in deeper layers in a similar way to CNNs. Notice that the footprints of sliding windows, which appear as a grid pattern, are also evident in all the components of Swin Factorizer.

\subsubsection{Ischemic Stroke Lesion Segmentation (ISLES)}  \label{sec: results and discussion: isles}

\paragraph{Quantitative Evaluation}
Similarly to Section \ref{sec: results and discussion: brats}, 5-fold cross-validation was performed, and pairwise Wilcoxon signed-rank tests were used for comparison purposes. The results on the ISLES'22 dataset are reported in \ref{tab: isles validation results} and illustrated by box plots in Figure \ref{fig: isles dice boxplot} and \ref{fig: isles hd boxplot}.

Swin Factorizer, with a Dice score of 76.49\% and HD95 of 11.96 mm, demonstrated the best performance, significantly outperforming nnU-Net and all the Transformer-based models with a p-value of $ <0.001 $ for the Dice score and $ <0.01 $ for the HD95 value. Moreover, Swin Factorizer showed improved performance compared to Res-U-Net, which has over three times more parameters and needs over four times more FLOPs. 

Despite having over 95\% fewer parameters and taking over 50\% fewer FLOPs, Local Factorizer yielded a Dice score of 74.28\%, which is higher than that of nnFormer, the best-performing Transformer-based model. Local Factorizer also demonstrated a smaller HD95 value with 80\% fewer FLOPs than Res-U-Net, the best baseline overall. Finally, Global Factorizer outperformed its Transformer-based model, Performer, by a large margin, whereas it still has the advantage of lower computational cost. As a side note, our Factorizer-based model trained on all the three modalities (i.e., DWI, ADC, and registered FLAIR) and submitted to the ISLES'22 MICCAI challenge ranked among the top three in the final leaderboard, which further verifies the potential of Factorizer as an effective alternative for 3D medical image segmentation (please see \href{https://isles22.grand-challenge.org/isles22/}{https://isles22.grand-challenge.org/isles22/}).

\begin{table}[t]
	\caption{Comparison of different models on the ISLES22 dataset. Scores are obtained by 5-fold cross-validation. The best results are in \textbf{boldface} and second best ones are \underline{underlined}.}
	\label{tab: isles validation results}
	\newcolumntype{P}[1]{>{\centering\arraybackslash}p{#1}}
	\def\arraystretch{1}
	\resizebox{\textwidth}{!}{	
		\begin{tabular}{m{.2\textwidth} | P{.15\textwidth} P{.15\textwidth} | P{.15\textwidth} P{.15\textwidth}}		
			\toprule
			
			Model             & \#Params & FLOPs  & Dice (\%) \textuparrow & HD95 (mm) \textdownarrow \\
			
			\hline
			
			nnU-Net           & 28.7M     & 143.1G & 69.71     & 26.59     \\
			Res-U-Net         & 28.9M    & 145.1G & \underline{75.41}     & 16.51     \\
			
			\hline
			
			Performer         & 8.2M     & 33.0G  & 71.30     & 22.18     \\
			TransBTS          & 31.0M    & 66.1G  & 72.44     & 19.35     \\
			UNETR             & 104.8M   & 137.8G & 61.43     & 41.62     \\
			Swin UNETR        & 62.2M    & 194.6G & 70.41     & 31.67     \\
			nnFormer          & 149.2M    & 60.6G  & 73.79     & \underline{12.99}     \\
			
			\hline
			
			Global Factorizer & 7.4M     & 28.5G  & 72.51     & 18.06     \\
			Local Factorizer  & 7.4M     & 28.5G  & 74.28     & 16.14     \\
			Swin Factorizer   & 7.4M     & 29.1G  & \textbf{76.49}     & \textbf{11.96}   \\
			
			\bottomrule
		\end{tabular}
	}
\end{table}

\begin{figure}[t]
	\newcommand{\imagewithdice}[2]{
		\begin{tikzpicture}[every node/.style={inner sep=0, outer sep=0}]
			\node (image) at (0,0) {\includegraphics[width=1.0\textwidth]{#1}};
			\node[white, anchor=west] (dice) at (-1.45, -1.45) {\fontfamily{phv}\scriptsize\textbf{}};
		\end{tikzpicture}
		\vspace*{-15pt}	
	}
	\centering
	
	\begin{subfigure}{1.0\textwidth}
		\centering
		\includegraphics[width=.6\textwidth]{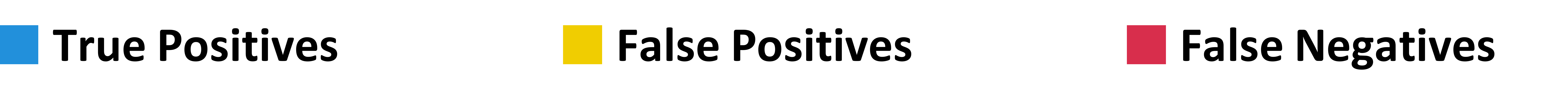}
	\end{subfigure}

	\begin{subfigure}{.2\textwidth}
		\centering
		\imagewithdice{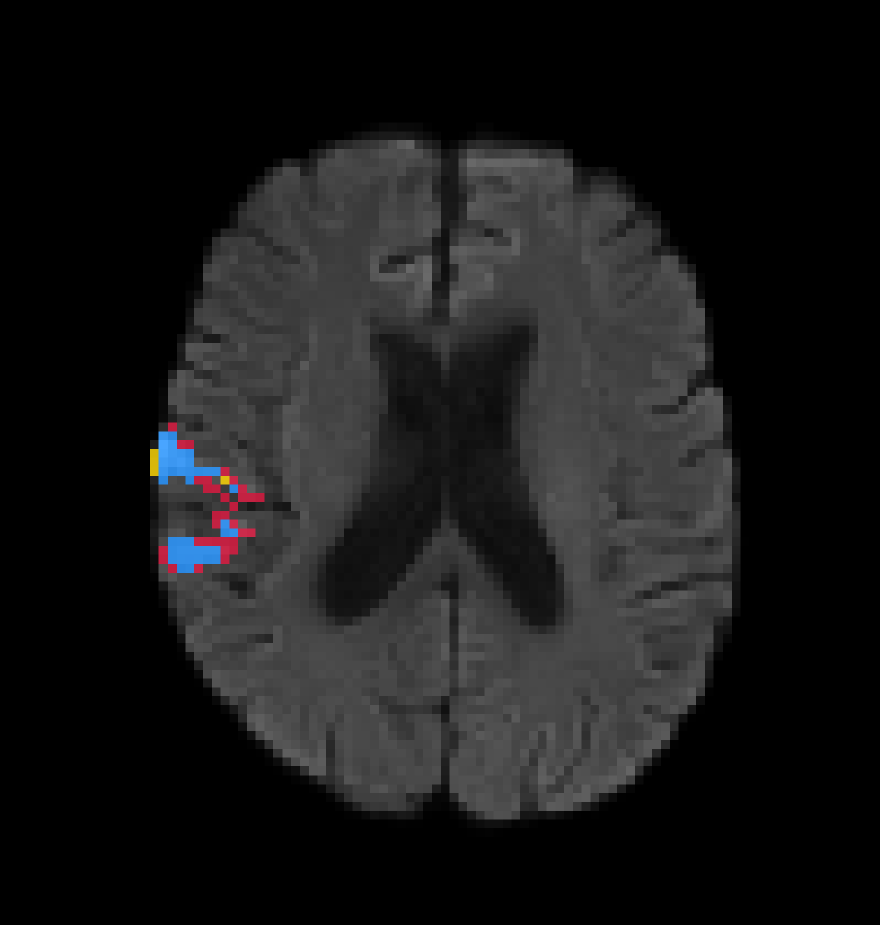}{86.2}
	\end{subfigure}%
	\begin{subfigure}{.2\textwidth}
		\centering
		\imagewithdice{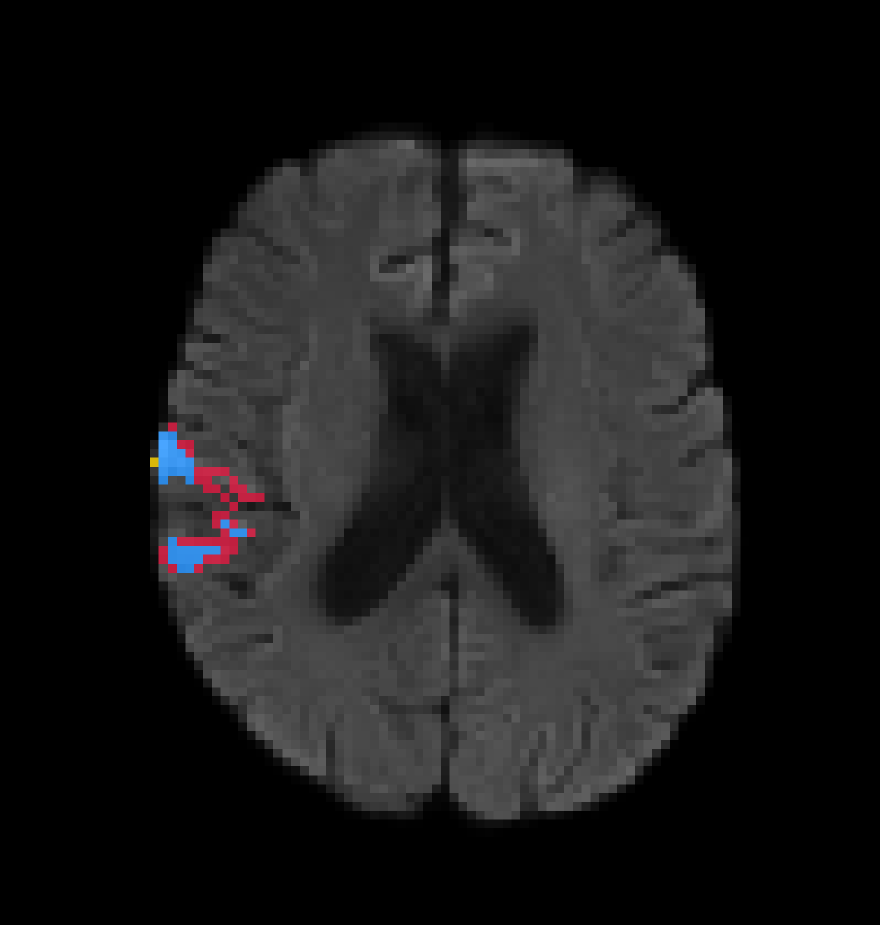}{80.9}
	\end{subfigure}%
	\begin{subfigure}{.2\textwidth}
		\centering
		\imagewithdice{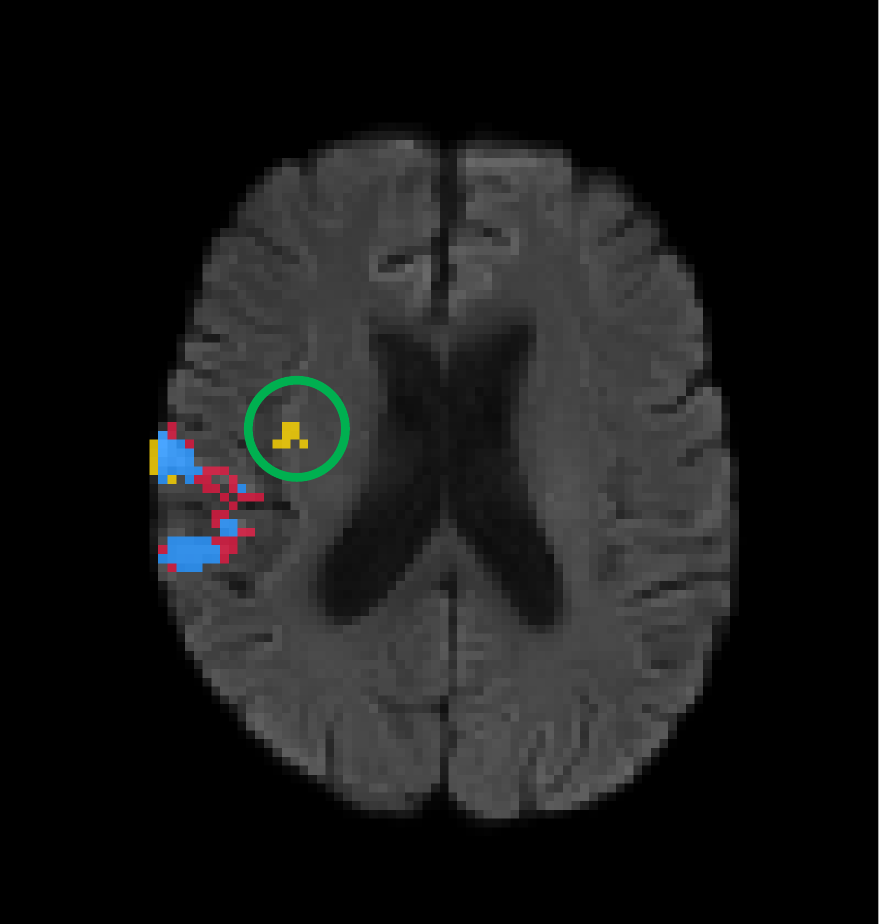}{79.3}
	\end{subfigure}%
	\begin{subfigure}{.2\textwidth}
		\centering
		\imagewithdice{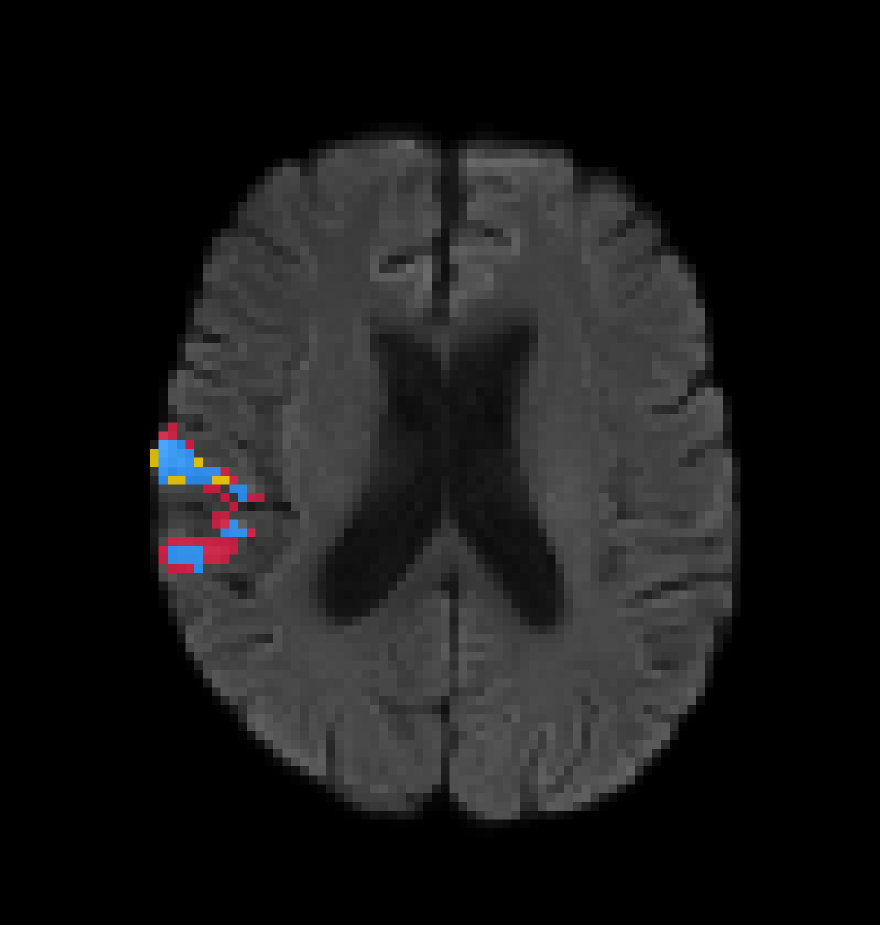}{78.9}
	\end{subfigure}%
	\begin{subfigure}{.2\textwidth}
		\centering
		\imagewithdice{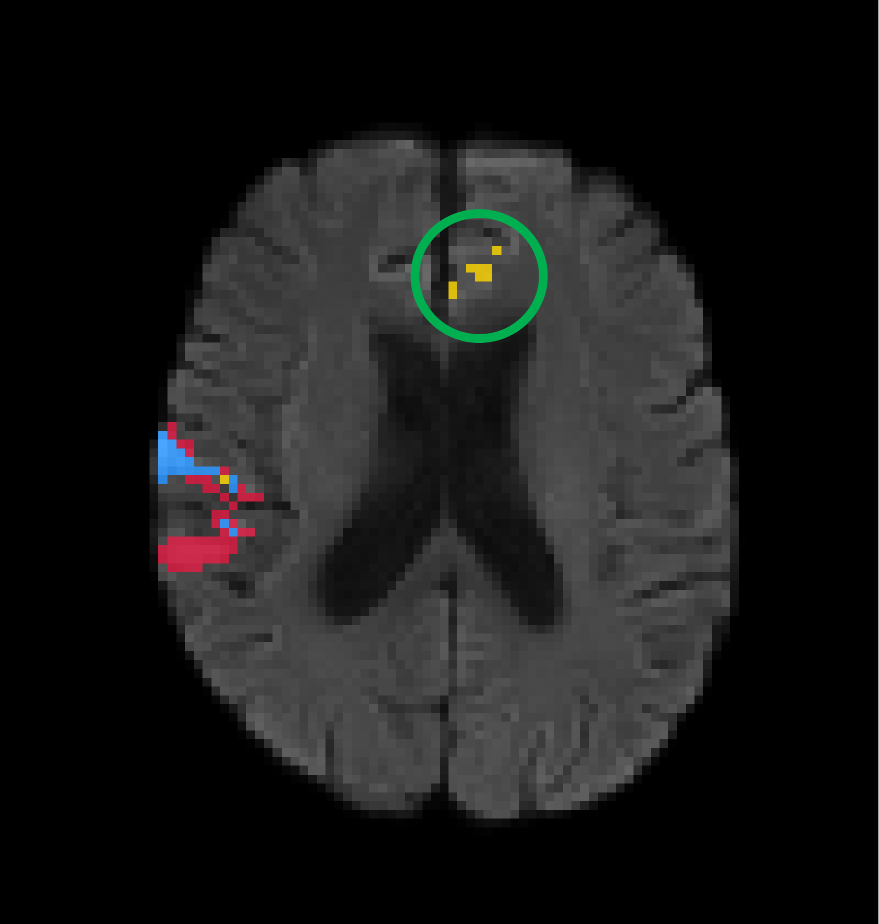}{64.4}
	\end{subfigure}

	\begin{subfigure}{.2\textwidth}
		\centering
		\imagewithdice{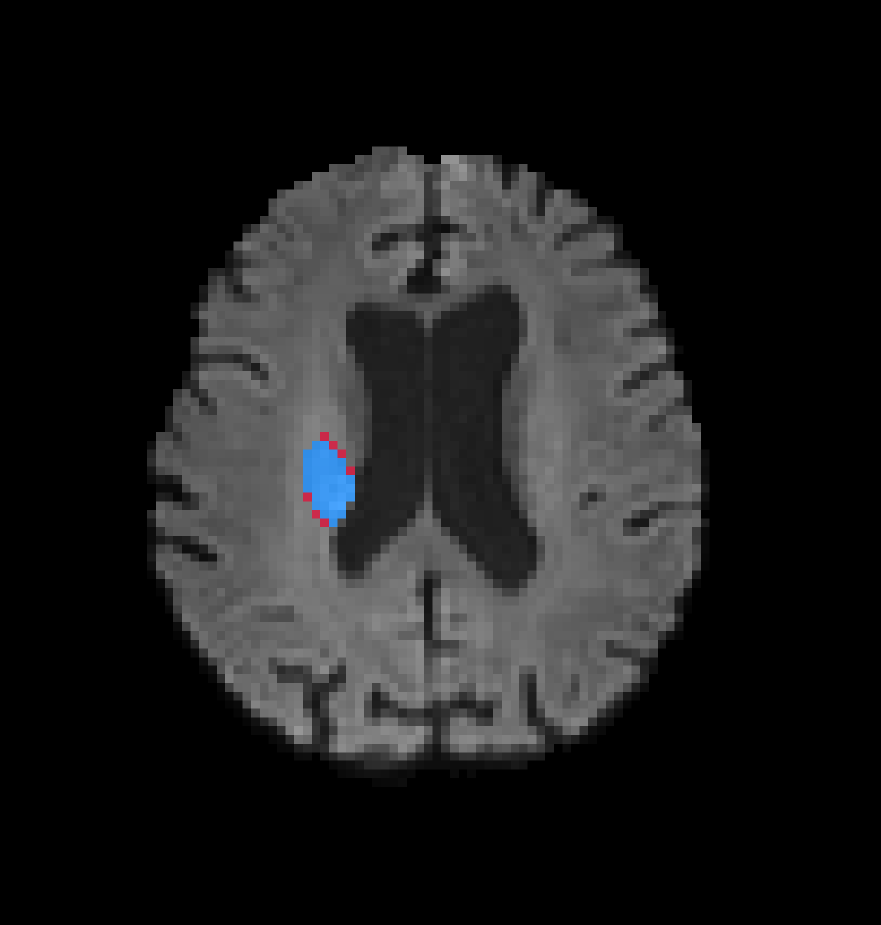}{90.8}
	\end{subfigure}%
	\begin{subfigure}{.2\textwidth}
		\centering
		\imagewithdice{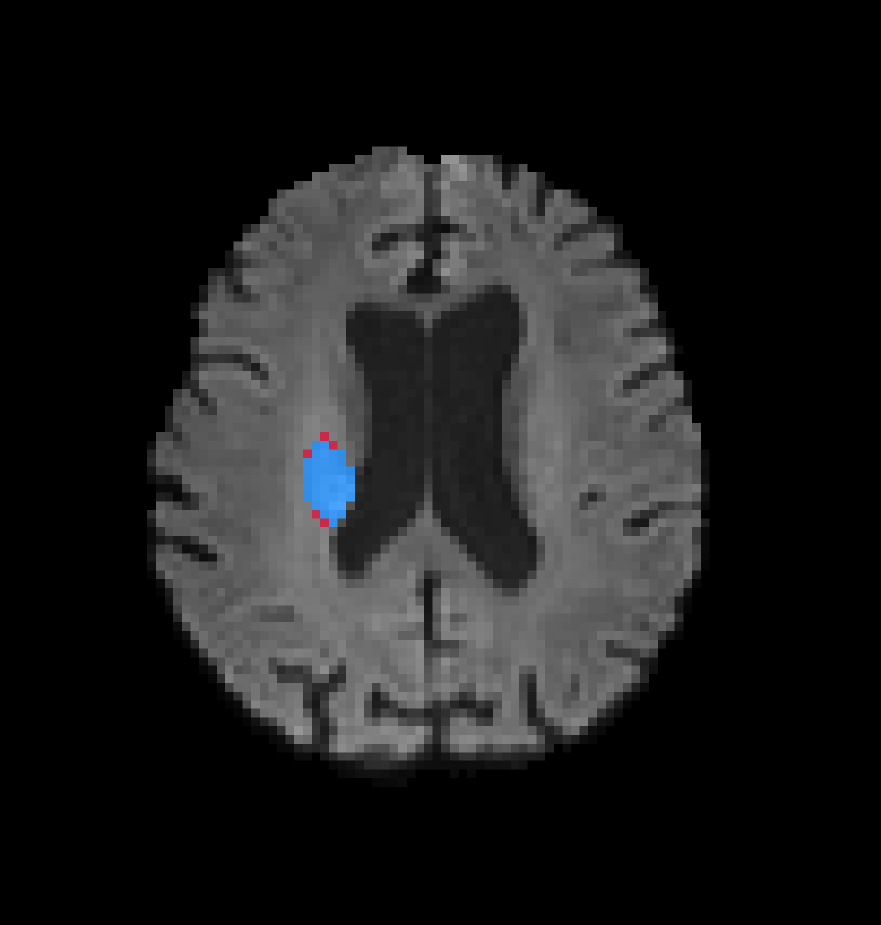}{90.7}
	\end{subfigure}%
	\begin{subfigure}{.2\textwidth}
		\centering
		\imagewithdice{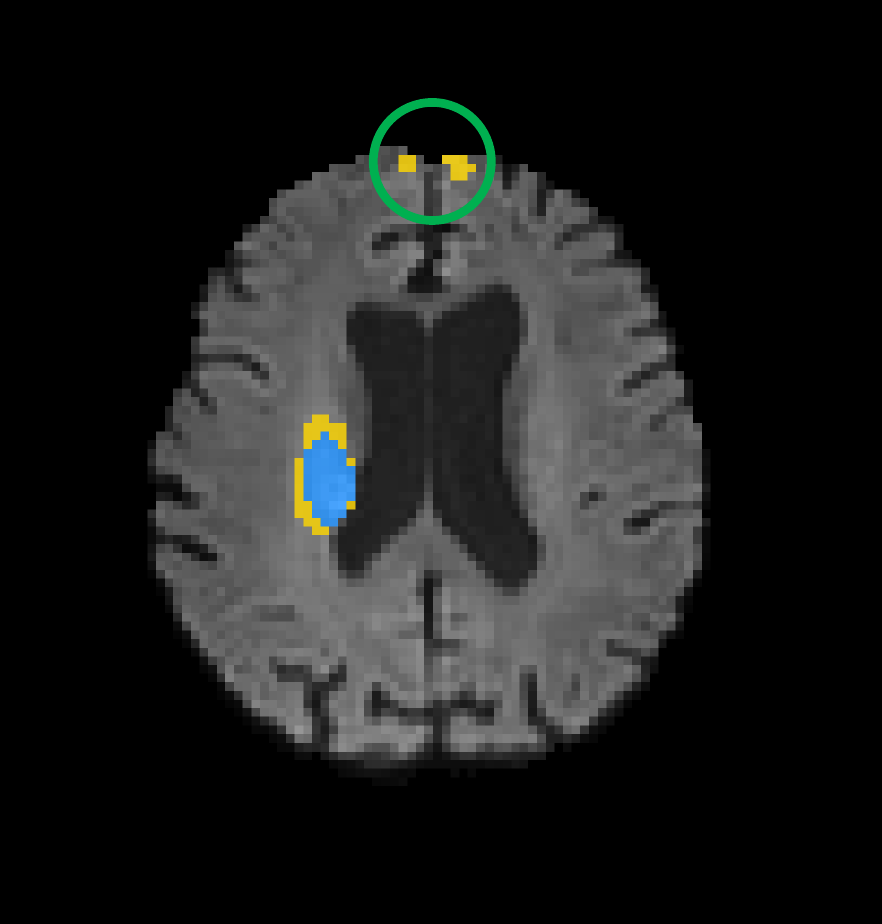}{50.8}
	\end{subfigure}%
	\begin{subfigure}{.2\textwidth}
		\centering
		\imagewithdice{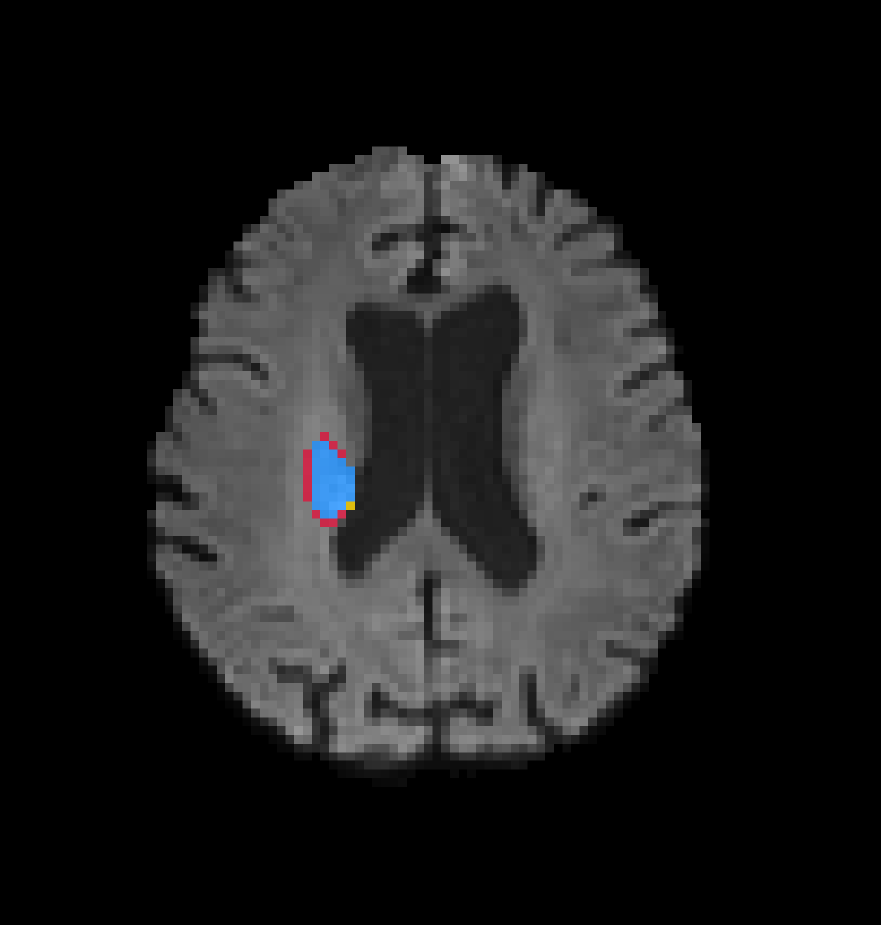}{89.1}
	\end{subfigure}%
	\begin{subfigure}{.2\textwidth}
		\centering
		\imagewithdice{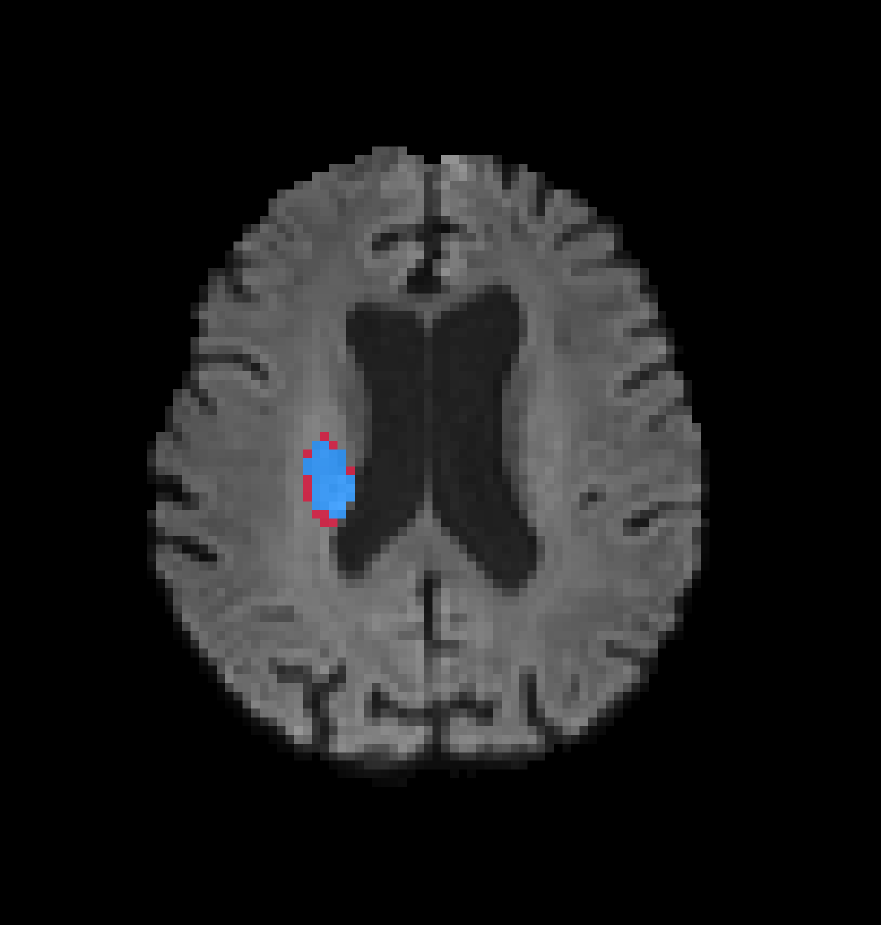}{89.2}
	\end{subfigure}

	\begin{subfigure}{.2\textwidth}
		\centering
		\imagewithdice{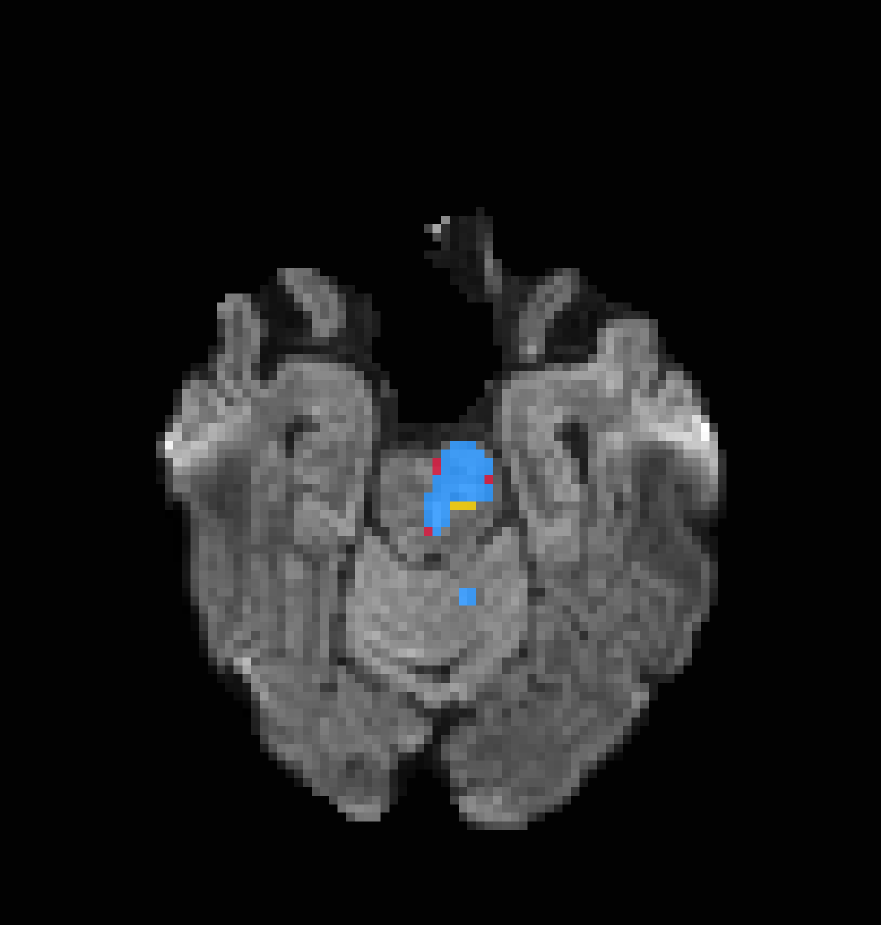}{81.1}
		\caption*{\scriptsize\textbf{Swin Factorizer}}
	\end{subfigure}%
	\begin{subfigure}{.2\textwidth}
		\centering
		\imagewithdice{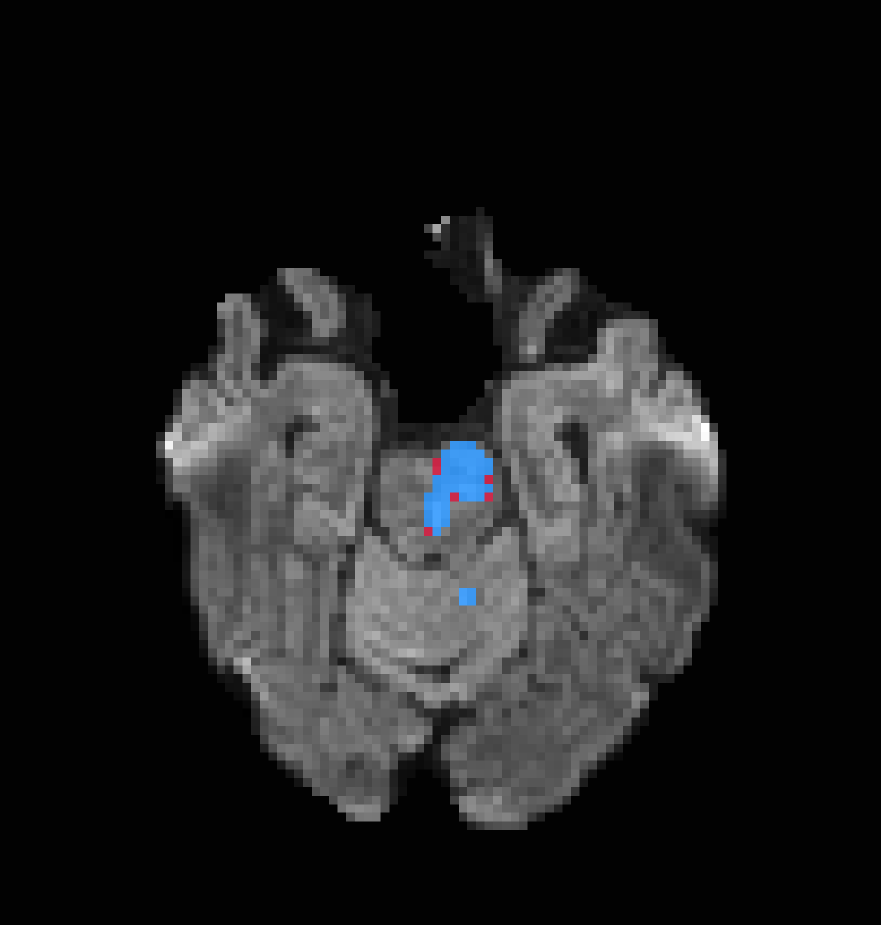}{78.5}
		\caption*{\scriptsize\textbf{Global Factorizer}}
	\end{subfigure}%
	\begin{subfigure}{.2\textwidth}
		\centering
		\imagewithdice{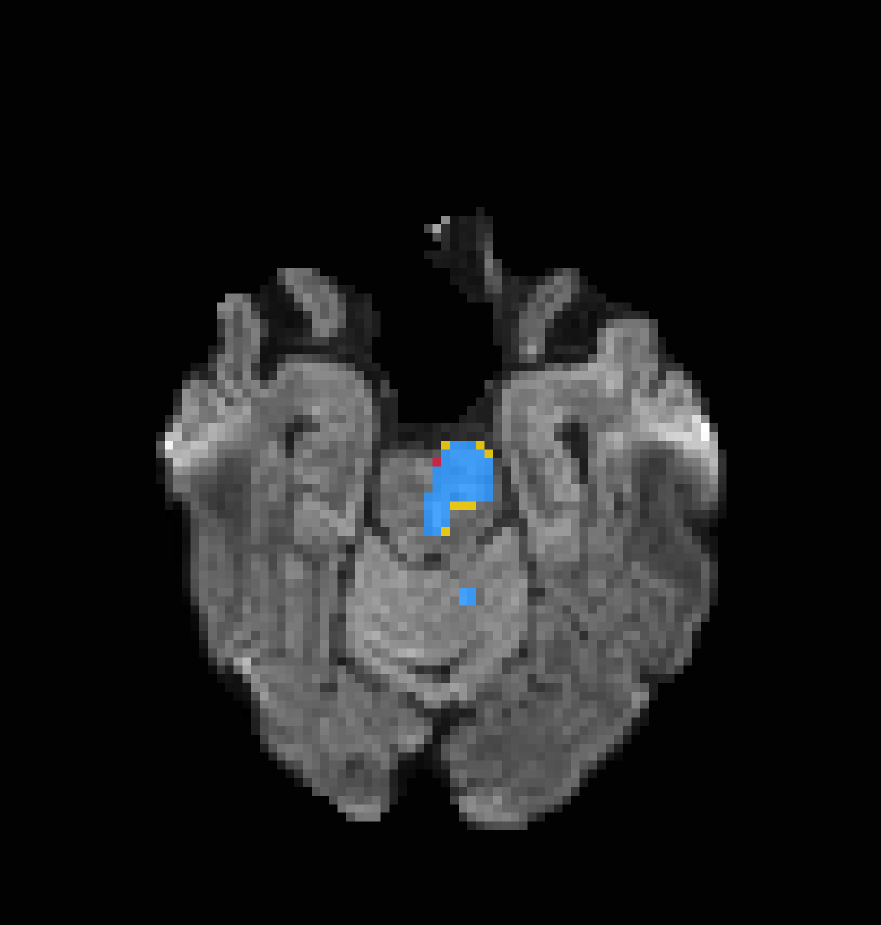}{76.8}
		\caption*{\scriptsize\textbf{UNETR}}
	\end{subfigure}%
	\begin{subfigure}{.2\textwidth}
		\centering
		\imagewithdice{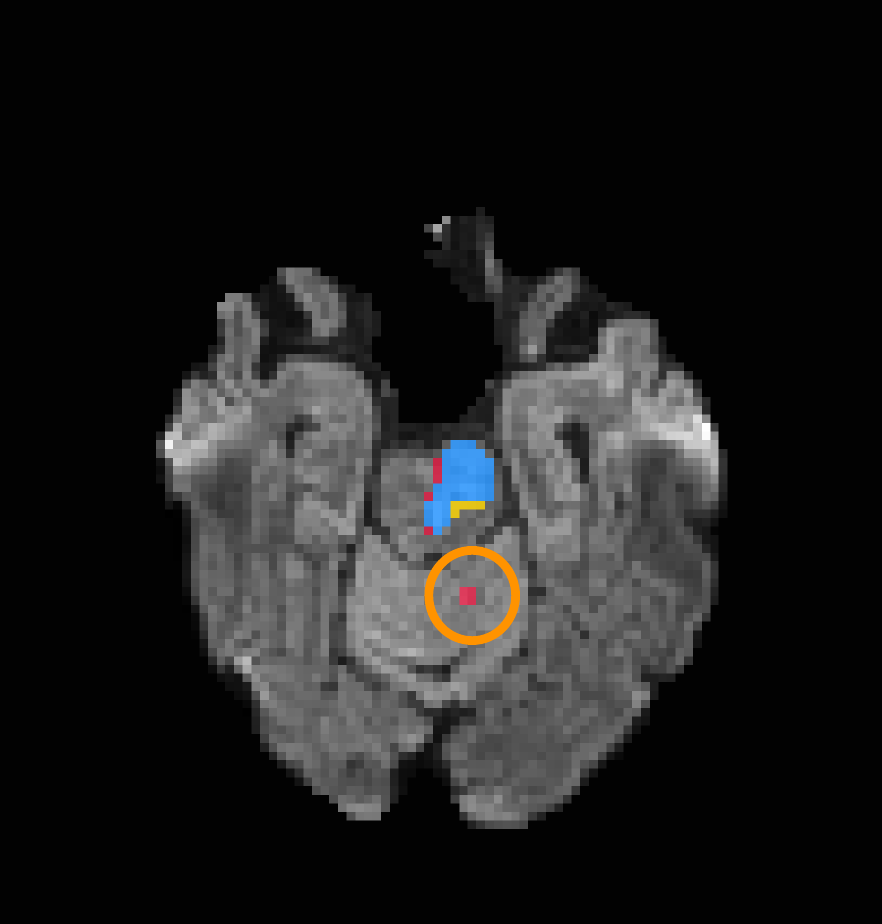}{73.6}
		\caption*{\scriptsize\textbf{nnFormer}}
	\end{subfigure}%
	\begin{subfigure}{.2\textwidth}
		\centering
		\imagewithdice{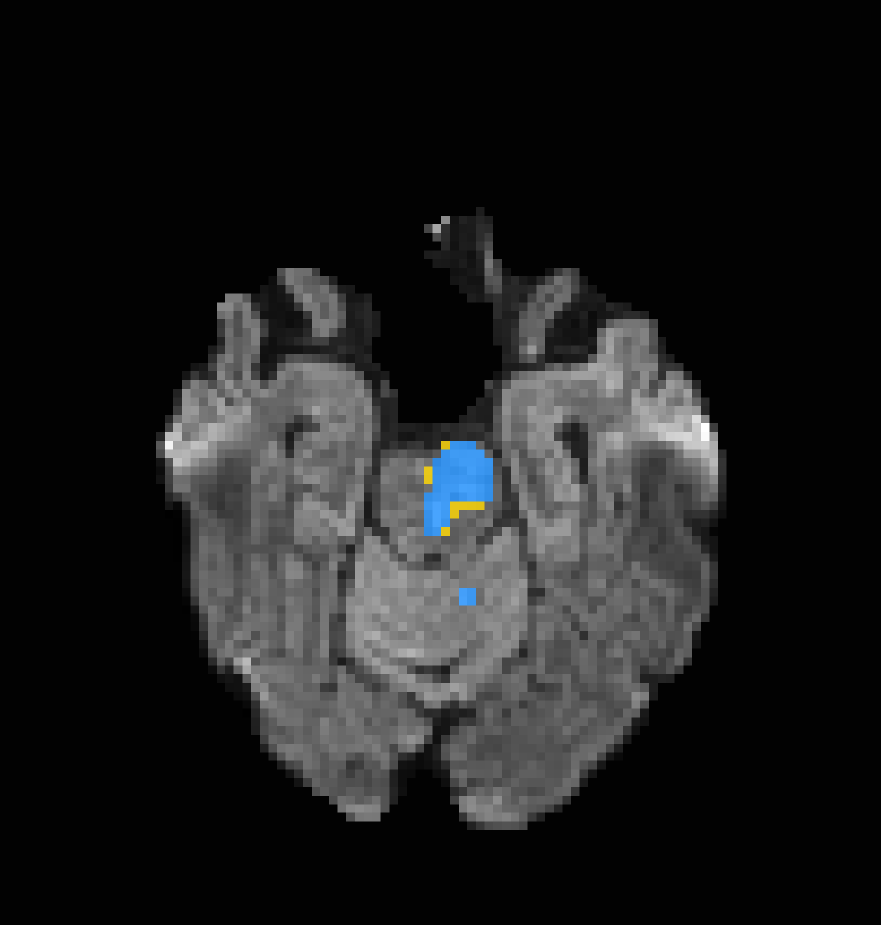}{60.3}
		\caption*{\scriptsize\textbf{nnU-Net}}
	\end{subfigure}
	
	\caption{Qualitative results on stroke lesion segmentation. All the examples are from the validation sets of the 5-fold cross-validation. UNETR and nnU-Net produce the false positive lesions circled in green (row 1 and 2), and nnFormer fails to detect the lesion circled in orange (row 3).}
	
	\label{fig: isles qualitative results}
\end{figure}

\paragraph{Qualitative Comparisons}
Qualitative comparisons of stroke lesion segmentation models are presented in Figure \ref{fig: isles qualitative results}. Compared to UNETR and nnU-Net, our Swin Factorizer and Global Factorizer models substantially reduce false positives, as observed in row 1.  Both UNETR and nnU-Net produce large regions of incorrect lesions circled in green, but nnU-Net suffers from fewer false positives than UNETR, as evident in row 2. 

Although nnFormer seems to yield relatively small false positive regions compared to UNETR and nnU-Net, Swin Factorizer not only produces even slightly fewer false positives but also enjoys more favorable results when it comes to false negatives. This is exemplified by row 3, where Swin Factorizer successfully captures both the stroke lesions, whereas nnFormer fails to detect the lesion circled in orange. Overall, Swin Factorizer displays very competitive segmentation results, superior to those of UNETR, nnFormer, and nnU-Net in most cases. These results verify the potential of Factorizer as an alternative to state-of-the-art models, such as nnFormer and nnU-Net.

\subsection{Ablation Studies: Training}  \label{sec: training ablation studies}

We conducted ablation studies on Factorizer to further investigate the effects of Factorizer subblocks and positional embedding. In this section, models with an ablated layer were trained from scratch on the BraTS dataset.

\paragraph{Factorizer Subblock} 
Ablation results of subblocks on brain tumor segmentation are reported in Table \ref{tab: block ablation results}. When we removed NMF subblocks from a Factorizer model, the performance substantially dropped while Factorizer models without MLP subblocks demonstrated less deterioration in results. Local Factorizer without MLP blocks yielded an average Dice of 82.67\%, still outperforming nnU-Net, and Swin Factorizer without MLP blocks yielded an average Dice of 83.57\%, which is significantly greater than that of nnU-Net and comparable with that of Res-U-Net. These results indicate the effectiveness of the NMF layer in improving the models.    

\begin{table}[t]
	\caption{The impact of each subblock on the performance of Factorizers on BraTS.}
	\label{tab: block ablation results}
	\newcolumntype{P}[1]{>{\centering\arraybackslash}p{#1}}
	\def\arraystretch{1.3}
	\resizebox{\textwidth}{!}{	
		\begin{tabular}{P{.12\textwidth} P{.15\textwidth} P{.15\textwidth} | *{4}{P{.06\textwidth}} | *{4}{P{.05\textwidth}}}		
			\toprule
			
			\multicolumn{1}{c}{\multirow{2}{*}{MLP Subblock}} & \multirow{2}{*}{NMF Subblock} & \multirow{2}{*}{Matricize} & \multicolumn{4}{c|}{Dice (\%) \textuparrow}           & \multicolumn{4}{c}{HD95 (mm) \textdownarrow}                 \\ \cline{4-11} 
			\multicolumn{1}{c}{}                           &                            &                            & ET & TC & \multicolumn{1}{c|}{WT} & Avg. & ET & TC & \multicolumn{1}{c|}{WT} & Avg. \\
			
			\hline   		
			\large\ding{51} &  \ding{55}                           &        \ding{55}                    &  77.02  &  80.89   & \multicolumn{1}{c|}{ 87.36}   &  81.76   &   6.55 &   8.90    & \multicolumn{1}{c|}{11.29}   &  9.02    \\
			\ding{55} &    \large\ding{51}                        &           Global                 &   77.51 &    80.78    & \multicolumn{1}{c|}{87.85}   &   82.05   &  8.82  &  11.24  & \multicolumn{1}{c|}{16.18}   &    12.33  \\
			\ding{55} &       \large\ding{51}                     &            Local                &    77.79 &    81.67 &    \multicolumn{1}{c|}{88.56} &     82.67  &   5.50 &   8.67 &   \multicolumn{1}{c|}{9.44} &    8.04 \\
			\ding{55}   &          \large\ding{51}               &           SW      &       78.60 &    82.61 &    \multicolumn{1}{c|}{89.51} &     83.57 &   5.00 &   7.47 &   \multicolumn{1}{c|}{7.86} &    6.87  \\
			
			\large\ding{51}   &          \large\ding{51}               &           SW  &        79.33 &    83.14 &    90.16    & \multicolumn{1}{|c|}{84.21}     &   4.91 &   7.31 &   \multicolumn{1}{c|}{8.23}  &  6.89  \\
			
			\bottomrule
		\end{tabular}
	}
\end{table}

\paragraph{Positional Embedding}
Table \ref{tab: positional embedding ablation results} shows the results of the ablation study on positional embedding. In all the cases, adding positional embedding to the bridge of a network improved the performance. Particularly, the average Dice score of Global Factorizer increased from 83.09\% to 83.24\% significantly with $ \text{p-value} = 0.031 $, and the average HD95 fell from 10.12 mm to 9.71 mm although this improvement is not statistically significant. Local and Swin Factorizers without positional embedding yielded an average Dice of 83.40\% and 83.99\%, respectively, slightly underperforming compared to their counterparts with positional embedding. Like Transformers, Factorizers lack any notion of voxel position, and therefore typically benefit from a positional embedding mechanism, which is generally consistent with our results. 

\begin{table}[t]
	\caption{The impact of positional embedding on the performance of Factorizers on BraTS (the asterisks indicate $\text{p-value} < 0.05 $ in the pairwise Wilcoxon signed-rank test between a model with and without positional embedding).}
	\label{tab: positional embedding ablation results}
	\newcolumntype{P}[1]{>{\centering\arraybackslash}p{#1}}
	\def\arraystretch{1.3}
	\resizebox{\textwidth}{!}{	
		\begin{tabular}{m{.2\textwidth} P{.17\textwidth} | *{4}{P{.07\textwidth}} | *{4}{P{.06\textwidth}}}		
			\toprule
			
			\multicolumn{1}{l}{\multirow{2}{*}{Model}} & \multirow{2}{.17\textwidth}{Positional Embedding} & \multicolumn{4}{c|}{Dice (\%) \textuparrow}           & \multicolumn{4}{c}{HD95 (mm) \textdownarrow}                 \\ \cline{3-10} 
			\multicolumn{1}{c}{}        &                      & ET & TC & \multicolumn{1}{c|}{WT} & Avg. & ET & TC & \multicolumn{1}{c|}{WT} & Avg. \\
			
			\hline
			
			%
			
			Global Factorizer &       \large\ding{55}    &  77.98  &  82.54  & \multicolumn{1}{c|}{88.76}   &   83.09   &  7.57  &  9.91   & \multicolumn{1}{c|}{11.73}   &   10.12   \\
			Global Factorizer   &          \large\ding{51}   &  78.20 &    82.86 &    \multicolumn{1}{c|}{88.65}    & \textbf{83.24}$^*$   &   6.67 &   9.53 &  \multicolumn{1}{c|}{11.69}  &  \textbf{9.71} \\
			
			\hline
			
			Local Factorizer &       \large\ding{55}    &  78.24  &  82.59  & \multicolumn{1}{c|}{89.37}   &   83.40   &  5.87  &  8.75  & \multicolumn{1}{c|}{9.98}   &   8.31     \\
			Local Factorizer   &          \large\ding{51}     &   78.67 &    82.85 &    \multicolumn{1}{c|}{89.30}    &  \textbf{83.61}  &  5.29 &   7.77 &   \multicolumn{1}{c|}{8.91}  &  \textbf{7.41}      \\

			\hline
			
			Swin Factorizer &       \large\ding{55}    &  78.63  &  83.18  & \multicolumn{1}{c|}{90.16}   &   83.99   &  5.50  &  7.24   & \multicolumn{1}{c|}{8.03}   &  6.97    \\
			Swin Factorizer   &          \large\ding{51}  &   79.33 &    83.14 &    \multicolumn{1}{c|}{90.16}    & \textbf{84.21}  &   4.91 &   7.31 &   \multicolumn{1}{c|}{8.23}    &  \textbf{6.89} \\
			
			\bottomrule
		\end{tabular}
	}
\end{table}

\subsection{Ablation Studies: Inference}  \label{sec: inference ablation studies}

Since the output of an NMF layer is a low-rank approximation of the input, it makes sense to perform an ablation study in the inference phase, for instance by short-circuiting an NMF layer. For all the experiments in this section, we ablated some NMF layers or changed their settings in the inference phase after training the model on the BraTS dataset. 

\begin{figure}[t]
	\centering
	\begin{subfigure}{.5\textwidth}
		\centering
		\includegraphics[width=.95\textwidth]{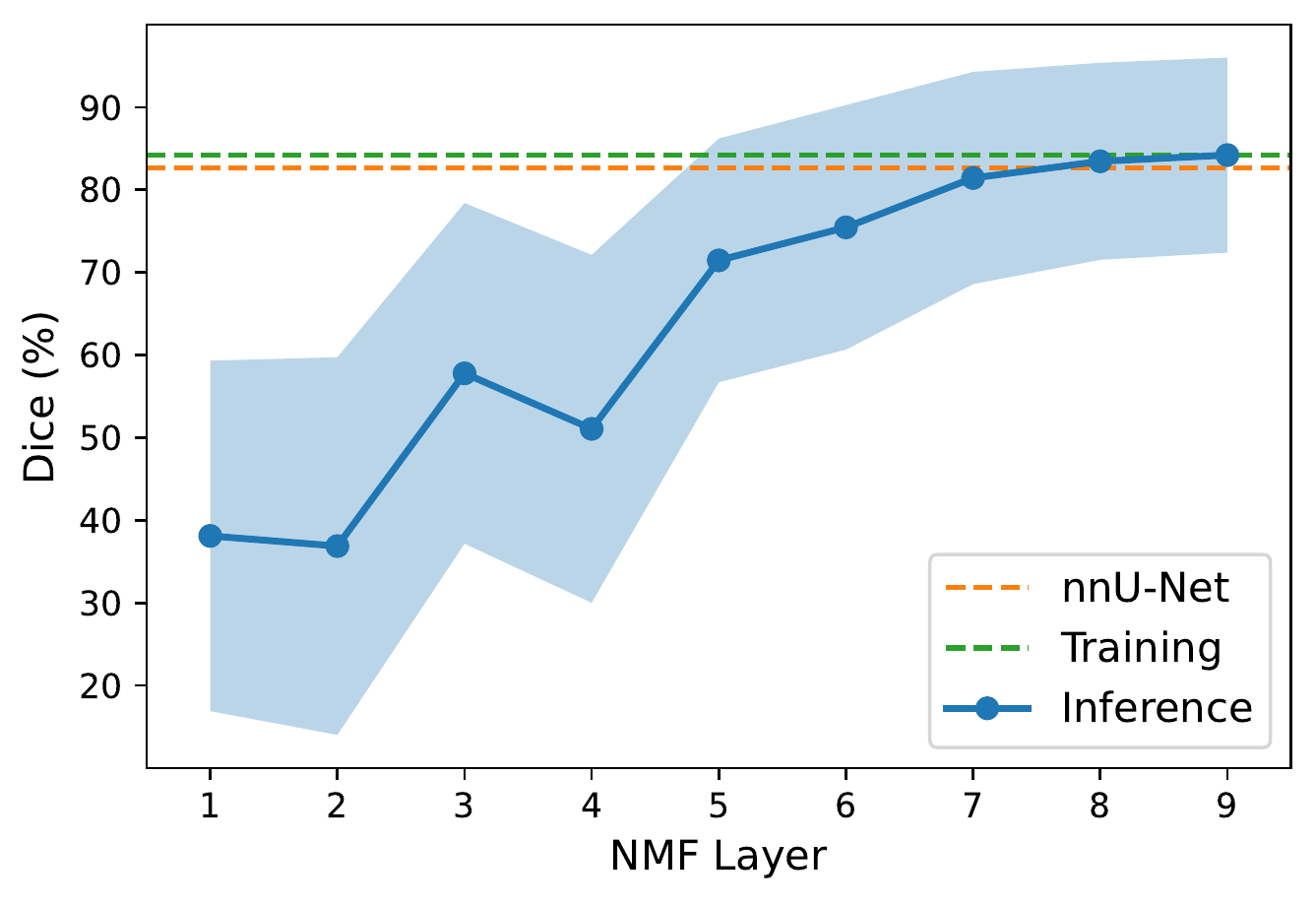}
		\caption{}
		\label{fig: dice_vs_keeptill}
	\end{subfigure}%
	\begin{subfigure}{.5\textwidth}
		\centering
		\includegraphics[width=.95\textwidth]{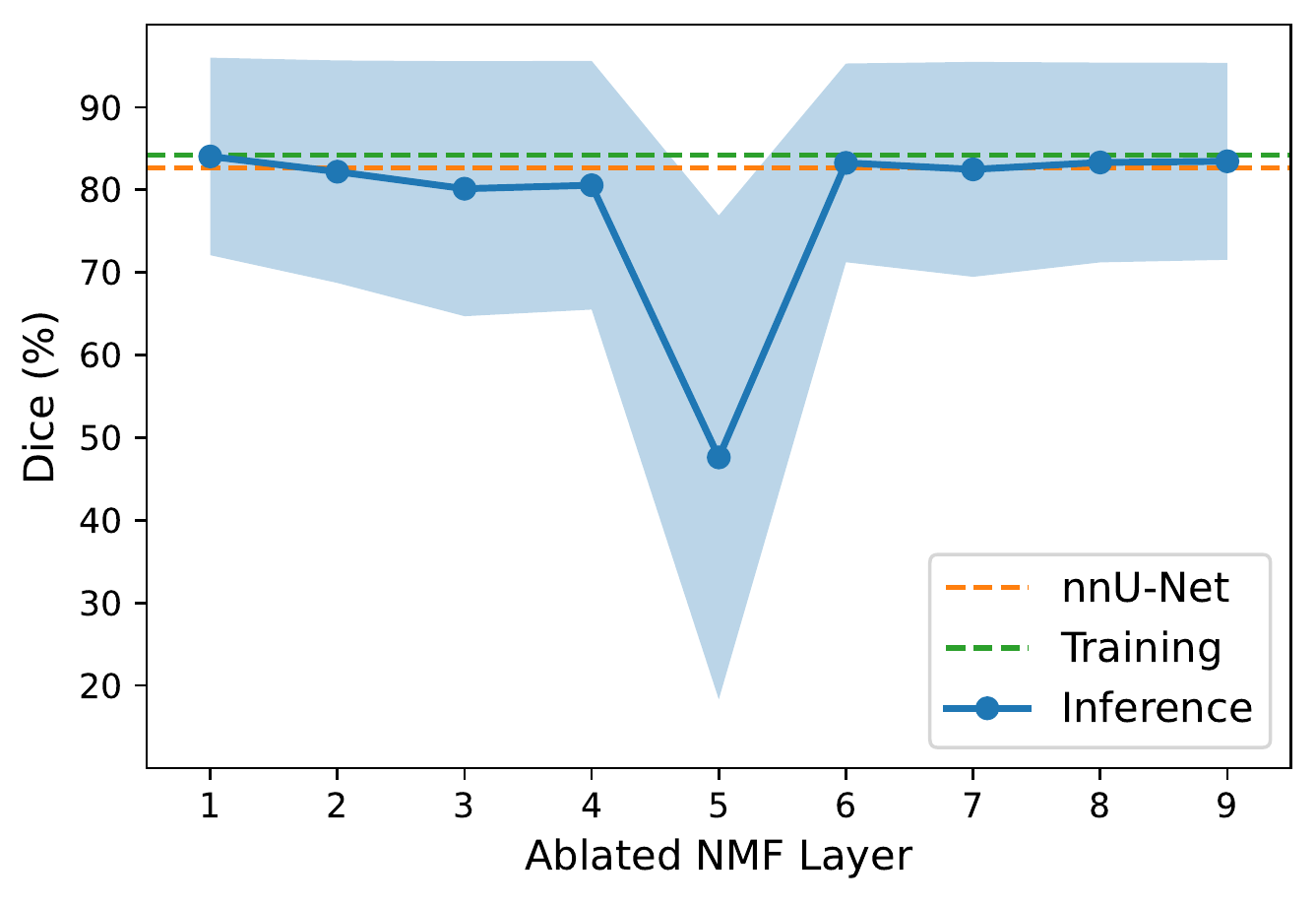}
		\caption{}
		\label{fig: dice_vs_rm}
	\end{subfigure}
	
	\begin{subfigure}{.5\textwidth}
		\centering
		\includegraphics[width=.95\textwidth]{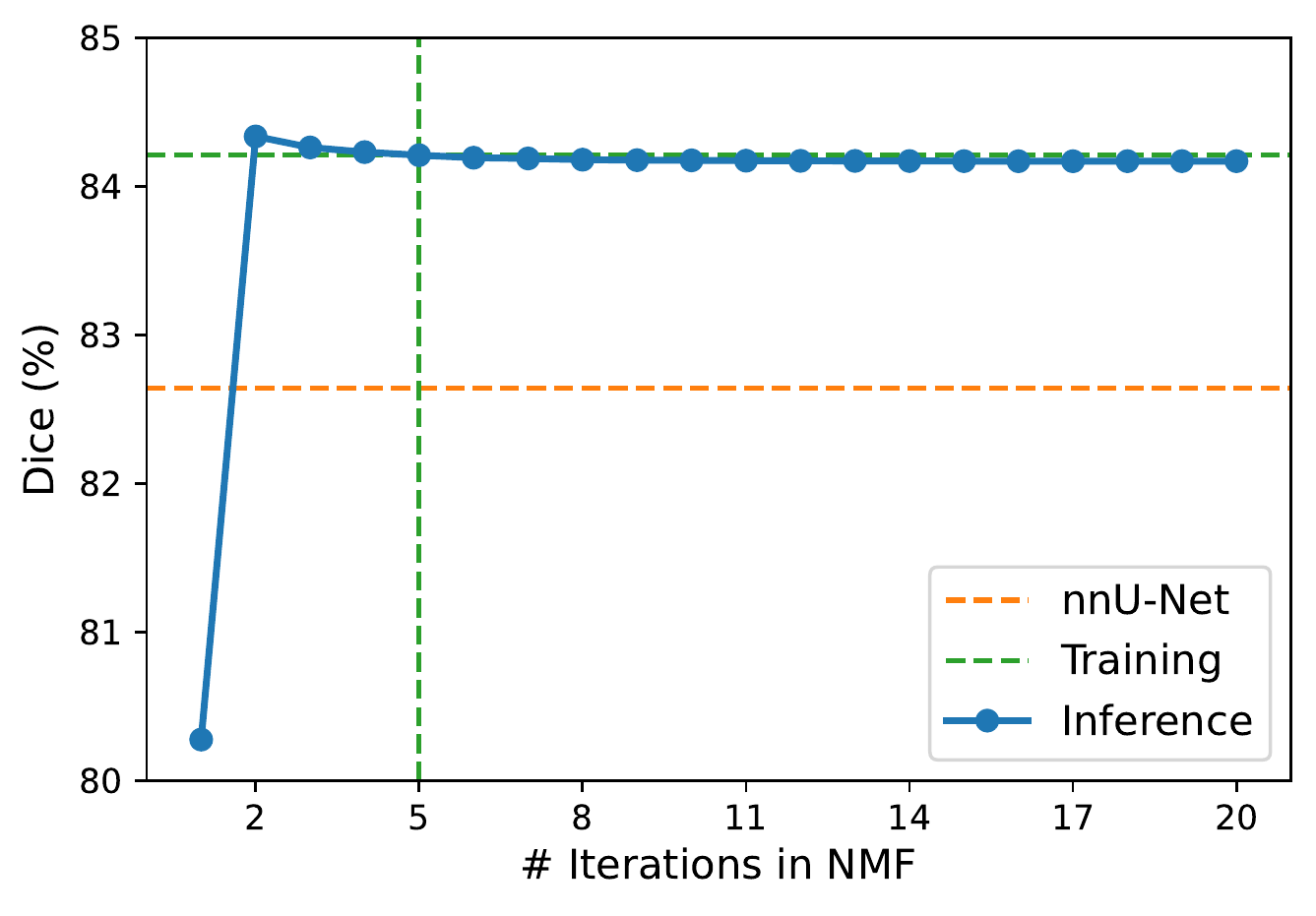}
		\caption{}
		\label{fig: dice_vs_iters}
	\end{subfigure}%
	\begin{subfigure}{.5\textwidth}
		\centering
		\includegraphics[width=.95\textwidth]{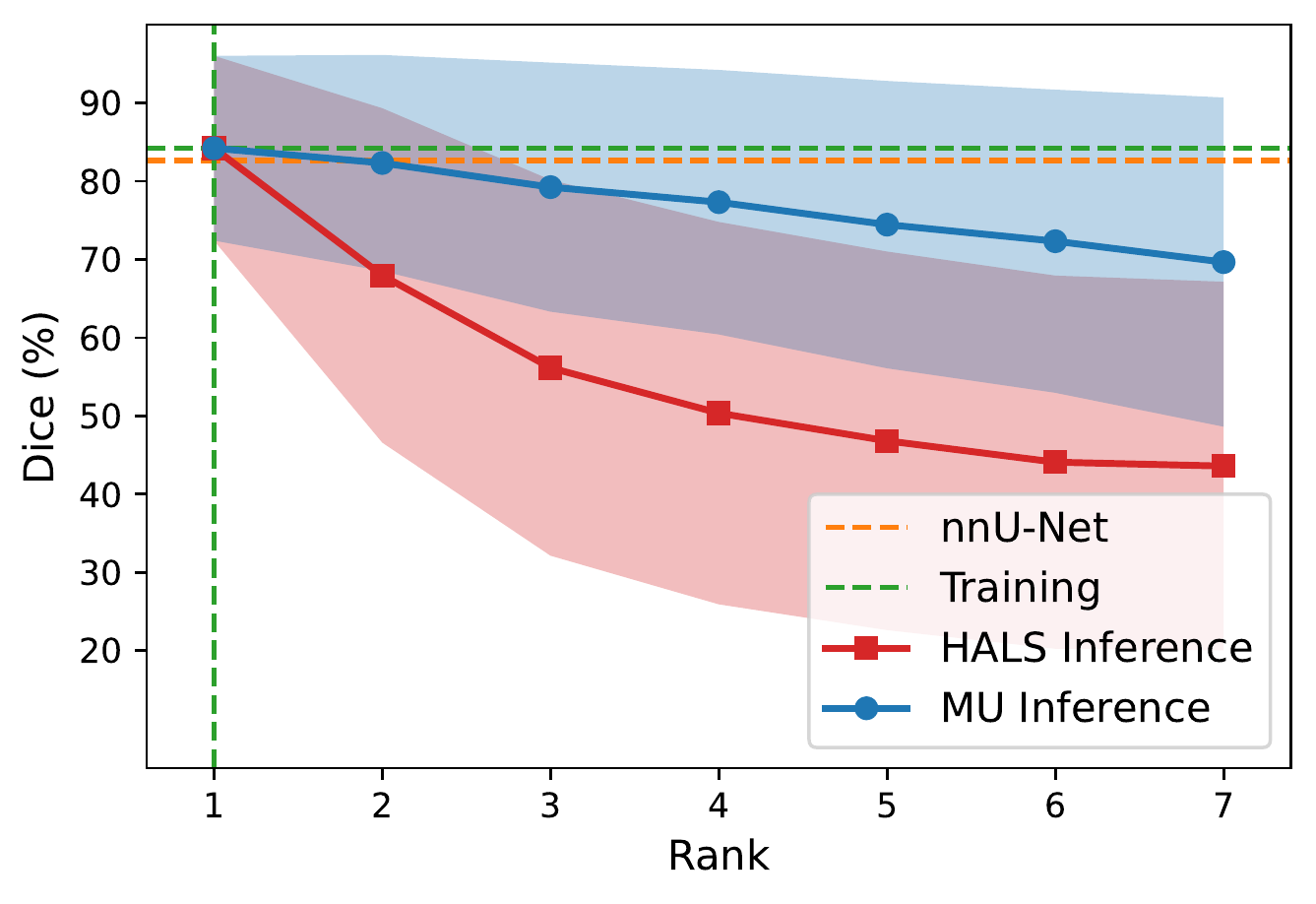}
		\caption{}
		\label{fig: dice_vs_rank}
	\end{subfigure}
	\caption{Inference-phase ablation experiments with Swin Factorizer pre-trained on BraTS. Panel (a) shows the average Dice score when only the first NMF layers are kept, while Panel (b) shows the Dice score plotted versus the NMF layer removed. Panel (c) shows the plot of Dice against the number of outer iterations in HALS. Panel (d) shows the impact of rank on performance in HALS and MU. The dashed orange lines indicate the nnU-Net Dice score, while the dashed green lines correspond to the pre-trained Swin Factorizer model.}
	\label{fig: inference ablation swin factorizer}
\end{figure}

\paragraph{NMF Layer} 
We investigated the impact of short-circuiting some NMF layers of the pre-trained Swin Factorizer model in the inference phase. In Figure \ref{fig: dice_vs_keeptill}, the average Dice score on BraTS is shown when we kept the first NMF layers and removed (or short-circuited) the rest. As expected, the more layers we kept, the higher the Dice score was achieved. We observed that most of the performance is achieved via the encoder, which includes the first five blocks. Figure \ref{fig: dice_vs_rm} shows the results of investigating the impact of each individual layer, where we kept all the layers except one at a time. Interestingly, we noticed that the NMF layer of the bridge block (layer 5) makes the greatest contribution to the performance. In fact, if an NMF layer except that of the bridge is ablated, the average Dice still stays above 80\%. Particularly, if an NMF layer in the decoder (layers 6 to 9) is ablated, the performance is still better than that of nnU-Net. Note that removing an NMF layer, especially those at higher-resolution stages of the network, can significantly reduce the computational complexity and speed up the inference time.   

\paragraph{NMF Solver Iterations} 
We investigate the effect of changing the number of outer iterations ($ T $) in HALS. Recall that all the Factorizer models were originally trained using HALS with $ T = 5 $. Figure \ref{fig: dice_vs_iters} shows the results of experimenting with $ T \in \{1, \dots, 20\} $ in the inference phase for the Swin Factorizer model pre-trained with $ T = 5 $. We noticed that for $ T > 1 $, the performance is very close to that of the original model ($ T = 5 $). Interestingly, $ T \in \{2, 3, 4\} $ yielded even higher average Dice scores compared to the original model although the improvement was not statistically significant. We attribute this to the possible regularization effect of reducing $ T $. Note that we proportionally decreased the computational cost of NMF layers by reducing $ T $ while preserving the accuracy. In fact, by training a Factorizer, we also obtain lightweight yet accurate versions of that for free without needing to re-train any model from scratch. All we need is to reduce $ T $ or ablate some costly NMF layers in the inference phase. When it comes to model speed-up, this brings a great advantage to Factorizers over CNNs and Transformers, which require much more complex mechanisms to achieve their faster versions.  

\paragraph{Rank} 
We investigated the impact of changing the rank of the pre-trained Swin Factorizer model in the inference phase. Recall that Swin Factorizers were trained with $ R = 1 $, in which both HALS and MU lead to the same update rule. Therefore, for $ R > 1 $ in inference, we experimented with both HALS and MU in order to make a comparison. In Figure \ref{fig: dice_vs_rank}, the average Dice score on BraTS is plotted as a function of rank. As observed, the more the rank deviated from $ R = 1 $ (the one used in training), the more significantly the Dice score dropped. In comparison to HALS, MU demonstrated a less dramatic reduction in performance as we increased the rank. This can be due to the fact that within the same number of outer iterations, HALS typically makes a larger decrease in the NMF objective (given in equation \eqref{eq: nmf objective}), causing the HALS-based model to deviate further from the original model than that of MU.

\section{Conclusion and Future Work} \label{sec: conclusion}

Vision Transformers, particularly those with hierarchical architectures, have recently achieved results comparable with state-of-the-art CNNs on various computer vision tasks. Nevertheless, the lack of locality inductive bias makes them underperform their CNN counterparts in low-data regimes, which is usually the case in medical image segmentation. Moreover, the quadratic complexity of attention makes existing Transformers apply self-attention layers only after somehow reducing the image resolution, and thus, fail to fully capture long-range contexts present at higher resolutions. Hence, this paper introduces a family of models, called Factorizer, which leverages the power of low-rank approximation for developing a scalable interpretable approach to context modeling by formulating NMF as a differentiable layer integrated into an end-to-end U-shaped architecture. Built upon NMF and shifted window idea, Swin Factorizer competed favorably with CNN and Transformer baselines in terms of accuracy and scalability. Swin Factorizer yielded state-of-the-art results on BraTS for brain tumor segmentation; with Dice scores of 79.33\%, 83.14\%, and 90.16\% for enhancing tumor, tumor core, and whole tumor, respectively; and on ISLES'22 for stroke lesion segmentation, with a Dice score of 76.49\%. Our experiments indicated that NMF components are highly meaningful, which gives a great advantage to Factorizers over CNNs and Transformers in terms of interpretability. Moreover, our ablation studies revealed a distinctive feature of Factorizers that allows the speed-up of inference for a pre-trained Factorizer with no extra steps and without sacrificing much accuracy.

Matrix factorization models are very flexible and versatile. In this work, we used an ordinary NMF model together with some matricization techniques to model local or global contexts. One possible extension of this work is to customize the NMF objective to simultaneously exploit both local and global contexts. Moreover, it would be useful to explore some ideas for automated selection of the rank hyperparameter, for example, by taking a greedy approach to NMF, a.k.a. Nonnegative Matrix Underapproximation, where the components are constructed and added one by one (sequentially) until some criteria are met. This can benefit especially Local Factorizer as different regions happen to have different optimal ranks depending on their contexts. We will explore the effectiveness of other NMF variants, such as Semi-NMF and Convex-NMF, which work on mixed-signed data matrices and relax the need for the ReLU activation function before factorization. Finally, while this paper focuses on the segmentation of 3D medical images, Factorizer may also potentially serve as an effective approach for efficiently processing high-resolution 2D medical or natural images, which can be further investigated in the future.

\section*{Declaration of Competing Interest}

The authors declare that they have no known competing financial interests or personal relationships that could have appeared to influence the work reported in this paper.

\section*{CRediT Authorship Contribution Statement}

\textbf{Pooya Ashtari}: Investigation, Conceptualization, Methodology, Software, Validation, Visualization, Writing - original draft, Writing - review \& editing. \textbf{Diana Sima}: Supervision, Writing - review \& editing. \textbf{Lieven De Lathauwer}: Supervision, Writing - review \& editing. \textbf{Dominique Sappey-Marinier}: Supervision, Writing - review \& editing. \textbf{Frederik Maes}: Supervision, Writing - review \& editing. \textbf{Sabine Van Huffel}: Supervision, Conceptualization, Writing - review \& editing, Funding acquisition.

\section*{Acknowledgments}

The research leading to these results has received funding from EU H2020 MSCA-ITN-2018: INtegrating Magnetic Resonance SPectroscopy and Multimodal Imaging for Research and Education in MEDicine (INSPiRE-MED), funded by the European Commission under Grant Agreement \#813120. This research also received funding from the Flemish Government (AI Research Program). Sabine Van Huffel, Frederik Maes, and Pooya Ashtari are affiliated to Leuven.AI - KU Leuven institute for AI, B-3000, Leuven, Belgium.

\appendix
\section{Data Augmentation Pipeline} \label{app: data augmentation pipeline}

The MONAI library \citep{monai} was used for data augmentation. The following random transforms were applied on the fly during training in the following order:
\begin{enumerate}	
	\item  \textbf{Affine transform}. An affine transform was applied with a probability of 0.15, where the angles of rotation (in degrees) and the scaling factor were drawn from $ \mathcal{U}(-30, 30) $ and $ \mathcal{U}(0.7, 1.3) $, respectively.
	
	\item \textbf{Flip}. All patches were flipped with a probability of 0.5 along each spatial dimension.
	
	\item \textbf{Gaussian Noise}. Each voxel of a patch was independently subject to additive white Gaussian noise with a variance drawn from $ \mathcal{U}(0, 0.1) $. This augmentation was applied with a probability of 0.15. 
	
	\item \textbf{Gaussian Smoothing}. Each modality of a patch was filtered independently via a symmetric Gaussian kernel whose width (in voxels) was drawn from $ \mathcal{U}(0.5, 1.5) $. This augmentation was applied with a probability of 0.15. 
	
	\item \textbf{Scale Intensity}. Voxel intensities were scaled with $ s \sim \mathcal{U}(0.7, 1.3) $ with a probability of 0.15.
	
	\item \textbf{Shift Intensity}. Voxel intensities were shifted with offsets drawn from $ \mathcal{U}(-0.1, 0.1) $ with a probability of 0.15.
	
	\item \textbf{Adjust Contrast}. Gamma correction with $ \gamma \sim \mathcal{U}(0.7, 1.5) $ was applied voxel-wise with a probability of 0.15.
\end{enumerate}

\section{Matricize Implementation Details} \label{app: matricize implementation details}

Algorithm \ref{alg: local-swin-matricize} presents PyTorch implementations of Local and SW Matricize modules. Here, we use the \texttt{einops} library to reshape tensors simply by declaring the corresponding Einstein notations.

\begin{algorithm}[t!]
	\caption{PyTorch pseudocode of Local and SW Matricize modules. \label{alg: local-swin-matricize}}
	\vspace{5pt}
	\small
\begin{minted}{python}
import torch
from torch import nn
from einops.layers.torch import Rearrange


class LocalMatricize(nn.Module):
    def __init__(self, size, head_dim=8, patch_size=(8, 8, 8)):
        super().__init__()
        C, H, W, D = size
        c2, (h2, w2, d2) = head_dim, patch_size
        c1, h1, w1, d1 = C // c2, H // h2, W // w2, D // d2
        dims = dict(c1=c1, h1=h1, w1=w1, d1=d1, c2=c2, h2=h2, w2=w2, d2=d2)

        eq = "b (c1 c2) (h1 h2) (w1 w2) (d1 d2) -> (b c1 h1 w1 d1) c2 (h2 w2 d2)"
        self.unfold = Rearrange(eq, **dims)

        eq = "(b c1 h1 w1 d1) c2 (h2 w2 d2) -> b (c1 c2) (h1 h2) (w1 w2) (d1 d2)"
        self.fold = Rearrange(eq, **dims)

    def forward(self, x):
        # x: (B, C, H, W, D)
        return self.unfold(x)

    def inverse_forward(self, x):
        # x: (B, C, L)
        return self.fold(x)


class SWMatricize(nn.Module):
    def __init__(self, size, head_dim=8, patch_size=(8, 8, 8), shifts=(4, 4, 4)):
        super().__init__()
        self.local_matricize = LocalMatricize(size, head_dim, patch_size)
        self.shifts = shifts
        self.shifts_inv = tuple(-s for s in patch_size)

    def forward(self, x):
        # x: (B, C, H, W, D)
        out1 = self.local_matricize(x)
        out2 = torch.roll(x, self.shifts, (2, 3, 4))
        out2 = self.local_matricize(out2)
        return torch.cat((out1, out2))

    def inverse_forward(self, x):
        # x: (B, C, L)
        B = x.shape[0]
        out1 = self.local_matricize.inverse_forward(x[: (B // 2)])
        out2 = self.local_matricize.inverse_forward(x[(B // 2) :])
        out2 = torch.roll(out2, self.shifts_inv, (2, 3, 4))
        return 0.5 * (out1 + out2)
\end{minted}
\end{algorithm}

\clearpage
\bibliography{ref}
	
\end{document}